\documentclass[11pt]{article}
\usepackage{jheparxiv}
\usepackage{hyperref}
\hypersetup{
    colorlinks,
    citecolor=black,
    filecolor=black,
    linkcolor=black,
    urlcolor=black
}

\usepackage[latin1]{inputenc}
\usepackage{amsmath}
\usepackage{dsfont}
\usepackage{mathrsfs}
\usepackage{graphicx}
\usepackage{amssymb}
\usepackage{enumerate}
\usepackage[usenames,dvipsnames,svgnames,table,x11names]{xcolor}
\usepackage{cleveref}
\usepackage[colorinlistoftodos]{todonotes}
\usepackage{blkarray}
\usepackage{tikz}
\usepackage{booktabs}
\usetikzlibrary{calc}
\usepackage{graphicx}
\usepackage{booktabs,multirow,tabularx} 
\usepackage{slashed}
\usepackage{comment}
\usepackage{upgreek}
\usepackage{diagbox}
\usetikzlibrary{calc}
\usepackage{booktabs,multirow,tabularx} 
\usepackage{nicefrac}

\numberwithin{equation}{section}

\newcommand{\half}{\nicefrac{1}{2}}
\newcommand{\A}{\mathbb{A}}
\newcommand{\C}{\mathbb{C}}
\newcommand{\HH}{\mathbb{H}}
\newcommand{\CP}{\mathbb{CP}}

\newcommand{\PT}{\mathrm{P}\mathbb{T}}
\newcommand{\PA}{\mathrm{P}\mathbb{A}}

\renewcommand{\P}{\mathbb{P}}

\newcommand{\M}{\mathbb{M}}

\newcommand{\T}{\mathbb{T}}
\newcommand{\Z}{\mathbb{Z}}
\newcommand{\p}{\partial}
\newcommand{\dbar}{\bar\partial}
\newcommand{\e}{\mathrm{e}}

\newcommand{\cE}{\mathcal{E}}

\newcommand{\cI}{\mathcal{I}}

\newcommand{\cN}{\mathcal{N}}
\newcommand{\cO}{\mathcal{O}}

\newcommand{\cZ}{\mathcal{Z}}
\newcommand{\fZ}{\mathcal{Y}}

\renewcommand{\P}{\mathbb{P}}
\newcommand{\SL}{\mathrm{SL}}
\newcommand{\SO}{\mathrm{SO}}
\newcommand{\Spin}{\mathrm{Spin}}

\newcommand{\SU}{\, \mathrm{SU}}

\newcommand{\diag}{\, \mathrm{diag}}
\newcommand{\rd}{\, \mathrm{d}}

\newcommand{\be}{\begin{equation}\label}
\newcommand{\ee}{\end{equation}}
\newcommand{\bea}{\begin{eqnarray}\label}
\newcommand{\eea}{\end{eqnarray}}

\newcommand{\la}{\langle}
\newcommand{\ra}{\rangle}

\newcommand{\sA}{{\scalebox{0.6}{$A$}}}
\newcommand{\sB}{{\scalebox{0.6}{$B$}}}
\newcommand{\sC}{{\scalebox{0.6}{$C$}}}

\definecolor{light-gray}{gray}{0.95}
\definecolor{light-grayII}{gray}{0.85}

\newcommand{\Tstrut}{\rule{0pt}{3.5ex}}         
\newcommand{\Bstrut}{\rule[-2ex]{0pt}{0pt}}   

\newcommand{\tikzmark}[1]{\tikz[overlay,remember picture] \node (#1) {};}
\newcommand{\DrawBox}[3][]{%
    \tikz[overlay,remember picture]{
    \draw[black,#1]
      ($(#2)+(-0.5em,2.0ex)$) rectangle
      ($(#3)+(0.75em,-0.75ex)$);}}

\newcommand{\Tr}{{\mathrm{Tr}}}
\newcommand{\ad}{{\mathrm{ad}}}

\title{\Large Massive ambitwistor-strings; twistorial models}
\author[a]{Giulia Albonico,}
\author[b]{Yvonne Geyer}
\author[a]{\& Lionel Mason}

\affiliation[a]{The Mathematical Institute, University of Oxford, Oxford OX2 6GG, UK}

\affiliation[b]{Department of Physics, Faculty of Science, Chulalongkorn University\\Thanon Phayathai, Pathumwan, Bangkok 10330, Thailand}
      
\emailAdd{giulia.albonico@maths.ox.ac.uk}
\emailAdd{yjgeyer@gmail.com}
\emailAdd{lmason@maths.ox.ac.uk}

\date{\today}
\abstract{Ambitwistor-strings are chiral strings whose targets are spaces of complex massless particles, and  whose correlation functions directly lead to simple, compact formulae for scattering amplitudes and loop integrands for massless gauge and gravity theories. 
This article extends this framework to worldsheet models for massive particles in 4d, obtained via a symmetry reduction of a higher dimensional massless model. The target space of the resulting models turns out to be the phase space of 4d massive particles in a twistorial representation, and so the worldsheet theory agrees with the two-twistor string previously introduced by the authors. However, the paper has been written so as to be largely self-contained.  We discuss two interesting classes of massive theories in detail.  
For gauge theories, the reduction procedure is explicitly adapted to  supersymmetric gauge theories on the Coulomb branch.   For supergravity theories, the reduction is adapted to give theories obtained via Cremmer, Scherk \& Schwartz  (CSS) reduction, with broken supersymmetry and massive multiplets.  The reduction procedure gives explicit and systematic rules to obtain amplitudes for all these theories and their amplitudes from two  compact master formulae that have their origins in 6d based on the polarized scattering equations; in the CSS case the formulae are new, and in both cases their derivation  is  systematic.
The freedom to include mass allows the definition of a loop insertion operator, thereby extending the formulae to 1-loop. Unlike the massless 4d twistorial models, these all display a perfect double copy structure, here   incorporating  massive particles in the relationship between gauge theory and CSS supergravity amplitudes.}

\begin{document}

\maketitle                  


\section{Introduction} 
The RNS ambitwistor-strings \cite{Mason:2013sva} are chiral string models whose targets are dimension agnostic  representations of the space of complex (super) light rays, \emph{ambitwistor space}. They very  directly lead to the compact  formulae for massless tree-level amplitudes of Cachazo, He and Yuan \cite{Cachazo:2013hca,Cachazo:2014xea} and loop integrands \cite{Geyer:2015bja, Geyer:2015jch}, and manifest  key amplitude structures such as the double copy and factorization.  They  are close cousins to the conventional string, sharing their critical dimensions and some of the worldsheet conformal field theory structures, but describe quantum field theories instead of the full string spectrum. 
While the RNS ambitwistor string produces beautifully compact formulae for bosonic amplitudes in any dimension, it poses some difficulties when it comes to the study of its fermionic sector and target space supersymmetry \cite{Adamo:2013tsa}.  It can however trace its origins to the twistor-strings of Witten \cite{Witten:2003nn}, Berkovits \cite{Berkovits:2004hg} and Skinner  \cite{Skinner:2013xp} that use twistorial representations of (ambi-) twistor space in 4d. These models, including the alternative models of \cite{Geyer:2014fka}, have the advantage that they incorporate fermions and supersymmetry very directly and lead to the RSVW formulae \cite{Roiban:2004yf, Cachazo:2004kj}, the Cachazo-Skinner formula \cite{Cachazo:2012pz,Cachazo:2012kg} 
that naturally manifest full supersymmetry, exploiting the spinorial nature of twistors.
However, despite the elegance of the tree-level amplitude formulae to which these twistorial models give rise, they have awkward aspects too. The anomalies of \cite{Geyer:2014fka} are not easy to eliminate, and those of \cite{ Witten:2003nn,Berkovits:2004hg,Skinner:2013xp} require extra worldsheet matter, the double copy structure is  obscured in all cases and they only exist for  a restricted range of massless theories in 4d. This obstructs, for example, the incorporation of loop amplitudes and mass into these models.  
See \cite{Geyer:2022cey} for a review and full references.

The incorporation  of mass into scattering amplitudes has become increasingly important, not only for particle physics applications, but also in the more recent applications of  amplitudes to gravitational wave computations; in these the black holes can be represented by massive particles or fields.   In \cite{Albonico:2022pmd} a massive ambitwistor model in 4d was introduced that includes  massive particles by changing the target to the space of (complex) massive geodesics.  This was represented twistorially by a pair of twistors following a construction by Penrose from the 1970s \cite{Penrose:1972ia}, see \cite{Kim:2021rda} for a recent discussion of this twistorial representation of the spinning massive particle.   This paper focuses on an alternative derivation of these models  by dimensional reduction from the twistorial models in 5d  in \cite{Geyer:2020iwz}. \footnote{These in turn were constructed from  formulae in 6d as described in \cite{Geyer:2018xgb, Albonico:2020mge} 
but which are not based on coherent 6d models.}  This allows us to follow the reduction of these models to provide the basic ingredients such as vertex operators and establish the anomalies and amplitude formulae.  We will see in particular that the anomalies behave much more like those of the RNS ambitwistor models described above, having their origins in consistent theories in 10 dimensions.  Furthermore, we will see that, unlike their massless counterparts, these 4d theories manifest the double copy, even for amplitudes with massive particles.

In an accompanying paper, we  describe symmetry reductions and their implementation in the context of the RNS ambitwistor string yielding dimension agnostic formulae, but without such easy representations for fermions or supersymmetry. In both papers we use the same strategy of performing the reductions by gauging  a combination of $P_4$, the  generator  on the worldsheet for the translation symmetry in the 5th dimension, with a further symmetry generator $j^H$ constructed from fields  from the rest of the worldsheet model;  gauging the combination  $P_4-j^H$  has the effect of assigning  a momentum in the 5th dimension to some charge associated to the symmetry $j^H$.  This leads to a  reduction prescription that gives a consistent mass assignment for internal propagating particles as the modulus of the corresponding charges.  The strategy is flexible and quite general, and we show that it  applies both to give the Coulomb branch for $\cN=4$ super-Yang-Mills in four dimensions  and the supergravity reductions of Cremmer, Schwarz and Scherk according to the choice of starting theory and symmetry generator.  Our formulae for the latter theories are new, whereas there are by now a number of formulae for Coulomb branch amplitudes in the literature \cite{Cachazo:2018hqa,Wen:2020qrj}.
Even for these, performing these reductions on the worldsheet gives an algorithmic procedure to write down amplitudes for the symmetry  reduced theories as correlators of  the gauged ambitwistor-string models.  In reducing massless worldsheet formulae from higher dimension to give massive   amplitudes in 4d as in \cite{Cachazo:2018hqa,Wen:2020qrj}, details such as  the signs of the higher momenta, that 
are resolved by guessing and checking.  Our construction here provides a systematic framework for identifying these.

The  flexibility to include off-shell propagators also allows us to introduce a \emph{gluing operator} that generates formulae for loop integrands at one loop for all these various theories. These were problematic to define for twistorial models because the gluing operator of \cite{Roehrig:2017gbt} relies on relaxing the constraint that the square $P^2$ of the  ambitwistor-string momentum operator $P_\mu$ should vanish.  However, $P^2=0$ identically in the twistorial models.  Instead we relax the constraint that gives the vanishing of $P_4$; indeed, this is already part of the construction of the massive amplitude formulae.

The plan of this paper is as follows.  In \S\ref{Background} we give sufficient background to make the paper self-contained.  We review the spinor-helicity and twistor and ambitwistor geometry,  together with the 4d twistorial realisation of the massless ambitwistor strings of  \cite{Geyer:2014fka} in \S\ref{4d-models-intro}, and those in 6d and 5d of \cite{Geyer:2020iwz} in \S\ref{subsec:6dmodels}. In \S\ref{sec:2TT-SymRed} we give the general framework of symmetry reduction via gauging currents on the worldsheet and re-derive the massive models in four dimensions of \cite{Geyer:2022cey} via this process.  This then leads to the corresponding  models, vertex operators, BRST gauge fixing and amplitude formulae as correlators.  \Cref{CB-section} gives the details of the models and reduction for Coulomb branch amplitudes, where $j^H$ is taken to be an element of the worldsheet current algebra that generates the gauge group.  From the path integral of these models, we derive the  compact formulae \eqref{final-formula}, supplemented by \eqref{measure} and \eqref{integrands}, supported on a massive version \eqref{massive-PSE} of the polarised scattering equations and with manifest supersymmetry for appropriate gauge and gravity  theories including massive particles.  A number of worked example amplitudes are included.  We further explain how these can be derived from each other by supersymmetric Ward identities. \Cref{sectionRsymred} discusses the reductions that arise when $j^H$ has been taken to be an $R$-symmetry generator for supersymmetric gauge and gravity theories.  In particular, we explain how this gives rise to the supergravity theories  with massive particles found by Cremmer, Scherk and Schwarz. In their original form, these arose by considering a massless reduction to 5d, and then the further reduction to 4d combined the translation reduction from 5d to 4d with a symmetry of the 5d internal manifold, analogous to our choice of $j^H$.  For us, $j^H$ is taken to be an $R$-symmetry generator, which can be the generator of a symmetry of the 5d internal manifold, but which can in practice give us more choice.
Having both gauge theory and gravity reductions allows us to fit the two sets of theories with massive particles into the double copy described in table \ref{table:DC} providing a worldsheet version of some of the discussion of \cite{Johansson:2019dnu}, although the worldsheet provides more routes to the double copy than momentum space.

\Cref{chapter:loops} explains how the ideas introduced earlier provide the necessary ingredients  to define one-loop amplitudes from a \emph{gluing operator}.  This extends a  technique introduced in \cite{Roehrig:2017gbt} for the RNS string that derives  one-loop amplitudes from a nodal Riemann sphere by defining them as worldsheet correlators that include a gluing operator at the node. 
We end the main part of the paper with a summary and discussion of a number of key issues.

\section{Twistorial ambitwistor strings}\label{Background}
We begin with a review of the twistorial ambitwistor string models. These rely on parametrisations of ambitwistor space that explicitly solve the $P^2=0$ constraint, whereas in the RNS model this constraint is gauged. As such, they are formulated in the language  of spinor-helicity variables, which we review in section \ref{SH-section-intro}: these are spinorial variables for on-shell momenta, i.e. for massless particles they solve $k^2=0$. In four, six and five dimensions, we  review how these variables are related to twistors and ambitwistors, allowing us to write actions for the ambitwistor string that rely on this parametrisation. In the following sections, we will use the five dimensional models as the starting point from which massive models can be obtained via a symmetry reduction.

\subsection{Spinor-helicity formalism}\label{SH-section-intro}
Spinor-helicity decompositions exploit the accidental isomorphisms of spin groups, and they are therefore specific to different dimensions. For both massless and massive particles in 4d, this has become an essential tool of modern amplitude methods and there are now a number of excellent reviews in the literature, e.g. \cite{Elvang:2013cua,Arkani-Hamed:2017jhn}.

\paragraph{Massless particles in four dimensions.}
In four dimensions we have the isomorphism $\Spin(4,\C)\simeq\SL(2,\C)\times\SL(2,\C)$.
Positive and negative chirality spinors transform under this group in the $(\half,0)$ and $(0,\half)$ representation respectively. Undotted and dotted indices label the left and right handed representations and can be raised and lowered with the Levi-Civita symbols $\varepsilon^{\alpha\beta}$ and $\varepsilon_{\dot \alpha\dot \beta}$ defining inner products:
\begin{equation}
    \la\lambda_1\lambda_2\ra=\varepsilon^{\alpha\beta}\lambda_{1\,\alpha}\lambda_{2\,\beta}=-\la\lambda_2\lambda_1\ra
    \qquad
    [\tilde\lambda_1\tilde\lambda_2]=\varepsilon_{\dot \alpha\dot \beta}\tilde{\lambda}_1^{\dot \alpha}\tilde{\lambda}_2^{\dot \beta}=-[\tilde\lambda_2\tilde\lambda_1]\,.
\end{equation}
The four-momentum $k^\mu$ transforms in the $(\half,\half)$ representation and can thus be mapped to an object carrying two spinor indices, one of each chirality:
\begin{equation}
    k_{\alpha\dot\alpha}=\sigma^\mu_{\alpha\dot\alpha}k_\mu\,.
\end{equation}
Then the on shell condition imposes $k^2=\det(k_{\alpha\dot\alpha})=m^2$. We will focus here on the $m=0$ case and discuss massive particles via dimensional reduction from higher dimension.
For a massless particle the momentum $k_{\alpha\dot\alpha}$ is a hermitian matrix of rank $1$ and can therefore be decomposed as the outer product of two complex chiral spinors:
\begin{equation}
    k_{\alpha\dot\alpha}=\kappa_\alpha\tilde\kappa_{\dot\alpha}\,.
\end{equation}
From the decomposition one can see that the little group $\SL(1,\mathbb{C})$ acts as:
\begin{equation}\label{LG-transf-4d-0}
    \kappa\rightarrow w^{-1}\kappa\qquad\tilde{\kappa}\rightarrow w\tilde{\kappa}\,.
\end{equation}
Polarization data corresponds to irreducible representations of the little group. For massless particles of helicity $h$, these are objects that scale as $w^{2h}$ under a little group transformation \eqref{LG-transf-4d-0}. Weyl spinors have polarization data $\epsilon\kappa$ and $\tilde \epsilon\tilde\kappa$. The Maxwell field strength $F_{\alpha\dot\alpha\beta\dot\beta}$ splits into an antiself-dual $F_{\dot\alpha\dot\beta}=\tilde{\epsilon}^2\tilde\kappa_{\dot\alpha}\tilde\kappa_{\dot\beta}$ and a self-dual $F_{\alpha\beta}=\epsilon^2\kappa_\alpha\kappa_\beta$ component corresponding to helicity $\pm1$ states and scaling accordingly.

\paragraph{Massless particles in six dimensions. } 
The spin group of the complexified Lorentz group in six dimensions is $\Spin(6,\C)\simeq \SL(4,\C)$. This group has fundamental $(\mathbf{4})$ and dual antifundamental $(\bar{\mathbf{4}})$ representations, giving two independent Weyl spinor representations. The simplest $\SL(4,\C)$ invariant is given by the singlet in $(\mathbf{4}\otimes\mathbf{\bar{4}})$:
\begin{equation}\label{SL4c-reps}
    \mathbf{4}\,:\,\nu_A\qquad\bar{\mathbf{4}}\,:\,\pi^A\qquad \mathbf{1}\,:\,\nu_A\pi^A\,.
\end{equation}
The only non-trivial invariant tensor is the four index object $\epsilon_{ABCD}$, which can be used to raise pairs of skew indices and to construct invariants:
\begin{equation}
    \la \kappa_1\kappa_2\kappa_3\kappa_4\ra=\kappa_{1A}\kappa_{2B}\kappa_{3C}\kappa_{4D}\epsilon^{ABCD}\qquad[ \kappa_1\kappa_2\kappa_3\kappa_4]=\kappa_{1}^{A}\kappa_{2}^{B}\kappa_{3}^{C}\kappa_{4}^{D}\epsilon_{ABCD}\,.
\end{equation}

A  six-vector $K_\mu$ in the fundamental $\mathbf{6}$ of $\SO(6,\mathbb{C})$ can be expressed as the antisymmetric product of two fundamentals or equivalently of two antifundamentals of the spin group. This identification, established through the chiral (skew) Pauli matrices $\sigma^\mu_{\sA\sB}$ is an isomorphism, a fact that no longer follows in higher dimensions. Because the matrix $K_{\sA\sB}$ is skew, it has even rank.  If $K^2=0$ it doesn't have full rank, so if non-zero it must have rank $2$ with basis $\kappa^a_A$ if antifundamental or $\kappa_{\dot a}^A$ if fundamental and these can be normalized up to $\SL(2,\C)$ by 
\begin{equation}
K^{\sA\sB}= \varepsilon ^{\dot a\dot b} \kappa_{\dot a}^\sA\kappa_{\dot b}^\sB\equiv\left[\kappa^\sA\kappa^\sB\right]\,,\qquad K_{\sA\sB}= \kappa^{ a}_\sA \kappa^{b}_\sB \varepsilon_{ a b}\equiv\left\langle\kappa_\sA\kappa_\sB\right\rangle\,.\label{eq:momenta}
\end{equation}
 These definitions hold up to two distinct $\SL(2,\mathbb{C})$ actions on the undotted and dotted indices of the Weyl spinors, so that these represent the little group $\mathrm{SO}(4, \C) \cong \mathrm{SL}(2, \C) \times \mathrm{SL}(2, \C) / \mathbb{Z}_{2}$. The $\varepsilon$s in \eqref{eq:momenta} define the skew inner products in each of the two copies of $\SL(2,\C)$ and lead to little group contractions denoted by $[\cdot,\cdot]$ and $\langle\cdot,\cdot\rangle$. \footnote{ This notation will also be used in 4d for contractions of chiral spinors too, but the distinction should be clear from the context.}

   As usual, irreducible representations of the little group provide  the polarization states for  plane wave solutions to massless   chiral and anti-chiral fields  with momentum $K_\mu$ in 6d. As described in 
\cite{Cheung:2009dc},
  A Dirac particle has polarization data $\epsilon_\sA=\epsilon_a\kappa^a_\sA$ because the Dirac equation gives $K^{\sA\sB}\epsilon_\sA=0$ and the $\kappa^a_\sA$ span the solutions.  A Maxwell 2-form field strength is represented in spinors by $F^\sA_\sB$, with $F_\sA^\sA=0$. For a momentum eigenstate, the Maxwell equations require $K_{\sA\sB}F_\sC^\sA=0=K^{\sA\sB}F^\sC_\sB$, so that we can write 
\begin{equation}\label{eq:pol}
F^\sA_\sB=\epsilon_\sB\epsilon^\sA\, , \qquad
\epsilon^\sA=\epsilon_{\dot a}\kappa^{\sA\dot a}\, , \qquad \epsilon_\sA=\epsilon_a\kappa_{\sA}^a\, .
\end{equation}
Thus the  polarization data is encoded in the product $\epsilon_a\epsilon_{\dot a}$.
\footnote{Note that $\epsilon_a$ and $\epsilon_{\dot a}$ cannot be taken to be real in Lorentz signature as the little group is $SU(2)\times SU(2)/\Z_2$.}

\paragraph{Massless particles in five dimensions.}
In order to dimensionally reduce to five dimensions, we pick a fixed non-null unit six-vector, $\Omega^{AB}$ in spinor form, and considers the five-dimensional plane $\mathbb{C}^5$ that is orthogonal to it, \cite{Cachazo:2018hqa,Albonico:2020mge,Geyer:2020iwz}. The choice of $\Omega$ breaks the spin group $\mathrm{SL}(4,\mathbb{C})\rightarrow \mathrm{Sp}(4,\mathbb{C})$, isomorphic to $\mathrm{Spin}(5,\mathbb{C})$, and allows one to raise and lower spinor indices using $\Omega^{AB}$ and $\Omega_{AB}=\frac{1}{2}\epsilon_{ABCD}\Omega^{AB}$.

Five-vectors then have the same spinor helicity decomposition as in six dimensions, with the additional constraint:
\begin{equation}\label{five-vector-orthogonality}
\Omega\cdot K = \Omega^{AB}(\kappa_A\kappa_B)=0\,.
\end{equation}
Because the fundamental and antifundamental representations are equivalent, the little group is now a single copy of  $\SL(2,\mathbb{C})$ and we denote little group contractions by $(\cdot,\cdot)$.

\paragraph{Massive and massless particles in 4d from massless particles in 5d.}\label{spinor-helicity-54d}
 We pick another fixed non-null unit vector $\Omega_2$ ($\Omega_1\cdot\Omega_2=0$). The reduction we seek constrains five dimensional momenta to obey:
\begin{equation}\label{5to4-momentum-constr}
\Omega_2\cdot K = M\,,
\end{equation}
where $M$ is a parameter related to the mass (although we allow for any sign of $M$) that will be determined  by the worldsheet model as we will explain in the next section.
The choice of $\Omega_2$ breaks the spin group $\mathrm{Sp}(4,\mathbb{C})\rightarrow\mathrm{SL}(2,\mathbb{C})\times\mathrm{SL}(2,\mathbb{C})\simeq\mathrm{Spin}(4,\mathbb{C})$, each factor acting separately on positive and negative chirality spinors. The index $A$ of $\mathrm{Sp}(4,\mathbb{C})$ decomposes accordingly into $(\alpha,\dot\alpha)$, one for each factor of the spin group. In the language of symmetry reduction, the component $\Omega_2\cdot K$ is the \textit{internal} momentum $\upkappa$. By the discussion above, the five dimensional momentum also satisfies $\Omega_1\cdot K=0$ so that we can pick a frame such that:
\begin{equation}\label{embedding-momentum}
K_{AB}=\left(\begin{array}{cc}
K\cdot\Omega_2\epsilon_{\alpha\beta} & k_{\alpha}^{\,\dot\beta} \\
k^{\dot\alpha}_{\,\beta} & K\cdot\Omega_2\epsilon^{\dot\alpha\dot\beta}
\end{array}\right),
\end{equation}
where $k_{\alpha\dot\beta}=k_{\alpha}^{\,\dot\gamma}\varepsilon_{\dot\gamma\dot\beta}$ is the spinorial form of the four dimensional massive momentum with mass $m=|\Omega_2\cdot K|=|M|$, as expected in the $(2,2)$ representation of the spin group. 

The spinor helicity decomposition of the massive momentum, as in \cite{Arkani-Hamed:2017jhn}, follows from the decomposition of $K_{AB}$:
\begin{equation}\label{LG-kappa}
k_{\alpha\dot\alpha}=\kappa_\alpha^a\tilde{\kappa}_{\dot\alpha a}\quad\operatorname{det} \kappa=\frac{1}{2}(\kappa_\alpha,\kappa^\alpha)=M=K\cdot\Omega_2= \operatorname{det} \tilde{\kappa}\,,
\end{equation}
where now $a=1,2$ is the single copy of  $\SL(2,\C)$ that we find for the massive little group.  Indices are raised and lowered by $\epsilon_{ab}=\epsilon_{[ab]}$, $\epsilon_{12}=1$. 
These spinor helicity variables consistently require $k^{2}=m^{2}=\operatorname{det}\left(k_{\alpha \dot{\alpha}}\right)=\operatorname{det} \kappa \operatorname{det} \tilde{\kappa}=M^2$. As before we denote little group contractions by $(\cdot\,,\,\cdot)$ and contractions of undotted and dotted 4d spinor indices as $\la\cdot\,,\,\cdot\ra$ and $[\cdot\,,\,\cdot]$ as is standard for four dimensional spinor-helicity.

Coming from higher dimension, it is natural to build representations out of the four dimensional Dirac spinor representation labelled by $A$ (rather than the chiral $\kappa_\alpha$).
We understand this as a reflection of the fact that massive particles are not chiral, so that for physical states Weyl spinors double up with their conjugates.
The polarization states of massive particles transform under irreducible representations of the massive little group, so spin-$s$ massive particles transform as the symmetric part of rank $2s$ tensors  with polarization data $\epsilon_{a_1\ldots a_{2s}}=\epsilon_{(a_1\ldots a_{2s})}$.
In particular we can write 
\begin{equation}
\psi_\sA=(\epsilon \kappa_A)\e^{ik\cdot x}=\epsilon_a\kappa_{\sA}^a\,\e^{ik\cdot x}\,, \qquad\kappa_{\sA a}=(\kappa_{\alpha a},\tilde\kappa_{\dot\alpha a})\,,
\end{equation}
for a massive Dirac field momentum eigenstate with polarization  $\epsilon_a$.
Similarly, a massive spin-1 field has polarization given by a symmetric $\epsilon_{(ab)}$ often written in factorized form as:
\begin{equation}
F_{\sA\sB}=\epsilon_{(a}\tilde{\epsilon}_{b)}\kappa^a_\sA\kappa^b_\sB\,\e^{ik\cdot x}\,.
\end{equation}
For generic spin-$s$ massive fields we will take the decomposition in Dirac indices:
\begin{equation}
\Psi_{\sA_1\dots \sA_{2s}}=\epsilon_{(a_1\dots a_{2s})}\kappa^{a_1}_{\sA_1}\cdots\kappa^{a_{2s}}_{\sA_{2s}}\,\e^{ik\cdot x}\,.
\end{equation}

Massless particles are naturally embedded in this description by taking $M=m=0$, which can be achieved with the above spinor helicity variables by restricting to $\kappa_{2\alpha}=0=\kappa_1^{\dot\alpha}$, so that the little group indices $a=1,2$ correspond to self-dual and anti-self-dual polarizations.

\subsection{Twistors and ambitwistors in four dimensions}\label{twistors-AT-4d-sec}
In four dimensions, twistor space $\PT$ is an open subset of $\C\P^3$, which we parametrise by homogeneous coordinates $Z^A$. These coordinates carry a natural action of $\SL(4,\C)$, the conformal group in 4d, and are closely related to the 6d spin group of the last section, since the 4d conformal group can be taken as (the double cover of) $\SO(6,\C)$ acting on the 6d projective light cone via the embedding formalism. The connection with space-time is established by a geometric correspondence between the space of lines in $\C\P^3$ and the complexified compactified Minkowski space-time represented as a quadric $Q$ in $\C\P^5$. It is usual to represent a twistor as a pair of spinors
\begin{equation}
    Z^A=(\lambda_\alpha, \mu^{\dot{\alpha}})\,.
\end{equation}
A point $x^{\alpha\dot\alpha}$ in finite Minkowski space $\C^4$ corresponds to the line in twistor space given by  the incidence relations
\begin{equation}
    \mu^{\dot{\alpha}}=ix^{\alpha\dot{\alpha}}\lambda_\alpha\,.
\end{equation}
  We can similarly define dual twistors $\tilde{Z}\in\PT^*$ as a pair of spinors by $ Z^A=(\tilde\lambda_{\dot\alpha}, \tilde\mu^{{\alpha}})$ with dual incidence relation
 \begin{equation}
   \tilde \mu^{\alpha}=-ix^{\alpha\dot{\alpha}}\tilde\lambda_{\dot\alpha}\,.
\end{equation}
and
 with the duality expressed by the inner product
\begin{equation}
    Z \cdot \tilde{Z}:=\lambda_\alpha \tilde{\mu}^\alpha+\mu^{\dot{\alpha}} \tilde{\lambda}_{\dot{\alpha}}\,.
\end{equation}
Twistors and dual twistors can be used to give a parametrization of ambitwistor space; the phase space of null rays. Given a null momentum $P$, we can seek a spinor-helicity decomposition of a null momentum $P_{\alpha\dot\alpha}=\lambda_\alpha\tilde\lambda_{\dot\alpha}$. Then given a null geodesic with momentum $P=\lambda\tilde\lambda$ going through a point $x$, we can introduce:
\begin{equation}
    Z=(\lambda_\alpha, ix^{\alpha\dot\alpha}\lambda_\alpha)\in\PT\qquad
    \tilde Z=(\tilde\lambda_{\dot\alpha},-ix^{\alpha\dot\alpha}\tilde\lambda_{\dot\alpha})\in\PT^*\,.
\end{equation}
These are both incident by the incidence relations and don't change if $x$ is translated by $P$, so is independent of point $x$ on the geodesic.   
\begin{equation}
    Z\cdot\tilde Z=0\,,
\end{equation}
and one can check that these equations are also sufficient for the existence of a corresponding null geodesic (possibly at infinity).
The twistor and dual twistor have two independent scalings. When combined, they corresponds to the scale of the null geodesic, i.e., that of $P$. However, the difference generated by $\Upsilon=Z\cdot\partial_Z-\tilde{Z}\cdot\partial_{\tilde{Z}}$ is redundant and so can be quotiented out. Ambitwistor space is therefore expressed as a quotient of the product of twistor and dual twistor space:
\begin{equation}\label{AT-space-4d}
    \A=\left\{(Z,\tilde Z)\in\T\times\T^*|Z\cdot\tilde{Z}=0\right\}\big/\,\Upsilon\, .
\end{equation}
This is a symplectic quotient with symplectic potential represented by $Z\cdot d\tilde Z - \tilde Z \cdot d Z$. Quotienting by the scale of the null geodesic further reduces to $\PA\subset \PT\times\PT^*$. The Penrose transform relates massless on-shell fields on spacetime to $H^1$-cohomology classes on twistor and dual twistor space. On ambitwistor space, such  cohomology classes can be pulled back from $\PT$ and $\PT^*$. We will discuss concrete representatives in \cref{sec:2TT-SymRed}.

\subsection{Four dimensional ambitwistor string}\label{4d-models-intro}

The bosonic action for the four-dimensional ambitwistor string of \cite{Geyer:2014fka} is based on the parametrisation of ambitwistor space as a quadric in the product of twistor and dual twistor space \eqref{AT-space-4d}. When constructing the model, twistors and dual twistors become fields on the worldsheet and they must be taken to have value in some line bundles $\mathcal{L}$ and $K_\Sigma\otimes\mathcal{L}^{-1}$ on $\Sigma$. Contrary to the twistor string \cite{Witten:2003nn,Skinner:2010cz}, which includes a sum over all positive degrees of the line bundle $\mathcal{L}$, the four-dimensional ambitwistor string fixes $\mathcal{L}= K_\Sigma^{\nicefrac{1}{2}}$ so that both the twistor and dual twistors are valued in $K_\Sigma^{\nicefrac{1}{2}}$, see \cite{Geyer:2016nsh,Geyer:2022cey} for a more detailed comparison. Beside this choice, both models are built on the action:
  \begin{equation}\label{action-4d-bos}
S_{\scalebox{0.7}{bos}}=\frac{1}{2 \pi} \int_{\Sigma} \tilde{Z} \cdot \bar{\partial}_e Z-Z \cdot \bar{\partial} \tilde{Z}+a\, Z \cdot \tilde{Z}\,,
\end{equation}
where the Lagrange multiplier $a\in\Omega^{(0,1)}(\Sigma)$ imposes the constraint $Z\cdot\tilde{Z}=0$ and gauges the transformations generated by $\Upsilon$. Identifying $K_\Sigma^{\nicefrac{1}{2}}$ with both the pullback to the worldsheet of the line bundle $\mathcal{O}_Z(1)$ on $\PT$ and $\cO_{\tilde{Z}}(1)$ on $\PT^*$ reduces the target space to projective ambitwistor space.
Worldsheet reparametrisations are gauged by $\bar\partial_e=\bar\partial+e\partial$, with $e\in T^{1,0}\Sigma\otimes \Omega^{0,1}_\Sigma$. Subsequently, we will gauge fix $e=0$, but the fermionic ghosts resulting from the BRST gauge fixing procedure will play an important role in the final amplitude formulae. 

Supersymmetry can be introduced into the bosonic model $S_{\scalebox{0.7}{bos}}$ by extending the twistor and dual twistor to their supersymmetric analogues. Here we employ a notation that will be natural in the context of higher dimensional models and dimensional reduction. We can repackage the degrees of freedom of both the twistor and dual twistor into one \textit{Dirac} supertwistor 
\begin{equation}
    \fZ=(\lambda_A,\mu^A,\eta^\mathcal{I})\, :\qquad \lambda_A=(\lambda_\alpha,\tilde \lambda_{\dot\alpha})\, , \quad \mu^A=(\mu^{\dot\alpha},\tilde\mu^{\alpha})\, , \quad \eta^{\mathcal{I}}=(\eta^I,\tilde \eta_I)\, ,\label{Dirac-twistor}
\end{equation} 
where $\lambda_A$ and $\mu^A$ are Dirac spinors made up of the homogeneous chiral and antichiral components of $Z$ and $\tilde{Z}$. In the fermionic components $\eta^\mathcal{I}=(\eta^I,\tilde \eta_I)$,  $I=1,\ldots,\mathcal{N}$   is the R-symmetry index, with $\mathcal{N}=4$ for maximal super-Yang-Mills. Clearly, this notation is most suitable for the \emph{ambitwistor} string, where now $\fZ\in\Omega^0(\Sigma,K_\Sigma^{\nicefrac{1}{2}})$.
The supersymmetric analogue of \eqref{action-4d-bos} is then given by:
\begin{equation}\label{action-4d-bos-Dirac}
S=\frac{1}{2 \pi} \int_{\Sigma} \Omega(\fZ ,   \bar{\partial} \fZ) +a\, \fZ \cdot \fZ\,,
\end{equation}
with the inner product $\fZ \cdot \fZ=Z \cdot \tilde{Z}+\eta^i\tilde\eta_i$ and skew product $\Omega( \fZ_1 ,\fZ_2)=\cZ_1\cdot \tilde\cZ_2-\cZ_2\cdot \tilde\cZ_1$. Whilst the liberty to choose the degree of the line bundle $\mathcal{L}$ in \eqref{action-4d-bos} is normal in the twistor-string, the fact that it is fixed for  the ambitwistor string is more natural  in the action \eqref{action-4d-bos-Dirac}; and we should consider this to be the action $S_{\mathbb{A}}$ of the ambitwistor string, with \eqref{action-4d-bos} only providing an explicit component expansion.

Constructing the BRST charge $Q$, one can verify that the obstructions to $Q^2=0$ vanish for maximal supersymmetry and a choice of worldsheet matter with central charge $\mathfrak{c}=14$ for super Yang-Mills, which we take to be a current algebra.

\subsection{Twistors and ambitwistors in six dimensions}\label{subsec:6dAT}
In six dimensions, twistors are pure spinors of the conformal group $\SO(8,\C)$. This group has three eight-dimensional representations: two spinorial ones with opposite chirality and the vector. \emph{Triality} (most naively seen as the 3-fold symmetry of the Dynkin diagram of $\SO(8,\C)$) permutes these three representations into each other.

Both chirality spinors in 8d can be employed to define twistor spaces in 6d. They can be represented as pairs of six-dimensional spinors:
\begin{equation}
Z^{\mathcal{A}}=\left(\mu^A, \lambda_A\right) \in\bar{\mathbf{4}} \oplus \mathbf{4} , \quad \tilde{Z}_{\mathcal{A}}=\left(\tilde{\mu}_A, \tilde{\lambda}^A\right) \in \mathbf{4} \oplus \bar{\mathbf{4}}\,,
\end{equation}
where $A$ labels the fundamental and antifundamental representations of $\SL(4,\C)$ as in \eqref{SL4c-reps}. Both chiralities have natural inner products with themselves and the twistor space is defined to be the space of null (or pure) twistors, i.e., the quadric  $Q=0$   in the projectivisation of the chiral spinor representation of $\SO(8,\C)$ defined by \cite{Penrose:1986ca,Mason:2011nw,Chern:2009nt,Mason:2012va}:
\begin{equation}\label{twistor-space-6d}
Q=\left\{[Z] \in \mathbb{C P}^{7} \mid Z \cdot Z=2 \mu^{A} \lambda_{A}=0\right\}\,.
\end{equation}
Similarly one can define primed twistor space $Q'$, built on antichiral spinors. Here twistor space and primed twistor space are each dual to themselves through the canonical inner product in \eqref{twistor-space-6d}, in contradistinction  to the four dimensional case where primed twistor space is the dual of twistor space.

Complexified compactified 6d space-time is also by the standard embedding formalism, a quadric $\M\subset\CP^7$; this three-fold appearance of quadrics in $\C\P^7$ as distinct homogeneous  spaces of $\SO(8,\C)$ is an expression of triality as mentioned above.  Pure twistors have a non-local correspondence with  $\M$ whereby points of $Q$ correspond to totally null self-dual $3-$planes in $\M$ via the incidence relations:
\begin{equation}
\mu^A=x^{AB}\lambda_B\,.\label{eq:incidence_6d}
\end{equation}
Primed twistors correspond to the anti-self-dual 3-planes.

Unlike in 4d, complex light rays $\mathcal{L}$ in $\M$ are obtained by the intersection of a pair of self-dual 3-planes $\mathcal{L}=\alpha_1\cap\alpha_2$ corresponding to  a pair of twistors\footnote{A twistor and a primed twistor only intersect at a point or along a 2-plane.}
 $(Z_1,Z_2)$.  Such twistors only intersect if $Z_1\cdot Z_2=0$ and then the line joining them lies entirely in twistor space $Q$ and not just in $\CP^7$. The same null geodesic is obtained if any pair of distinct points along  the null lines in $Q$ is chosen and so ambitwistor space can be defined by 
\begin{equation}
\mathbb{A}_{6}=\left\{\left[Z^{a}\right] \in \CP^{7} \mid Z^{a} \cdot Z^{b}=0\quad a,b=1,2\right\} \big/\, \SL(2, \C)\,.\label{eq:bos_ambitwistor_space}
\end{equation}
This again is a symplectic quotient with symplectic potential now $(Z \cdot d Z)$.

This description can be extended to superambitwistor space by replacing twistors $Z$ with supertwistors $\cZ$. Supertwistor space is defined as the super-quadric $Q_{N}$ in $\mathbb{CP}^{7|2N}$:
\begin{equation}
Q_{N}=\left\{[\mathcal{Z}] \in \mathbb{C P}^{7} \mid \mathcal{Z} \cdot \mathcal{Z}=2 \mu^{A} \lambda_{A}+\omega_{IJ}\eta^I\eta^J=0\right\}\,,
\end{equation}
parametrized by $[\mathcal{Z}]=\left[\mu^{A}, \lambda_{A}, \eta^{I}\right]$. Here $\omega_{IJ}$ is a skew $2N\times2N$ matrix, $N=\mathcal{N}^{(5d)}$ is the number of supercharges in six and five dimensions with $N=2$ for 5d maximal super-Yang-Mills.
The incidence relations
\begin{equation}
\mu^A=x^{AB}\lambda_B+\omega_{IJ}\theta^{AI}\eta^J\qquad \eta^{I}=\theta^{A I} \lambda_{A}\,,
\end{equation}
establish the correspondence with chiral Minkowski superspace $\mathbb{C}^{6|8N}$, here parametrized by $(x^{AB},\theta^{AI})$ with $I=1,\dots 2N$. Superambitwistor space $\mathbb{A}_{6}$ is then defined as the natural supersymmetric analogue of \eqref{eq:bos_ambitwistor_space}.

In the language of triality,  ambitwistor space is the central node of the $D_4$ Dynkin diagram, with  points of ambitwistor space corresponding to complex projective lines in $Q$,  $Q'$ and $\M$ respectively. We have seen the first of these correspondences -- between points in $\mathbb{A}_6$ and projective lines in $Q$ -- explicitly in \eqref{eq:bos_ambitwistor_space}, and the second one follows straightforwardly from interchanging self-dual with anti-self dual 3-planes in $\M$. Finally, the last correspondence encodes the relation between points in $\mathbb{A}_6$ and null geodesics on $\M$.  The correspondence between  points of $Q$ and totally null 3-planes in $\M$ operates in reverse also and between  $Q'$ and $\M$ and between $Q$ and $Q'$ \cite{Geyer:2020iwz}.

\subsection{Models in six and five dimensions} \label{subsec:6dmodels}\label{subsec:5dmodels}
In light of the geometry presented in the previous section, we can proceed as in four dimensions and write a twistorial model that solves the $P^2=0$ constraint explicitly. 
In \cite{Geyer:2018xgb} a bosonic action was formulated that was further studied in \cite{Albonico:2020mge,Geyer:2020iwz};
\begin{equation}\label{6dmodel-bos}
S_{6 d}=\int_{\Sigma} \frac{1}{2} \left(Z \cdot \bar{D} Z\right) + S_m\,,
\end{equation}
where $\bar{D} Z^{a}=\dbar_e Z^{a}+A_{b}^{a} Z^{b}$, and $A_{b}^{a} \in \Omega^{0,1}\left(\Sigma, \mathfrak{s l} _{2}\right)$ is a worldsheet $(0,1)-$form gauging the $\mathfrak{s l} _{2}$ little group. 
As in 4d, deformations of the worldsheet complex structure are gauged via  $\dbar_e=(\bar{\partial}+e \partial)$, and we will eventually gauge-fix the Beltrami differential $e$ to zero. 

Unlike in four dimensions, there is  no choice but to take the twistors to be spinors on the worldsheet, as the $\mathfrak{s l} _{2}$  gauging in  the action \eqref{6dmodel-bos}  cannot change the degree. Again, we identify $K_{\Sigma}^{1 / 2}$  with the pullback to the worldsheet of $\mathcal{O}(1)\rightarrow\CP^7$, reducing the target space to projective ambitwistor space.
One can include $(N, 0)$ supersymmetry by replacing the twistors $Z^a$ with supertwistors $\cZ^a$. Finally, $S_m$ denotes the action for a choice of worldsheet `matter' system, such as a current algebra $S_C$, which we will discuss in more detail below and in subsequent chapters. As  should be clear from the discussion of triality above, two other distinct models exist based on the alternative representations of ambitwistor space as null lines in primed twistor space $Q'$ or in space-time $\mathbb{M}$ in six dimensions.

Ambitwistor string models in five dimensions are obtained by implementing the condition \eqref{five-vector-orthogonality} on the worldsheet:
\begin{equation}\label{6to5-translations-twist}
S_{5 d}=\int_{\Sigma} \frac{1}{2} \left(Z \cdot \bar{D} Z\right)+a\Omega_1^{AB}(\lambda_{A} \lambda_{B})+S_{m}\,,
\end{equation}
The field $a$ is a Lagrange multiplier for the constraint $(\lambda^{A} \lambda_{A})$ and it acts as a gauge field for the transformations:
\begin{equation}
\delta a =\bar\partial\alpha\qquad\delta\mu^{A a}=\alpha\Omega_1^{AB}\lambda^a_B\qquad\delta\lambda^a_A=0\,.
\end{equation}
From the incidence relations, we see that these correspond to translations in the $\Omega_1$ direction.\footnote{We label this direction $\Omega_1$ anticipating the further reduction we will perform in section \ref{sec:2TT-SymRed}.} These are generated by the Hamiltonian vector field $\Omega_1^{A B} \lambda_{A}^{a} \partial / \partial \mu^{B a}$ associated to the constraint $(\lambda^A\lambda_A)=0$. The gauging of this constraint then reduces the target space to the space of null geodesics in five dimensions as the symplectic quotient:
\begin{equation}\label{ambitwistor-5}
\mathbb{A}_{5}=\left\{Z^{a} \in \mathbb{T} \times \mathbb{T} \mid Z^{a} \cdot Z^{b}=0,(\lambda^{A} \lambda_{A})=0\right\} \big/\,\{\mathrm{SL}(2, \mathbb{C}) \times \mathbb{C}\}\,,
\end{equation}
with the extra quotient by $\mathbb{C}$ accounting for the transformations gauged by $a$ in \eqref{6to5-translations-twist}. As described in \eqref{five-vector-orthogonality}, this description picks a fixed non-null six-vector, $\Omega_1^{AB}$ in spinor form, and considers null geodesics along null tangent vectors in the five-dimensional plane $\mathbb{C}^5$ that is orthogonal to it. The choice of $\Omega_1$ breaks the spin group $\mathrm{SL}(4,\mathbb{C})\rightarrow \mathrm{Sp}(4,\mathbb{C})$, isomorphic to $\mathrm{Spin}(5,\mathbb{C})$, and allows one to raise and lower spinor indices using $\Omega_1^{AB}$ and $\Omega_{1\,AB}=\frac{1}{2}\epsilon_{ABCD}\Omega_1^{AB}$.
For supersymmetric theories, the model is naturally extended by replacing $Z^a$ with $\mathcal{Z}^a$, thus obtaining the target space $\mathbb{A}_{5|2N}$.

\paragraph{Worldsheet matter} To construct critical models, we need to supplement the action $S_{5d}$ with a worldsheet matter system $S_m$. This can be achieved either via an internal CFT, e.g. a current algebra, or by coupling the 5d action to an additional fermionic system that plays a similar role to the worldsheet supersymmetry of the RNS models. Combinations of these two matter systems give rise to the bi-adjoint scalar,  maximally supersymmetric Yang-Mills theory and Einstein gravity, see \cite{Geyer:2020iwz,Albonico:2022pmd}.

The simplest example of a matter system is an internal CFT in the form of a current algebra, whose action we denote $S_C$.  We do not specify $S_C$ or its fundamental fields, as the only ingredients we will use are its currents $j^\mathfrak{a}\in\Omega^0(\Sigma,K_\Sigma\otimes\mathfrak{g})$ for some Lie algebra $\mathfrak{g}$, satisfying the defining relations
\begin{equation}\label{current-algebra}
j^{\mathfrak{a}}(\sigma) j^{\mathfrak{b}}(0) \sim \frac{l \delta^{\mathfrak{a} \mathfrak{b}}}{\sigma^{2}}+\frac{f_{\mathfrak{c}}^{\mathfrak{a} \mathfrak{b}} j^{\mathfrak{c}}}{\sigma}\,.
\end{equation}
Here $\mathfrak{a,\!b}$ are Lie algebra indices, $l\in\mathbb{Z}$ denotes the level and $f_{\mathfrak{c}}^{\mathfrak{a} \mathfrak{b}}$ the structure constants of $\mathfrak{g}$.

The second type of matter we need to consider in order to build gauge and gravity models is the system of worldsheet fermions $(\rho_A,\tilde\rho^A)\in\Pi\Omega^0(\Sigma,K_{\Sigma}^{1 / 2})$ with action given by:
\begin{equation}\label{action-WS-fermions}
S_{\rho}=\int_{\Sigma} \tilde{\rho}^{A} \bar{\partial} \rho_{A}+b_{a}\lambda^{Aa} \rho_{A}+\tilde{b}_{a} \lambda_{A}^{a} \tilde{\rho}^{A}\,,
\end{equation}
where $(b^{a}, \tilde{b}^{a})$ are $(0,1)$-forms on the worldsheet acting as fermionic Lagrange multipliers for the constraints $\lambda^{Aa} \rho_{A}=0=\lambda_{A}^{a} \tilde{\rho}^{A}$. These fermionic currents join with the bosonic currents of $S_{5d}$ to form the worldsheet gauge superalgebra $\mathfrak{sl}_2 \ltimes H(0, 2)$,\footnote{Here, $H(m_b, m_f)$ denotes the Heisenberg Lie superalgebra.} that plays a similar role in the twistorial models as the worldsheet supersymmetry in the RNS string.

Using  these two types of systems, we can construct 5d models with vanishing $\SL(2,\mathbb{C})$ anomalies for bi-adjoint scalars, gauge theory and gravity as follows:
\begin{align*}
 & \text{massive bi-adjoint scalar}
  && S^{\scalebox{0.7}{BAS}}_m \;= S_C + S_{\tilde C}\,,\\
   &\text{super Yang-Mills on the Coulomb branch}
 && S^{\scalebox{0.7}{CB}}_m \hspace{4pt}\;= S_\rho+S_C\,,\\
  &\text{super-gravity}
 && S^{\scalebox{0.7}{sugra}}_m = S_{\rho_1}+ S_{\rho_2}\,.
\end{align*}

Naively, similar models could have been formulated in 6d, but the form of the fermionic currents requires either a choice of $\Omega_1$ to raise and lower spinor indices on $\rho$, or the inclusion of both chiral and antichiral spinors (in contrast to the chiral model in 6d discussed above). In spite of these difficulties, the 6d `models' have inspired amplitude formulae that were presented in \cite{Geyer:2018xgb} and proved by BCFW recursion in \cite{Albonico:2020mge}. We presented the models here in five dimensions, in line with \cite{Geyer:2020iwz}, where they are well-defined since the fundamental and antifundamental representations are equivalent, and the aforementioned problems disappear.

A full treatment of the BRST gauge fixing and vertex operators for these 5d models can be found in \cite{Geyer:2020iwz}, and we will discuss the closely related calculations in the  four-dimensional massive models in the next section.

\section{The 4d two-twistor string  as a symmetry reduction}\label{sec:2TT-SymRed}
In this section we discuss the general procedure for  to obtaining massive worldsheet models in four dimensions via a symmetry reduction of the twistorial 5d models of \cite{Geyer:2020iwz} reviewed in the last section to obtain  massive models in four dimensions.  These massive models will  more easily describe fermions and incorporate various degrees of supersymmetry than the RNS models of the accompanying paper \cite{upcoming:part1},
but will no longer be dimension agnostic.

Symmetry reduction  in the target space of a worldsheet model is implemented by 
gauging the currents that generate the symmetry; this was the strategy used to reduce the 6d models based on \eqref{6dmodel-bos} to the 5d model \eqref{6to5-translations-twist} above where the current $\Omega_1^{AB}(\lambda_A\lambda_B)$ generates the translation in the 6th dimension, but  also has the effect of imposing the constraint that the momentum should vanish in that direction.  
Here we apply the same procedure to the five dimensional models \eqref{6to5-translations-twist} but now introduce a further current so as to prescribe some non-zero momentum in the 5th dimension.  This general framework can be used to derive families of 4d massive models from any 5d worldsheet theory with interesting symmetries. In particular, it 
gives the massive models in four dimensions described in \cite{Albonico:2022pmd}, as well as a class of interesting gauged supergravities.

\subsection{Massive models from symmetry reduction}\label{sec:4d-model}
The abstract framework for symmetry reduction is as follows.
We start with  a worldsheet sigma model of maps $X:\Sigma \rightarrow M$ where $M$ is a target complex manifold (our sigma models in this paper are all models of holomorphic maps).  Now assume  further that we   have a 
symmetry group $G$ acting on   $M$ generated by holomorphic vector fields $V_a(X)$ where $a$ is a Lie algebra index,  with quotient $N=M/G$.  One can relate the sigma model of maps $X:\Sigma\rightarrow M$ to that of maps $Y:\Sigma \rightarrow N$ by \emph{gauging} the symmetry generators.  The symmetries are generated in the model by worldsheet currents\footnote{These are obtained as the boundary term arising from the variation of the action along a symmetry generator. } $j_a\in \Omega^{1,0}_\Sigma \otimes \mathfrak{g}$, where  $\mathfrak{g}$ is the Lie algebra 
of $G$ and $a$ a Lie-algebra index.  By this, one means that one introduces worldsheet gauge fields $A^a\in \Omega^{0,1}_\Sigma \otimes \mathfrak{g}$ via a term in the sigma model action 
\begin{equation}
    \int_\Sigma  A^a\wedge j_a\, .
\end{equation}
This now has a worldsheet gauge symmetry 
\begin{equation}
    (\delta X,\,\delta A^a)= (\epsilon^a(\sigma,\bar\sigma)V_a(X), \dbar \epsilon^a(\sigma,\bar\sigma))\, .
\end{equation}
Classically, it is clear that quotienting by these gauge transformations is the same as considering the reduced model $Y:\Sigma\rightarrow N=M/G$.  Quantum mechanically  BRST gauge fixing gives a strategy to quantize the reduced model, which will often be more nonlinear and therefore less easy to quantize directly.  Examples of this already appear in the definition of the 4d and 5d ambitwistor strings above where we have gauged the little-groups.

In order to perform the symmetry reduction of the 5d models, we  gauge the current that generates translations in the 5th dimension.  On its own, this introduces the constraint that implements $\Omega_2\cdot K$ at the level of the worldsheet. If this were to vanish, then there would be no mass.  Because $P_5=\Omega_2^{AB}(\lambda_A\lambda_B)$ has worldsheet weight $(1,0)$, the `mass' $M$ cannot be a scalar, and needs to be implemented by a $(1,0)-$form on the worldsheet that we will take to be a current $j^H$ that  generates some symmetry in  the rest of the five-dimensional model we are reducing. In twistorial variables, the constraints to be implemented for the reduction\footnote{including the trivial reduction from six dimensions that was included in \eqref{6to5-translations-twist}} are
\begin{equation}\label{currentsH}
J_{\Omega_1}=\frac{1}{2}(\lambda^2-\tilde\lambda^2)\qquad J_{\Omega_2}=\frac{1}{2}(\lambda^2+\tilde\lambda^2)-j^H\,,
\end{equation}
where $\lambda^2=\frac{1}{2}(\lambda_\alpha\lambda^\alpha)=\det(\lambda_\alpha^a)$ and similarly for $\tilde{\lambda}$. Here $j^H$ denotes a current generator associated to some symmetry of the theory in five dimensions; for the Coulomb branch, we will take $H\in\mathfrak{g}$ in the Cartan subalgebra of the gauge group, represented by a current algebra on the worldsheet as described below. More generally it could be constructed out of any of the fields appearing in the model.

We can rearrange the currents in a more symmetric form by taking combinations $J_{\Omega_\pm}=J_{\Omega_2}\pm J_{\Omega_1}$, leading to:
\begin{equation}\label{symred-model}
S =\int_\Sigma \mathcal{Z}^a\cdot\dbar_e\mathcal{Z}_a+A_{ab}\mathcal{Z}^a\cdot\mathcal{Z}^b + a(\lambda^2-j^H)+ \tilde a (\tilde{\lambda}^2- j^{H})+S_m\,.
\end{equation}
The supertwistor fields $\mathcal{Z}^a$ are worldsheet spinors as in the five-dimensional models. Little group transformations are gauged by the fields $A_{ab}=A_{(ab)}\in\Omega^{0,1}(\Sigma,\mathfrak{sl}_2)$ and deformations of the worldsheet complex structure $\dbar_e=\dbar+e\partial$ are given by the Beltrami differential $e\in\Omega^{0,1}(\Sigma,T_{\Sigma})$. The fields $a, \tilde{a}\in\Omega^{0,1}(\Sigma)$ are Lagrange multipliers imposing the constraints $J_{\Omega_\pm}=0$ and effectively enforcing the symplectic reduction to the target space of the model. We recognize this as the phase space of the complexified massive particle by comparison with the construction presented in \cite{Albonico:2022pmd}. The action $S_m$ is again composed of matter systems such as the current algebra or the fermion systems $S_\rho$ reviewed in the last section.

\paragraph{Symmetry reduction} First let us take a closer look at the gauge fields $a$, $\tilde a$ and their role in implementing the symmetry reduction. From the action, we see that their gauge transformations combine translations along the $\Omega_2$ and $\Omega_1$ directions with the action of the symmetry group $G$ via $H$. These transformations are:
\begin{equation}\label{gauge-transf-a}
\delta a =\dbar\alpha\qquad\delta\mu^{Aa}=\alpha(\Omega_2^{AB}+\Omega_1^{AB})\lambda_B^a\qquad\delta\lambda_A^a=0\qquad\delta S_m=\int_\Sigma \dbar\alpha\, j^H\,,
\end{equation}
and
\begin{equation}\label{gauge-transf-atilde}
\delta \tilde a =\dbar\tilde\alpha\qquad\delta\mu^{Aa}=\tilde\alpha(\Omega_2^{AB}-\Omega_1^{AB})\lambda_B^a\qquad\delta\lambda_A^a=0\qquad\delta S_m=\int_\Sigma \dbar\tilde\alpha\, j^H\,.
\end{equation}
Without the need to specify the action of the symmetry on the matter fields in $S_m$, we can specify the variation of the action by the definition of the Noether current $j^H$. It is then precisely the symmetry of the theory specified by $S_m$ that guarantees the consistency of the reduction.
We note that the above constraints imply the original constraints $J_{\Omega_1,2}$, so the model is gauging translations in the $\Omega_1$ direction and combinations of translations in the $\Omega_2$ direction and transformations under $H$, as we expect in the sequence of reductions $6\text{d}\xrightarrow{}5\text{d}\xrightarrow{m}4$d.

Here the phase space of complexified massive particles described in \cite{Albonico:2022pmd} arises as a symplectic quotient of the five dimensional ambitwistor space parametrized as in \eqref{ambitwistor-5} with the additional constraints $J_{\Omega_{\pm}}=0$ and the gauging of transformations that combine translations along the $\Omega_{\pm}$ direction and the action of the symmetry element $H$. The five dimensional little group combines with the gauging along the two reduced dimensions to form the internal symmetry group of the complexified massive particle. We can then identify the six dimensional supertwistor $\mathcal{Z}$ with the Dirac supertwistor $\fZ$ we used to construct the four dimensional models in \S\ref{4d-models-intro}, c.f. \cite{Albonico:2022pmd}. As expected, these have $\mathcal{N}=\mathcal{N}^{(4d)}=2N$ fermionic components $\eta^I$. Through the reduction, we obtain the correspondence with four dimensional Minkowski superspace $\mathcal{M}_{4|4\mathcal{N}}$ parametrized by $(x_{\alpha\dot\alpha},\theta^{AI})$, with $\theta^{AI}=(\theta_{\alpha}^I,\tilde\theta^{\dot\alpha I})$ in the usual notation and $I=1,\dots\mathcal{N}$ via:
\begin{equation}
\mu^{A}=\left(\begin{array}{cc}
0 & x^{\alpha \dot{\beta}} \\
-x^{\dot{\alpha} \beta} & 0
\end{array}\right) \lambda_{B}+\omega_{I J} \theta^{A I} \eta^{J}, \quad \eta^{I}=\theta^{A I} \lambda_{A}\,,
\end{equation}

\paragraph{Algebra of constraints.} One might be tempted to generalize the models in \eqref{symred-model} by taking independent currents $j^H$ and $j^{\tilde H}$ in the constraints. This would correspond to performing a less trivial reduction directly from the six dimensional model, where translations are combined with the action of a symmetry in both extra dimensions. However, as we showed in \cite{Albonico:2022pmd}, in order to guarantee the closure of the algebra of constraints in the presence of worldsheet supersymmetry $S_\rho$ we need to take $j^H=j^{\tilde H}$. This reflects the absence of complete models for 2nd order gauge and gravity theories in six dimensions as described in \cite{Geyer:2020iwz}.

\paragraph{Gauge fixing.} The BRST gauge fixing of the model \eqref{symred-model} was discussed in \cite{Albonico:2022pmd} and we briefly review the key ingredients. We introduced fermionic ghosts associated to the gauge fixing of worldsheet diffeomorphisms, the little group and the internal symmetry group $\SL(2,\mathbb{C})\times\mathbb{C}\times\tilde{\mathbb{C}}$, as well as bosonic ghosts associated to the fermionic constraints in $S_{\rho}$. The gauge field $e$ can be fixed to zero and, in the absence of operator insertions, so can the fields $A_{ab},\,a,\,\tilde a,\,b^a,\,\tilde b^a$.

We can define the BRST charge as usual, and verify the possible obstructions to the vanishing of $Q^2$ at the level of the QFT. Such anomalies in general might arise from an $\mathfrak{sl}_{2, \mathbb{C}}$ anomaly or a conformal anomaly. The $\mathfrak{sl}_{2, \mathbb{C}}$ anomaly coefficient vanishes for maximally supersymmetric gauge theory and gravity. The central charge vanishes for a suitable choice of $S_j$ such that $c_j=28$ for Super Yang-Mills, $c_j=40$ for biadjoint scalars and the residual central charge in the gravitational theory can be understood as coming from six compactified dimensions, \cite{Albonico:2022pmd}.

\paragraph{Plane wave representatives}
We will consider scattering of plane wave representatives on ambitwistor space as in \cite{Geyer:2020iwz}. 
In a general 
ambitwistor space, scalar plane waves on space-time give  elements of $H^{1}(\mathbb{PA},\mathcal{O}(-2))$ of the form 
\begin{equation}
  \Phi_{\kappa}\left(Y_{a}\right):=  \bar\delta(k\cdot P) \e^{ik\cdot X}  \label{RNS-rep}
\end{equation} 
as described in \cite{Mason:1987,Mason:2013sva}.  It is not obvious from this formula that $\Phi_\kappa(Y_a)$ in fact only depends on the $Y_a$ coordinates.  To see that we note that in  our model $P_{AB}=(\lambda_A\lambda_B)$ and we can assume the 6d incidence relation \eqref{eq:incidence_6d} and it is then obvious in the support of the delta function it is invariant under translation along the null geodesic $\delta X\sim P$. To make this more explicit, as described in \cite{Geyer:2018xgb,Albonico:2020mge}, we can eliminate $X^{AB}$ by first observing that if we introduce a pair of little group parameters $(u_a, v_a)$, we have
\begin{equation}
  \exists (u_a, v_a)\neq0 \quad \mbox { such that }\quad   \cE_A:= (u\lambda_A)-(v\kappa_A)=0\, , \qquad \Leftrightarrow \qquad K\cdot P=0\, .
\end{equation}
Choosing a polarization spinor $\epsilon_a$ we can normalize the solution $(u_a,v_a)$ to $\cE_A=0$ 
on the support of $K\cdot (\lambda\lambda)=0$ by the condition $(v\epsilon)=1$.  On the support of \eqref{eq:incidence_6d} we then find $k\cdot X= (u\mu^A)\epsilon_A$.  Putting all this together we find the representative element of $H^{1}(\mathbb{PA},\mathcal{O}(-2))$ for  our twistorial representation of $\mathbb{PA}$ to be
\begin{equation}\label{h1-representative}
\Phi_{\kappa}\left(Y_{a}\right)=  \int d^{2}u\; d^{2} v\; \bar{\delta}^{4}\!\left(\left(u \lambda_{A}\right)-\left(v \kappa_{A}\right)\right) \bar{\delta}((v\, \epsilon)-1) \exp \left(\left(u \mu^{A}\right) \epsilon_{A}\right)\,,
\end{equation}
where $u^a,v^a$ are  four auxiliary complex variables determined by $\cE_A=0$, which were referred to as the \emph{polarized scattering equations} in \cite{Geyer:2018xgb,Albonico:2020mge}.  Although this formula is of course much more complicated than \eqref{RNS-rep}, the $(u_a,v_a)$ play a key role in representing Fermions on supersymmetry.  
On reduction to 4d, as noted in \cite{Albonico:2022pmd}, the components $\det(\kappa)$ and $\det(\tilde\kappa)$ of the external kinematics are unconstrained by the Penrose transform, and are only fixed by the model via BRST to give the signed mass of the state.

\paragraph{Supersymmetric extension}
In order to describe supersymmetric theories, we take the extension of \eqref{h1-representative} following \cite{Geyer:2020iwz,Albonico:2022pmd}. When considering momentum eigenstates of momentum $K_{AB}=(\kappa_A\kappa_B)$, the supersymmetry algebra $\{Q_{AI},Q_{BJ}\}=2\Omega_{IJ}P_{AB}$ of six and five dimensions reduces to the little group as:
\begin{equation}
Q_{A I}=\kappa_{A}^{a} Q_{a I}, \quad\left\{Q_{a I}, Q_{b J}\right\}=2 \Omega_{I J} \varepsilon_{a b}
\end{equation}
This leads to the four dimensional supersymmetry algebra with a central extension $Z_{IJ}=2\det(\kappa)\Omega_{IJ}$. For the simplest case of a symmetry reduction describing gauge theory on the Coulomb branch, the symplectic form $\Omega_{IJ}$ is preserved and the action of the supercharges organizes the states in massive and massless supermultiplets as we will detail in \S\ref{CB-section}. For now, let it suffice to say that both kinds of multiplets are annihilated by half of the supercharges so that on shell superspace can be parametrized by $\mathcal{N}=4$ fermionic supermomenta $q_I,\,I=1,\cdots4$. These are taken to be eigenvalues of an anticommuting subset of the supercharges $Q_{aI}$, thus necessarily breaking the action of either the little group or the R-symmetry. In line with previous work, we employ the R-symmetry preserving representation and define supermomenta $q_I$ as:
\begin{equation}\label{supermomenta}
Q_{a I} \tilde{\mathscr{F}}(\kappa, q)=\left(\xi_{a} q_{I}+\epsilon_{a} \Omega_{I J} \frac{\partial}{\partial q_{J}}\right) \tilde{\mathscr{F}}(\kappa, q)\,,
\end{equation}
where $(\epsilon_a,\xi_a)$ define a basis of the little group fundamental representation and $\tilde{\mathscr{F}}(\kappa, q)$ is a function on on-shell superspace.

We will detail later how the states of multiplets on the Coulomb branch of $\mathcal{N}=4$ SYM are encoded in the exterior powers of the supermomenta.  One approach is to express everything in terms of superfields  on superspacetime and establish a super Penrose transform to super-ambitwistor space as for example in \cite{Mason:2012va} for twistor space followed by the transform to ambitwistor space as in \cite{Geyer:2020iwz}. Here we follow the simpler approach of using the supersymmetry generators on super-ambitwistor space to present the supersymmetric extension of the plane wave representative \eqref{h1-representative} on superambitwistor space parametrized by $\mathcal{Y}_a=(\lambda_{Aa},\mu^A_a,\eta^I_a)$. On this space, the supercharges act geometrically as 
\begin{equation*}
Q_{A I}=\lambda_{A} \frac{\partial}{\partial \eta^{I}}+\eta^{J} \Omega_{J I} \frac{\partial}{\partial \mu^{A}}\,,    
\end{equation*}
 so that the function:
\begin{equation}\label{super-representative}
\Phi_{(\kappa, q)}(\mathcal{Y}_a)=\int d^{2} u\; d^{2} v\;\;  \bar{\delta}^{4}\!\left(\left(u \lambda_{A}\right)-\left(v \kappa_{A}\right)\right)\; \bar{\delta}((\epsilon v)-1) e^{i u_{a}\left(\mu^{A a} \epsilon_{A}+q_{I} \eta^{I a}\right)-\frac{1}{2}(\xi v) q^{2}}
\end{equation}
obeys the correct intertwining relations between superambitwistor space and on-shell momentum superspace.

\subsection{Vertex operators, BRST gauge fixing and pictures}\label{subsec-PCO}
In building correlators we will need to define various types of vertex operators, whose form is determined by how much residual gauge freedom they have left; they encode the same space-time fields, but relate to the gauge fixing in the correlator in different ways. 
Most well known are the effects of gauge fixing diffeomorphisms on the vertex operators, namely the distinction between \textit{fixed} and \textit{integrated} vertex operators. We will not detail how this distinction arises as it is well known in the literature and refer the reader to \cite{Witten:2012bh}, where integrated vertex operators are derived from the treatment of the moduli space of metrics. Intuitively, the fixed vertex operators remove any residual gauge freedom after the initial gauge fixing, while additional moduli integrals render the remaining vertex operators integrated.
At the level of the path integral, this is reflected in the requirement to saturate the zero modes of the $c$-ghosts associated to the field $e$ gauging chiral worldsheet diffeomorphisms: at tree-level, $c$ has three zero modes, reflecting the remaining M\"obius gauge freedom. Correlators are only non-vanishing when these zero-modes are saturated, and thus must contain three fixed and $n-3$ integrated vertex operators. The fixing of the insertion points of the three fixed vertex operators serves to remove the residual M\"obius gauge freedom at the same time as saturating the zero-modes of $c$.   There is  a similar story for the other symmetries that are gauged.
We will structure the discussion around vertex operators for symmetry reductions of maximal Super Yang-Mills in five dimensions. Because these involve both a current algebra and worldsheet fermions it is easy to carry the discussion over to the biadjoint scalar and supergravity case via the double copy.

On the worldsheet, we build $(1,1)$-form vertex operators by combining the plane wave representative \eqref{super-representative} with a theory-specific  $w\in\Omega^0(\Sigma,K_\Sigma^2)$ via a product $\circ$, which is intended to denote the fact that  $w$ may depend on $u$ so that the $u$-integration in \eqref{super-representative} should be understood to take place after multiplication by $w$. Gauge fixing worldsheet diffeomorphisms distinguishes fixed and integrated vertex operators as:
\begin{equation}
V=c\,w \circ \Phi_{(\kappa, q)}(\sigma)\qquad \mathcal{V}=\int d \sigma\, w \circ \Phi_{(\kappa,q)}(\sigma)\,,
\end{equation}
where $\Phi_{(\kappa, q)}(\sigma)\in H^1(\Sigma,T_\Sigma)$ is the pullback to the worldsheet of the plane wave representative \eqref{super-representative}. The theory-specific current $w$ will carry all further information concerning the scattered states and gauge fixing for the remaining symmetries.

\paragraph{Fixed vertex operators.}
A similar distinction between fixed and integrated vertex operators appears when gauge fixing the other gauge fields present in the models. We take here the perspective whereby fixed vertex operators are fundamental objects and the integrated ones are derived from the integration of moduli associated to the gauge field. We will say more generally that the fixed vertex operators are in picture $-1$ and the integrated ones in picture $0$, in analogy with the fermionic symmetries in ordinary string theory. There is one distinct picture number for each of the gauge fields we are fixing: $w_{(p_a,p_{\tilde{a}})}^{(p_b,p_{\tilde{b}})}$, with $p_a$ denoting the picture associated to gauge field $a$ and so on. Different picture numbers can all be obtained from the fully-fixed vertex operator,
\begin{equation}\label{pic1111}
    w_{\scalebox{0.7}{(-1,-1)}}^{\scalebox{0.7}{(-1,-1)}}=t\,\tilde t\,\delta\left((u\gamma)\right)\delta\left((u\tilde\gamma)\right)\,\mathfrak{t}_{\mathfrak{a}} j^{\mathfrak{a}}\,.
\end{equation}
Here $(s,t)$ are the ghosts associated with the gauging of the $a$  fields, while the ghosts $(\beta^a,\gamma^a)$ are associated with the fermionic gauge fields $b^a,\tilde b^a$, and similarly for the tilde'd fields.\footnote{These shouldn't be confused with the antighost $b$ for the gauging of worldsheet diffeomorphisms.} Because the vertex operator is automatically invariant under the $u$-projected fermionic currents $(u\lambda_A)\rho^A$ and $(u\lambda_A)\tilde\rho^A$, the vertex operator above is BRST invariant.

\paragraph{Picture changing operators.} Vertex operators with different ghost numbers -- i.e., in different pictures -- are most easily derived using the formalism of \emph{picture changing operators} (PCOs). These can be derived from a careful treatment of the 
 BRST gauge fixing of the fields $a,\tilde a$ as well as the fermionic $b^a,\tilde{b}^a$ in the presence of vertex operators.  For each of the gauge fields we introduce a gauge fixing term in the action of the form:
\begin{equation}\label{GF-term}
    \{Q_B,b\,F(\phi)\}\,,
\end{equation}
where $F(\phi)=\phi-\phi^{GF}$ is the gauge fixing condition and $b$ is the associated antighost (here not referring to diffeomorphisms).\footnote{The reader might notice that here we give a general prescription to gauge fix all residual gauge transformations after having gauge fixed worldsheet diffeomorphisms, but we do not discuss the gauge fixing of the little group via the fields $A_{ab}$ in the same manner as our vertex operators are already manifestly $\SL_2$ invariant.
}

Having already fixed worldsheet diffeomorphisms, the gauge transformations associated to $a,\tilde a$ are as in \eqref{gauge-transf-a} and \eqref{gauge-transf-atilde}, where the variations are required to vanish at the vertex operators insertion points. This means that we are not able to gauge fix the fields to zero and these are only allowed to vary within a cohomology class of $H^{0,1}\left(\Sigma, \mathcal{O}\left(-\sigma_{1}-\ldots-\sigma_{n}\right)\right)$. The gauge fixed fields can then be expanded in a basis ${h_i}$ of $(0,1)$-forms on the worldsheet that span this $n-1$ dimensional cohomology group:
\begin{equation}
    a^{\mathrm{GF}}=\sum_{i=1}^{n-1}h_ia_i\, ,\qquad \tilde a^{\mathrm{GF}}=\sum_{i=1}^{n-1}h_i\tilde a_i\,.
\end{equation}
The off-shell BRST transformations of the fields are:
\begin{equation}
\begin{aligned}
    \delta_B\,a=\dbar t\, ,\qquad\delta_B\,a_i&=\alpha_j\, ,\qquad\delta_B\,s=N\\
    \delta_B\,\alpha_j=0\, , \quad&\quad\delta_B\,N=0\,,
\end{aligned}
\end{equation}
where $N$ is the Nakanishi-Lautrup field that acts as a Lagrange multiplier for the gauge fixing condition.
Then the gauge fixing term for $a$ (and similarly for $\tilde{a}$) can be expanded as follows:
\begin{equation}
    \int_\Sigma\{Q_B,s\,(a-a^\text{GF})\}=\int_\Sigma N(a-a^{\text{GF}})+s\dbar t + \sum_{i=1}^{n-1}\alpha_i \int_\Sigma s\,h_i
\end{equation}
Integrating out the auxiliary field $N$ enforces the gauge fixing condition and produces a term of the form 
\begin{equation}
    \sum_{i=1}^{n-1}a_i \int_\Sigma h_i\,J_{\Omega_+}
\end{equation}
Then integrating out the fermionic and bosonic moduli $\alpha_i,a_i$ we obtain $n-1$ insertions of the \emph{picture-changing operators} $\Xi_i$ \cite{Adamo:2013tsa,Geyer:2020iwz},
\begin{equation}
    \Xi_i=\delta\left(\int_\Sigma h_i J_{\Omega_+}\right)\left(\int_\Sigma h_i\,s\right)\,,
\end{equation}
as well as analogous contributions $\tilde\Xi_i$ from $\tilde a$. As expected, the picture changing operators contain an insertion of the anti-ghost $s$ that will absorb the $t$ insertion of the fixed vertex operator. In addition, it contains a characteristic delta-function that will play an important part in fixing the mass of the scattered particle.

The treatment of the fermionic gauge fields is analogous and was presented in \cite{Geyer:2020iwz}. By the invariance of the vertex operators under half of the Heisenberg superalgebra, the components of $b^a,\tilde b^a$ that are parallel to $u$ can be gauge fixed to zero, while the orthogonal ones develop moduli and produce $n-2$ picture changing operators;
\footnote{The picture changing operators actually contain other terms involving mixed ghost products that are generated by the term $\{Q,\beta^a\}F_a(b^a)$, where $(\beta^a, \gamma^a)$ is the ghost system for the gauging of $b^a$. These however don't contribute to the scattering amplitude as they either vanish on the support of the delta functions or they have the wrong ghost number.}
\begin{equation}
\Upsilon\left(z_{l}\right)=\delta(\langle\hat{u} \beta\rangle)\left\langle\hat{u} \lambda_{A}\right\rangle \rho^{A}, \quad \tilde{\Upsilon}\left(z_{l}\right)=\delta(\langle\hat{u} \tilde{\beta}\rangle)\left\langle\hat{u} \lambda_{B}\right\rangle \tilde{\rho}^{B}\,,
\end{equation}
where $\hat u^a$, $u^a$ form a local basis for the $\C^2$ on which the little group acts.

From the form of these PCOs, we can now derive vertex operators in the remaining pictures.
As both the $\Xi_i,\,\tilde\Xi_i$ and $\Upsilon_j,\,\tilde\Upsilon_j$ come in pairs, we will only need additional vertex operators in pictures $(0,0;-1,-1)$ and $(0,0;0,0)$ in order to compute correlators. These are obtained as the limit as $\sigma\rightarrow\sigma_i$ of the OPE $\quad\text{PCO}(\sigma)\cdot w(\sigma_i),\quad$ and we obtain:
\begin{align}\label{vo-matter}
    &w_{\scalebox{0.7}{(0,0)}}^{\scalebox{0.7}{(-1,-1)}}=\delta\big( \mathrm{Res}_{\sigma_i}(\lambda^2-j^H)\big) \; \delta\big(\mathrm{Res}_{\sigma_i}(\tilde\lambda^2-j^H)\big)
    \delta\left((u\gamma)\right)\delta\left((u\tilde\gamma)\right)\,\mathfrak{t}_{\mathfrak{a}} j^{\mathfrak{a}}\,,\\
    &w_{\scalebox{0.7}{(0,0)}}^{\scalebox{0.7}{(0,0)}}=
    \delta\big( \mathrm{Res}_{\sigma_i}(\lambda^2-j^H)\big) \; \delta\big(\mathrm{Res}_{\sigma_i}(\tilde\lambda^2-j^H)\big)
    \left(\frac{\left\langle\hat{u} \lambda_{A}\right\rangle \epsilon^{A}}{\langle u \hat{u}\rangle}+\epsilon^{A} \epsilon_{B} \rho_{A} \tilde{\rho}^{B}\right)\,\mathfrak{t}_{\mathfrak{a}} j^{\mathfrak{a}}\,.
\end{align}
From this derivation we observe that the term
\begin{equation*}
Q_m:= \oint t (\lambda^2-j^H) +\tilde t (\tilde \lambda^2 -j^H)\, .
\end{equation*}
in the BRST operator is responsible for fixing the masses of the external particles via the delta functions in \eqref{vo-matter}. The mass is assigned as the residue of the current $j^H$ acting on the external state as an OPE.
This action depends on the choice of current $j^H$ and we will detail it later on as we consider specific theories.

\subsection{Massive amplitudes as correlators}\label{sec:massive_as_correl}
We compute scattering amplitudes as correlators in the models described above. Because the ghost zero modes need to be saturated\footnote{The $c$ ghosts have $3$ zero modes, the $t,\tilde t$ have one each, and the $\gamma^a, \tilde\gamma^a$ have two each.} and the residual gauge symmetry fixed, the only non trivial correlators with $n$ insertions must contain vertex operators in the various pictures as:
\begin{equation*}
\mathcal{A}_{n}=\left\langle V_{1\,\scalebox{0.7}{(-1,-1)}}^{\,\,\scalebox{0.7}{(-1,-1)}}
V_{2\,\scalebox{0.7}{(0,0)}}^{\,\,\scalebox{0.7}{(-1,-1)}}
V_{3\,\scalebox{0.7}{(0,0)}}^{\,\,\scalebox{0.7}{(0,0)}}
\prod_{i=4}^{n} \mathcal{V}_{i\,\scalebox{0.7}{(0,0)}}^{\,\,\scalebox{0.7}{(0,0)}}
 \right\rangle\,.
\end{equation*}
After gauge fixing, all the fields are free. The evaluation of the scattering amplitude is analogous to that in \cite{Geyer:2020iwz,Albonico:2022pmd}, so we only discuss here the main features of the resulting formulae:
\begin{equation}\label{final-formula}
\mathcal{A}_{n}=\prod_{i=2}^n\delta(\kappa_i^2-M_i)\; \delta(\tilde\kappa_i^2- Mi)\int d\mu_{n}^{\mathrm{pol}} \; \mathcal{I}_{n}\, e^{F_\mathcal{N}}\,,
\end{equation}
The $2\times( n-1)$ delta functions sitting in front of the formula fix the mass parameters $\kappa_i^2$ and $\tilde \kappa_i^2$ of the external particles to be equal to the eigenvalue $M_i$ under the group by which we are reducing. These are precisely the delta-functions originating from the picture-changing operators  $\Xi,\tilde\Xi$, but now evaluated in the path integral. this can be seen as follows:
When computing the path integral, the $\lambda\mu$ correlators can be evaluated exactly, localizing the path integral onto its classical solution to the equations of motion, \cite{Geyer:2020iwz},
\begin{equation}\label{sol-path-integral}
        \lambda_{A}^a(\sigma)=\sum_{i=1}^n\frac{u_i^a\epsilon_{iA}}{\sigma-\sigma_i}\,.
\end{equation}
On the support of this solution, as well as the delta-functions contained in the plane wave vertex operators, we can then extract the residue in \eqref{vo-matter}:
\begin{equation}\label{eq:delta_mass}
        \text{Res}_{\sigma_i}(\lambda^2-j^H)=\kappa_i^2-M_i\,.
\end{equation}
Here we denote by $M_i$ the eigenvalue of the external state under the action of the element $j^H$. It is important to note that $M_i$ is not the mass but rather a `signed mass' parameter: we will refer to the mass as $m_i=|M_i|$. In general, the vertex operator will carry a representation of the symmetry group so that:
    \begin{equation}\label{group-action-on-vo}
        j^H(\sigma)\, V_i(\sigma_i)\sim\frac{M_i}{\sigma-\sigma_i}V_i(\sigma_i)\,.
    \end{equation}
More specifically, in the case of the Coulomb branch, massive external states carry a factor $\mathfrak{m}_aj^a$ and $j^H$ is an element of the current algebra. We will discuss this in more detail around \eqref{group-action-CB}.
From the form of the constraint \eqref{eq:delta_mass}, we see that the delta-functions in vertex operators $V_{(0,0)}$ enforce the mass-shell condition, so $n-1$ particles in the correlator are on-shell. The remaining $n$-th mass-shell condition is then a consequence of overall momentum conservation as reduced from higher dimension, as we will discuss now.

The integration measure is built on the six- and five-dimensional polarized measure, with additional delta-functions enforcing the mass-shell constraint;
    \begin{equation}\label{measure}
    d\mu_{n}^{\mathrm{pol}}:=\frac{\prod_{j} d \sigma_{j}\, d^{2} u_{j} \,d^{2} v_{j}}{\operatorname{vol} \mathrm{SL}(2, \mathbb{C})_{\sigma} \times \mathrm{SL}(2, \mathbb{C})_{u}} \; \prod_{i=1}^{n}  \bar{\delta}^{4}\Big(( u_{i} \lambda_{A}(\sigma_{i}))-( v_{i} \kappa_{i A})\Big)
    \bar{\delta}\big((v_i\epsilon_i)-1\big)\,.
    \end{equation}
This measure localizes on solutions to the so-called \emph{polarized scattering equations} \cite{Geyer:2018xgb,Albonico:2020mge},
    \begin{equation}\label{massive-PSE}
        \mathcal{E}_{i\sA}=(u_i\lambda_{A}(\sigma_i))-(v_i\kappa_{iA})=\sum_{j\neq i}\frac{(u_iu_j)\epsilon_{jA}}{\sigma_i-\sigma_j}-(v_i\kappa_{iA})=0\,.
    \end{equation} 
Crucially, these equations imply \cite{Albonico:2022pmd} the massive scattering equations that arise in the symmetry reduced RNS models of the accompanying paper \cite{upcoming:part1} and that were originally conjectured by Naculich \cite{Naculich:2015coa} and Dolan $\&$ Goddard \cite{Dolan:2013isa}:
\begin{equation}\label{massive-SE-vectorial}
        E_i^{\scalebox{0.5}{SE}}=\sum_{j\neq i}\frac{k_i\cdot k_j-M_iM_j}{\sigma_i-\sigma_j}=0\,.
\end{equation}
Returning to the $5n$ polarized scattering equations, $5n-6$ delta-functions localise the ($u_i,v_i,\sigma_i)$-variables but, because three of the $u$'s and three of the $\sigma$'s are already fixed by the gauge, overall six delta functions remain after integration. These impose conservation of the six dimensional momentum \eqref{embedding-momentum} and ultimately lead to the consistency of the mass assignments. Indeed they give $\kappa_1^2=-\sum_{i=2}^n\kappa_i^2=-\sum_{i=2}^n M_i=M_1$, where the last equality is given by charge conservation. In particular, this  guarantees that the amplitudes vanish unless $\sum_{i=1}^n M_i=0$, \cite{upcoming:part1}. 

The worldsheet matter systems contribute the familiar ambitwistor integrands  \cite{Albonico:2022pmd};
    \begin{align}\label{integrands}
    &\mathcal{I}_{n}^{\scalebox{0.6}{BAS}}=\operatorname{PT}(\alpha)\operatorname{PT}(\beta)\,,
    && \mathcal{I}_{n}^{\scalebox{0.6}{CB}}=\operatorname{PT}(\alpha)\det{}'\HH\, ,  
    && \cI_n^{Grav}= \det{}'\HH\det{}'\tilde \HH\,,
    \end{align}
    where $\text{PT}(\alpha)$ denotes the Parke-Taylor factor arising from the current algebra correlator,\footnote{As usual, here we ignore the multi-trace terms that will correspond to some higher-order gravity mediated terms that arise in ambitwistor-string heterotic models as described in \cite{Witten:2003nn}.} and the reduced determinant is obtained from the evaluation of the $\rho\tilde\rho$ system (c.f. \cite{Geyer:2020iwz}):
    \begin{equation*}
    \det{}'\HH:=\frac{1}{(u_1u_2)}\det \HH^{[12]}_{[12]} \,,
    \end{equation*}
    where, the $n\times n$ matrix $H$ is defined by
    \begin{equation*}
\HH_{i j}=\frac{\epsilon_{i A} \epsilon_{j}^{A}}{\sigma_{i j}}, \qquad \HH_{i i}=-e_i^{\sA\sB}(\lambda_\sA\lambda_\sB)\left(\sigma_{i}\right)\,,
    \end{equation*}
 and the sub- and superscripts indicate that both the rows and the columns 1 and 2 have been removed.
 The exponential factors in the supersymmetric plane wave give rise to the term
    \begin{equation}\label{exponential-F}
    e^{F_\mathcal{N}}
    :=\exp\left(\sum_{j<k}\frac{(u_ju_k) q_j\cdot q_k}{\sigma_j-\sigma_k}-\frac{1}{2}\sum_{j=1}^n(\xi_jv_j)q_j^2\right)\,.
    \end{equation}
All the dependence on the supermomenta is contained in this factor and when expanding it in different powers of $q$ one can read off it the various component amplitudes as we will detail below. We note that, while in the reduction that leads to the Coulomb branch supersymmetry is preserved and all states in the original multiplets have the same mass, there are other ways of performing a reduction, such as the R-symmetry reduction described in \S\ref{sectionRsymred}, that break the original supersymmetry and give rise to smaller supermultiplets. In these cases we can still read component amplitudes off this formula, but we should keep in mind that states in the same \textit{higher dimensional} multiplet can have different masses. In the resulting formulae, this is implemented by promoting the mass parameters $M_i$ in the delta-functions to mass-operators that include derivatives in the fermionic supermomenta.

The final formula \eqref{final-formula} for massive scattering amplitudes is valid for a broad class of massive theories obtained via symmetry reduction from a 5d massless theory. The framework and derivation as a symmetry reduction of the ambitwistor string guarantees that the formula is consistent, and indeed describes massive amplitudes in well-defined theories. The specific mass assignment and the form of the integrands are however theory- and symmetry reduction-dependent, and we will describe two interesting cases in the next two sections: the Coulomb branch, and gauged supergravities.

\section{Coulomb branch}\label{CB-section}
One of the interesting theories that can be obtained via the symmetry reduction procedure is the Coulomb branch of $\mathcal{N}=4$ super Yang-Mills. In this section we justify this claim, showing that the Lagrangian theory can be defined via a symmetry reduction of $5$d $N=2$ SYM. This procedure imposes a specific dependence of fields on the extra dimension, which can be eliminated by a gauge transformation at the price of giving a vacuum expectation value to a scalar field, thus producing the more familiar formulation of this theory. 
We will begin withs a review of the usual description of the Coulomb branch, show the equivalence to the model derived via symmetry reduction and identify the spectrum of this theory of massive particles. We will then implement the symmetry reduction in the worldsheet model for maximal super Yang-Mills in five dimensions and derive amplitude formulae. 

\paragraph{Coulomb branch via VEV'd scalars}
The Coulomb branch of $\mathcal{N}=4$ SYM  with Lie algebra $\mathfrak{g}$ is usually described by assigning a vacuum expectation value to some of the scalar fields. The theory at the origin of the moduli space is a theory of massless particles describing a vector potential field $A_\mu$, six real scalars $\Phi^a$ transforming in the $\mathbf{6}$ of the $\SO(6)$ R-symmetry group and four Majorana spinors $\Psi_A^I$ in the fundamental of $\SU(4)\simeq \SO(6)$. All the fields transform in the adjoint representation of the gauge group.

In the simplest case, a gauge group $U(N+M)$ is spontaneously broken to $U(N)\times U(M)$ by the vacuum configurations of some of the scalars, e.g.:
\begin{equation}\label{vev-scalars}
\la\Phi^1\ra=iH=iv\,\mathrm{diag}(\mathds{1}_{N},0_M)\quad\quad\la\Phi^a\ra=0\quad a\neq6\,.
\end{equation}
Writing the scalars as the antisymmetric product of two fundamentals of $\SU(4)$, this is equivalent to:
\begin{equation}
\la\Phi_{IJ}\ra=\Omega_{IJ}H\,,
\end{equation}
where $\Omega$ is skew and reduces the $R$-symmetry to the $\mathrm{Sp}(4)\subset \SU(4)$ R-symmetry, preserving the bilinear form $\Omega$ on the Coulomb branch.

The fields are in the adjoint representation and are thus represented by $(N+M)\times(N+M)$ matrices. Under this symmetry breaking, the $(N+M)^2$ generators of the original gauge group reduce to those  for the residual $U(N)$ and $U(M)$, together with $2NM$ broken ones:
\begin{equation}\label{vector-decomposition}
\ad_{N+M}\rightarrow(\ad_N,1)\oplus(1,\ad_M)\oplus(\mathbf{N},\bar{\mathbf{M}})\oplus(\bar{\mathbf{N}},\mathbf{M})=\begin{pmatrix}
A_\mu^{ab} & W^{a\dot{b}} \\
\bar{W}^{\dot{a}b} & A_\mu^{\dot{a}\dot{b}}
\end{pmatrix}\,.
\end{equation}
The mass terms arise upon replacing $\Phi^I\rightarrow H+\phi^I$ in the Lagrangian of $\mathcal{N}=4$ SYM.
The decomposition \eqref{vector-decomposition} of the adjoint under 
the action of $H$ demonstrates that the fields in the residual $U(N)\times U(M)$ remain massless, whereas the broken generators $W$ and $\bar{W}$ acquire a mass proportional to $v$. The fermions and the scalars also live in the adjoint representation of the gauge group, and decompose similarly. In addition to this, however, they also transform non-trivially under the R-symmetry group, which is broken to $\mathrm{Sp}(4)$. Under this residual symmetry, the six massive scalars $w_{IJ}$ transform in a $\mathbf{5}$ plus a singlet, consisting of the trace $\Omega_{IJ}w^{IJ}$.\footnote{Here and below, $w_{IJ}$ denotes the massive scalars on the Coulomb branch. The distinction from $w(\sigma, u)$ in the vertex operators of the ambitwistor string should always be clear from the index structure and context.} This last component is absorbed by the gluons and becomes a polarization state of the massive spin-one field via the Higgs mechanism.
Then the massive scalars are $w_{12},w_{13},w_{34},w_{24}$ and the combination of $w_{14}$ and $w_{23}$ that is orthogonal to the longitudinal boson, i.e. $(w_{14}+w_{23})/\sqrt{2}$.

\paragraph{Spectrum.} The procedure outlined above leaves two types of states with respect to the color group. The first corresponds to elements $\mathfrak{t}\in \mathfrak{u}_N\times \mathfrak{u}_M$, that commute with $H$ and therefore correspond to massless states. The second are the off-diagonal blocks consisting of elements $\mathfrak{m} \in \left(\C^N\otimes (\C^M)^*\right)\oplus \left(\C^M\otimes (\C^N)^*\right)$ for which 
 $[H,\mathfrak{m}]=   M^H_\mathfrak{m}\,\mathfrak{m}\,$ so that they define massive states with mass $|M^H_\mathfrak{m}|$.

\paragraph{Supersymmetry.}
The action of the supercharges casts the massless states in a vector multiplet transforming in the adjoint of the residual gauge group:
\begin{equation}\label{N=4massless}
\mathscr{F}^0=(\phi_{IJ}=\phi_{[IJ]}\,,\Psi^I_\alpha, \tilde\Psi_{I\dot \alpha}\,,F_{\alpha\beta},F_{\dot\alpha\dot\beta})\, .
\end{equation}
For these multiplets the R-symmetry is enhanced to a full SU$(4)$, so that the fundamental indices can no longer be raised and lowered. The multiplet contains the two familiar ${\pm1}$ helicity states of the massless spin-1, six real massless scalars  $\phi_{IJ}$ and eight massless gluino states via the chiral parts of $\Psi_\alpha^I, \tilde \Psi_{I \dot\alpha}$.

The massive supermultiplets are the so-called $1/2$-BPS, ultrashort massive representations of $\mathcal{N}=4$ with central extension $Z_{IJ}=2M\Omega_{IJ}$, with $\mathrm{Sp}(\mathcal{N})$ R-symmetry, with skew form $\Omega_{IJ}$ and indices $I,J=1,\ldots,\cN=4$.
They are bifundamentals of $U(N)\times U(M)$, composed of a massive W-boson ($3$ bosonic d.o.f.) $F_{AB}$, five massive scalars $w_{IJ}$ and the fermionic partners, four massive Weyl-Majorana spinors $\Psi_A^I$ ($8$ fermionic d.o.f.):
\begin{equation}\label{N=4massive}
\mathscr{F}^m=(w_{IJ}=w_{[IJ]},\Psi_{I}^ A\,,F^{AB}=F^{(AB)})\, , \qquad  w_{IJ}\Omega^{IJ}=0\, .
\end{equation}

\paragraph{Coulomb branch as a symmetry reduction.}\label{CB-sym-red-section} 
Equivalently, the Coulomb branch of $\mathcal{N}=4$ SYM can be described as a symmetry reduction of five dimensional maximally supersymmetric Yang-Mills, which can be seen as follows.
The gauge field $\mathcal{A}^{(4)}$ on the Coulomb branch of $\mathcal{N}=4$ SYM can be embedded in a five dimensional gauge field $\mathcal{A}^{(5)}$, where the extra component is the vev'd scalar $\Phi^1$, so that $\mathcal{A}^{(5)}=\mathcal{A}^{(4)}+(iH+\phi)\mathrm{d}x^4$. The field $\phi$ has vanishing vev and neither $\phi$ nor $\mathcal{A}^{(4)}$ have any dependence on the coordinate $x^4$, i.e. in this gauge $\partial_4\mathcal{A}^{(5)}=0$.

This 5d embedding turns out to be equivalent to a symmetry reduction from 5d to 4d, up to a gauge transformation. In particular, consider the gauge transformation $\mathcal{U}=\exp(iHx^4)$, under which the connection transforms as
\begin{equation}\label{GTconnection}
    \mathcal{A}^{(5)'}=\mathcal{U}\mathcal{A}^{(5)}\mathcal{U^\dagger}+\mathcal{U}\partial_4\mathcal{U}^\dagger dx^4 =\mathcal{U}\mathcal{A}^{(5)}\mathcal{U^\dagger}-iH\mathrm{d}x^4=\mathcal{U}\mathcal{A}^{(4)}\mathcal{U^\dagger}+\mathcal{U}\phi\mathcal{U^\dagger}\mathrm{d}x^4.
\end{equation}
This gauge transformation has eliminated the non zero vev of the scalar. The price is the introduction of an explicit dependence on the $x_4$ coordinate:
\begin{equation}\label{CBsymred}
\partial_4 \mathcal{A}^{(5)'} = \partial_4 \mathcal{U}\mathcal{A}^{(5)}\mathcal{U^\dagger} + \mathcal{U}\mathcal{A}^{(5)}\partial_4 \mathcal{U^\dagger}
= i[H,\mathcal{U}\mathcal{A}^{(5)}\mathcal{U^\dagger}]= i[H,\mathcal{A}^{(5)'}]\,.
\end{equation}
This equation defines a symmetry reduction from $N=2$ SYM in five dimensions to the Coulomb branch. In this description the mass terms are derived from the kinetic terms of the five-dimensional theory via \eqref{CBsymred} and the dependency of the fields on $x_4$ is fixed in such a way that this drops out of the action. This is sketched in appendix \ref{CB-SR-appendix}.

\paragraph{From space-time Lagrangian to ambitwistor model.} As we have just seen, the Coulomb branch of $\mathcal{N}=4$ super Yang-Mills can be described equivalently as a symmetry reduction \eqref{CBsymred} from 5d maximal super Yang-Mills. We can thus combine the results of the two preceding sections to construct a massive ambitwistor string model for the Coulomb branch: start from the 5d model for sYM of \S\ref{subsec:5dmodels}, and reduce it to 4d as discussed in \eqref{symred-model}. The resulting worldsheet model then describes scattering amplitudes on the Coulomb branch. In particular, we can understand the spectrum as follows:  
The elements $\mathfrak{t}\in\mathfrak{u}_N\times \mathfrak{u}_M$ and $\mathfrak{m}\in \big(\C^N\otimes (\C^M)^*\big)\oplus \big(\C^M\otimes (\C^N)^*\big)$ of the broken algebra described above define states in the worldsheet theory via the current algebra generators: massless $\mathfrak{t}\cdot j(\sigma)$ and massive $\mathfrak{m}\cdot j(\sigma)$.
This is consistent with the mass assignments for these states as the OPEs \eqref{group-action-on-vo} of the respective currents take the following form:
\begin{align}\label{group-action-CB}
& j^H(\sigma)\; \mathfrak{t}\cdot j(\tilde\sigma)\sim 0,
&&j^H(\sigma)\;\mathfrak{m}\cdot j(\tilde\sigma)\sim\frac{M^H_\mathfrak{m}}{\sigma-\tilde\sigma}\mathfrak{m}\cdot j\,,
\end{align}
as can be seen from the definition of $j^H$ as a current in the Cartan.
We can thus identify vertex operators built from currents $\mathfrak{t}\cdot j$ with the massless vector multiplet transforming in the adjoint of the residual $U(N)\times U(M)$ gauge group, whereas vertex operators built from $\mathfrak{m} \cdot j$ describe the massive vector multiplet.

Expanded on the on-shell superspace we described in \eqref{supermomenta}, the supermultiplets are organized as follows:
\begin{equation}\label{CB-massive-massless-multiplet}
\begin{aligned}
\tilde{\mathscr{F}}_{(\kappa, q)}^{(m)} &=F^{\epsilon \epsilon}(\kappa)+q_{I} \Psi^{\epsilon I}(\kappa)+q^{2} F^{\epsilon \xi}(\kappa)+\frac{1}{2} q_{I} q_{J} \Phi^{I J}(\kappa)+q^{2} q_{I} \Psi^{\xi I}(\kappa)+q^{4} F^{\xi \xi}(\kappa) \\
\tilde{\mathscr{F}}_{(\kappa, q)}^{(0)}&=g^h(\kappa)+q_{I} \Psi^{\epsilon I}(\kappa)+\frac{1}{2} q_{I} q_{J} \varphi^{I J}(\kappa)+q^{2} q_{I} \Psi^{\xi I}(\kappa)+q^{4} g^{-h}(\kappa)\,,
\end{aligned}
\end{equation}
with $q^4= (q_Iq_J)(q^Iq^J)$ and $(q^3)^I_a=\p q^4/\p q^a_I$. This tells us that the leading term in the amplitude involves $n$ (massive or massless) bosons, which is what we expect from the way we constructed vertex operators. Component amplitudes can be read off the exponential factor in the superamplitude by matching the corresponding powers in the multiplets.

\subsection{Amplitudes}
Extending the results of \cref{sec:massive_as_correl} for generic massive correlators obtained as a symmetry reduction, the formula for the superamplitude of $\mathcal{N}=4$ SYM on the Coulomb branch is now given by:
\begin{equation}
 \mathcal{A}_n^{\scalebox{0.6}{$\mathrm{CB}$}}(\alpha,\{k_i\},\{M_i\},\{q_i\}) = \int \rd\mu_n\; \mathrm{PT}(\alpha) \det{}'\HH^{\scalebox{0.6}{$\mathrm{CB}$}}\;\e^{\scalebox{0.7}{$F_{\mathcal{N}}$}}\,.
\end{equation}
We begin by considering component amplitude involving only the leading vector component of either the massive or massless supermultiplet \eqref{CB-massive-massless-multiplet}, i.e. $ \mathcal{A}_n^{\mathrm{CB}}(\alpha,\{k_i\},\{M_i\},\{q_i=0\})$. This is either a massive $W$ boson ($W^{\epsilon\epsilon}=\epsilon_a\epsilon_bW^{(ab)}$) or a gluon of helicity $h$ dictated by the polarization $\epsilon_i$. In this case the amplitude is simply:
\begin{equation}
\int \rd\mu_n\; \mathrm{PT}(\alpha) \det\mathbb{H}^{\scalebox{0.6}{$\mathrm{CB}$}}\;=\delta\left(\sum_jk_j\right)\delta\left(\sum_j M_j\right)\sum_{i=1}^{(n-3)!}\mathrm{PT}(\alpha) \det{}'H^{\scalebox{0.6}{$\mathrm{CB}$}}\,J\,,
\end{equation}
with the Jacobian $J=(\sigma_{ij}\sigma_{jk}\sigma_{ki})^2(\det\Phi_{ijk}^{ijk})^{-1}$, and
\begin{equation}
|\Phi_{ij}|:=\left|\partial \mathcal{E}_{i} / \partial \sigma_{j}\right|=\left\{\begin{array}{ll}
\frac{(k_i+k_j)^2-(M_i+M_j)^2}{\sigma_{i j}^2} \quad i \neq j \\
\sum_{k\neq i}\frac{(k_i+k_k)^2-(M_i+M_k)^2}{\sigma_{i k}^2}, & i=j
\end{array}\right.
\end{equation}
\paragraph{Example: Four vector bosons.}
At four points, and specifying the colour ordering $\alpha=(1234)$, the expression above reads:
\begin{equation}\label{4ptformula}
A_4=\left.\frac{1}{(u_1u_2)(u_3u_4)} \frac{\sigma_{12} \sigma_{34}}{((k_1+k_2)^2-(M_1+M_2)^2)}\left(\epsilon_{1 A} \epsilon_{3}^{A} \epsilon_{2 B} \epsilon_{4}^{B}-\frac{\sigma_{31} \sigma_{42}}{\sigma_{41} \sigma_{32}} \epsilon_{1 A} \epsilon_{4}^{A} \epsilon_{2 B} \epsilon_{3}^{B}\right)\right|_{*}
\end{equation}
where $*$ indicates that we are evaluating the expression on the unique solution:
\begin{equation}
\sigma_{1}=[(1,0)]\quad \sigma_{2}=[(1,1)]\quad \sigma_{3}=\left[\left(1,-\frac{((k_1+k_3)^2-(M_1+M_3)^2)}{((k_1+k_2)^2-(M_1+M_2)^2)}\right)\right] \quad\sigma_{4}=[(0,1)]
\end{equation}
\begin{equation}
(u_1u_2)=-\frac{\varepsilon^{A B C D} k_{1AB} \epsilon_{3 C} \epsilon_{4 D}}{\varepsilon^{A B C D} \epsilon_{1 A} \epsilon_{2 B} \epsilon_{3 C} \epsilon_{4 D}}\qquad (u_3u_4)=-\frac{\varepsilon^{A B C D} k_{3AB} \epsilon_{1 C} \epsilon_{2 D}}{\varepsilon^{A B C D} \epsilon_{1 A} \epsilon_{2 B} \epsilon_{3 C} \epsilon_{4 D}}
\end{equation}
We obtain the generic formula for amplitudes involving gluons and $W$ bosons\footnote{The details of this evaluation can be found in the derivation of the four point amplitude in $6d$ SYM, in section 5.2 of \cite{Albonico:2020mge}.}:
\begin{equation}\label{4pt-leading}
\mathcal{A}_4=\frac{(\varepsilon^{ABCD}\epsilon_{1A}\epsilon_{2B}\epsilon_{3C}\epsilon_{4D})^2}{((k_1+k_2)^2-(M_1+M_2)^2)((k_1+k_4)^2-(M_1+M_4)^2)}\,.
\end{equation}
From this expression one can extract four point amplitudes for specific states by assigning the correct kinematics, polarization and mass to the external particles. $W$-bosons have massive momenta, decomposed into massive spinor-heliciy variables, and generic polarization $\epsilon^a$, together with the following assignment of mass parameters\footnote{One should keep in mind that $M^H$ denotes the eigenvalue under the symmetry by which we are reducing and it corresponds to a \textit{signed} mass parameter. Here $m$ is the (positive) mass of the $W$-bosons.}:
\begin{equation}
M^W=m\qquad M^{\bar W}=-m
\end{equation}
Gluons, on the other hand, have massless momenta, whose spinor helicity variables are embedded in the massive ones as explained in section \eqref{spinor-helicity-54d}. They have $M^g=0$, and their polarization data depends on the helicity of the gluon;
\begin{equation}
\epsilon_a^{+1}=(1,0)\qquad\epsilon_a^{-1}=(0,1)\,.
\end{equation}
From the discussion in \S\ref{sec:massive_as_correl}, it is clear that the amplitude vanishes unless $\sum_iM_i=0$, i.e. unless $W$ and $\bar W$ come in pairs, with any number of gluons. Let us consider for example the amplitude for a $W\bar W$ pair and two negative helicity gluons. From \eqref{4pt-leading} we get:
\begin{equation}\label{amp-2W2g}
    A_4(W,\bar{W},g^-,g^-)=\epsilon_{1a}\epsilon_{1b}\epsilon_{2c}\epsilon_{2d}\frac{[1^a2^c][1^b2^d]\la34\ra^2}{s_{12}(s_{14}-m^2)}\,,
\end{equation}
where $s_{ij}=(k_i+k_j)^2$.
In line with our previous work, the amplitudes are contracted into arbitrary polarisation data and one can deduce the amplitude in the standard spinor-helicity form with free little group indices by stripping off the $\epsilon_i$:
\begin{equation}
    A_4(W^{ab},\bar{W}^{cd},g^-,g^-)=\frac{[1^{a}2^{c}][1^{b}2^{d}]\la34\ra^2}{s_{12}(s_{14}-m^2)}+(a \leftrightarrow b)+(c\leftrightarrow d)\,.
\end{equation}
A particularly compact notation was introduced in \cite{Arkani-Hamed:2017jhn}, which we will employ from here on. Massive spinor-helicity variables are written in \textbf{bold}, to indicate that they carry completely symmetrized little group indices. We can then rewrite \eqref{amp-2W2g} one more time:
\begin{equation}
    A_4(W,\bar{W},g^-,g^-)=\frac{[\mathbf{1}\mathbf{2}]^2\la34\ra^2}{s_{12}(s_{14}-m^2)}\,.
\end{equation}
Similarly one can obtain expressions for different orderings and helicity assignments:
\begin{equation*}
    A_4(W,\bar{W},g^-,g^+)=\frac{(\la\mathbf{1}3\ra[\mathbf{2}4]-\la\mathbf{2}3\ra[\mathbf{1}4])^2}{s_{12}(s_{14}-m^2)}\qquad
    A_4(W,g^-,\bar{W},g^-)=\frac{[\mathbf{1}\mathbf{3}]^2\la24\ra^2}{(s_{12}-m^2)(s_{14}-m^2)}\,.
\end{equation*}
Finally, for four $W$ bosons:
\begin{equation*}
    A_4(W,\bar{W},W,\bar{W})=\frac{1}{s_{12}s_{14}}\cdot\Big(\la\mathbf{1}\mathbf{2}\ra[\mathbf{3}\mathbf{4}]+[\mathbf{1}\mathbf{2}]\la\mathbf{3}\mathbf{4}\ra-\la\mathbf{1}\mathbf{3}\ra[\mathbf{2}\mathbf{4}]-[\mathbf{1}\mathbf{3}]\la\mathbf{2}\mathbf{4}\ra+\la\mathbf{1}\mathbf{4}\ra[\mathbf{2}\mathbf{3}]+[\mathbf{1}\mathbf{4}]\la\mathbf{2}\mathbf{3}\ra\Big)^2
\end{equation*}
See also  \cite{Cachazo:2018hqa} for similar formulae from dimensional reduction and  \cite{Craig_2011,Herderschee_2019} from BCFW recursion.

Comparing our results to the $n$-point formula obtained by recursion in \cite{Craig_2011,Herderschee_2019,Ochirov_2018} for a pair of $W \bar W$ bosons and $n-2$ same-helicity gluons, we could further verify numerically agreement for 5 particles;
\begin{equation}
A_5(W,\bar W,g^+,g^+,g^+)=\frac{\la\mathbf{1}\mathbf{2}\ra^2[3|(m^2+(k_4+k_5+k_1)(k_2+k_3+k_4))|5]}{\la34\ra\la45\ra((k_2+k_3)^2-m^2)((k_2+k_3+k_4)^2-m^2)}\,.
\end{equation}
The existence of compact expressions at $n$ points suggests that simplifications might occur in our formulae when this particular set of polarisation data is chosen with many $g^+$ particles as in MHV amplitudes in four dimensions. For massless amplitudes, this is well established, and relies (among other properties) on the \emph{refinement} of the scattering equations by MHV sector. However, this property does not extend to massive amplitudes, which are supported on all $(n-1)!$ solutions of the scattering equations. This clearly poses a challenge for attempting to make contact with the compact $n$-particle results of \cite{Craig_2011,Herderschee_2019,Ochirov_2018}, and we will return to this point briefly in \S\ref{sec:disc-chap5}.

\paragraph{Massive scalars and gluons}
In order to obtain amplitudes for states further down the multiplet, one needs to consider the expansion in supermomenta \eqref{CB-massive-massless-multiplet} and take the corresponding derivatives of the exponential factor. The massive and massless supermultiplet have a similar structure, with the main distinction being that in the massless multiplet the component $\sim q^2\Omega_{IJ}$ is a scalar state. For example,  the amplitude for two massive scalars and two gluons is extracted from the superamplitude as follows:
\begin{equation}
A_{4}\left(w_{IJ},g,g,\bar w_{KL}\right)=\left.\mathcal{A}_4 \frac{\partial}{\partial q_{1}^{I}} \frac{\partial}{\partial q_{1}^{J}} \frac{\partial}{\partial q_{4}^{K}} \frac{\partial}{\partial q_{4}^{L}} e^{F_{\mathcal{N}}}\right|_{q_{i}=0}\,,
\end{equation}
where $\mathcal{A}_4 $ is the leading amplitude \eqref{4pt-leading} and the exponential is given by \eqref{exponential-F}. The only term contributing is the quadratic one in the expansion of the exponential. Since for massive scalar states $\Omega_{IJ}w^{IJ}=0$, the terms $\sim q_j^2$ do not contribute and the derivatives bring down a factor of $U_{14}^2(\Omega_{IK}\Omega_{JL}+\Omega_{IL}\Omega_{JK})$  in  front of the amplitude, with $U_{ij}=\frac{u_{ij}}{\sigma_{ij}}$. Evaluated on the solution to the scattering equations, this gives:
\begin{equation}
A_{4}\left(w_{IJ},g,g,\bar w_{KL}\right)
=(\Omega_{IK}\Omega_{JL}+\Omega_{IL}\Omega_{JK})\frac{(\varepsilon^{ABCD}\epsilon_{ab}\kappa^a_{1A}\kappa^b_{1B}\epsilon_{2C}\epsilon_{3D})^2}{(s_{12}-m^2)s_{14}}\,.
\end{equation}
We can now evaluate the amplitude for different helicity assignments:
\begin{equation}
\begin{aligned}
A_{4}\left(w_{IJ},g^+,g^-,\bar w_{KL}\right)&=(\Omega_{IK}\Omega_{JL}+\Omega_{IL}\Omega_{JK})\frac{\la3|k_1|2]^2}{(s_{12}-m^2)s_{14}}\,,\\
A_{4}\left(w_{IJ},g^+,g^+,\bar w_{KL}\right)&=(\Omega_{IK}\Omega_{JL}+\Omega_{IL}\Omega_{JK})\frac{m^2[23]^2}{(s_{12}-m^2)s_{14}}\,.\\
\end{aligned}
\end{equation}
These expressions match the results obtained in \cite{Forde:2005ue,Arkani-Hamed:2017jhn}.

\paragraph{Massive quarks} We can perform the symmetry reduction to generate the symmetry breaking $SU(N+1)\rightarrow SU(N)\times U(1)$. From the discussion in the previous section, we know that the massive states are in the fundamental of $SU(N)$. At tree level, when looking at amplitudes involving only gluons and massive fermions, the truncation of the theory is consistent with the standard model description of massive quarks in QCD, and the massive ambitwistor string generates amplitudes that match the results in the literature \cite{Ochirov_2018};
\begin{equation*}
    A_4(\psi,\bar\psi,g^+,g^+)=m\frac{\la\mathbf{12}\ra[34]^2}{s_{12}(s_{14}-m^2)}\, ,\qquad
    A_4(\psi,\bar\psi,g^-,g^+)=\frac{\la3|k_1|4]([\mathbf{1}4]\la\mathbf{2}3\ra-[\mathbf{2}4]\la\mathbf{1}3\ra)}{s_{12}(s_{14}-m^2)}\,.
\end{equation*}

\subsection{Supersymmetric Ward identities.}\label{sec:ward-id}
At $n$-points, while we cannot simplify the formula further, we can establish relations between the different component amplitudes for one pair of massive particles and $(n-2)$ gluons of positive (resp.\ negative) helicity. These relations rely on supersymmetric Ward identities, as well as the simple structure of the polarized scattering equations for this particular configuration. Indeed, for this specific configuration, the $\alpha$ ($\dot\alpha$) components of the polarized scattering equations are particularly simple because only the massive particles contribute to it, so it decouples from the rest of the system. Taking the massive particle indices to be $i=1,2$, these components take the form
\begin{equation}\label{SE_special_case}
    U_{12}\epsilon_{2\alpha}=(v_1\kappa_{1\alpha})\, , \qquad U_{12}=\frac{m}{\la12\ra}\, .
\end{equation}
On the other hand, supersymmetric Ward identities relate the different component amplitudes, e.g. for W-bosons and massive fermions,
\begin{equation*}
    A_n(\psi,\bar\psi,g^+,\dots g^+)=A_n(W,\bar W,g^+,\dots g^+)\partial_{q_1}\partial_{q_2}\exp(\mathcal{F}_\mathcal{N})=U_{12}A_n(W,\bar W,g^+,\dots g^+)\, .
\end{equation*}
This  follows from the expansion of the supermultiplets on on-shell superspace \eqref{CB-massive-massless-multiplet}. Using the simple expression for $U_{12}$ derived above from the polarized scattering equations for these kinematics, this establishes a supersymmetry Ward identity,
\begin{equation}
    A_n(\psi,\bar\psi,g^+,\dots g^+)=\frac{m}{\la12\ra}A_n(W,\bar W,g^+,\dots g^+)\,,
\end{equation}
which is consistent with the $n$-point formulae obtained in \cite{Forde:2005ue} and \cite{Ochirov_2018}. Similar relations hold for scalars:
\begin{equation}
    A_n(w,\bar w,g^+,\dots g^+)=\left(\frac{m}{\la12\ra}\right)^2A_n(W,\bar W,g^+,\dots g^+)\,,
\end{equation}
which agrees with\cite{Lazopoulos:2021mna}.

Instead of deriving these Ward relations from the requirements of supersymmetry and the support of the polarized scattering equations, we can alternatively derive them from supersymmetry alone (since supersymmetric invariance of the amplitude implies the polarized scattering equations in this worldsheet formalism). To see this, note first that supersymmetry of the amplitude implies
\begin{equation}
    \mathcal{A}_n=A_n(W,\bar W,g^+,\dots g^+)\big(1+\sum_{j<k}U_{jk}q_j\cdot q_k-\frac{1}{2}\sum_j(\xi_jv_j)q_j^2+\mathcal{O}(q^4)\big)\,,
\end{equation}
for two massive and $n-2$ massless states. Then by the definition of on-shell superspace;
\begin{equation}
\begin{aligned}
    0&=[Q_{AI},\mathcal{A}_n]=\sum_i[Q_{iAI},\mathcal{A}_n]\\
     & =A_n(W,\bar W,g^+,\dots g^+)\sum_i\big((\kappa_{iA}\xi_i)q_{iI}-\epsilon_{iA}(\xi_iv_i)q_{iI}+\sum_{j\neq i}\epsilon_{jA}U_{ji}q_{jI}+\mathcal{O}(q^3)\big)\,.
\end{aligned}
\end{equation}
Then at order $\mathcal{O}(q_{1})$, supersymmetry imposes:
\begin{equation}
    (\kappa_{1A}\xi_1)+\sum_{k\neq 1}U_{k1}\epsilon_{kA}-(\xi_1v_1)\epsilon_{1A}=0\,,
\end{equation}
Contracting into $(\epsilon_{1}^\alpha,0)$, we see that this is equivalent to the polarised scattering equation \eqref{SE_special_case}, using that $\xi$ is normalised against $\epsilon$ as $(\epsilon_i\xi_i)=1$, and we can derive the Ward identity following the same steps as above.

The above discussion reveals two interesting cases where relations between component amplitudes can be derived \emph{without} explicitly solving all of the (polarized) scattering equations, but where their simplicity can be used to easily find important quantities such as $U_{12}$ above. It would be interesting to search for more of these cases, see also the discussion section \ref{sec:disc-chap5}.

\section{R-symmetry reduction}\label{sectionRsymred}
Our symmetry reduction procedure provides a general framework, requiring only (i) a massless theory in $d+1$ dimensions, and (ii) the identification of a symmetry to generate the mass. It can thus be applied to give less familiar theories with massive  particles from any choice of higher-dimensional symmetry. We illustrate this here by performing a reduction of maximally supersymmetric gauge and gravity theories using a generator of the R-symmetry to provide the masses. This gives massive scalars and spinors that are in nontrivial $R$-symmetry representations, and consequently breaks maximal supersymmetry. In the supergravity literature, such reductions are known as (Cramer)-Scherk-Schwarz (CSS) reductions and have been shown to generate gauged supergravities in four dimensions with massive particles. We will present a few examples of these theories in our formalism and discuss a peculiar instance of double copy at the level of the worldsheet.

\subsection{Reducing SYM}\label{SYM-Rsymred}
Let us first investigate the simpler of these cases; the R-symmetry reduction of maximially supersymmetric Yang-Mills theory. Following the general procedure of \cref{sec:2TT-SymRed}, we perform a symmetry reduction by associating one component of momentum to charges $H_I^K$ in the adjoint of the $R-$symmetry group $\mathrm{Sp}(4)_R$;
\begin{equation}\label{Rsymred}
\frac{\partial}{\partial x^4}(A_m,\Phi_{IJ},\Psi_A^{I})=(0,H_{[I}^K\Phi_{J]K},H_I^K\Psi_{AK})\,.
\end{equation}
Expanding the kinetic terms of the action under \eqref{Rsymred}, one obtains mass terms for some of the fermions and scalars as well as some interaction terms. The reduced theory contains one massless vector $A_\mu$, two massless scalars, four massive scalars and four massive Majorana fermions. Details are given in appendix \ref{R-sym-red-appendix}.

\paragraph{Spectrum}
For $\mathrm{Sp}(4)$, in the Cartan-Weyl basis we can write a linear combination of the two Cartan elements in the fundamental as:
\begin{equation}\label{Cartan-R-sym}
    H=\mathrm{diag}(m_1,m_2,-m_1,-m_2)\,,
\end{equation}
where $m_s$ are the mass parameters of the linear combination (\emph{not} the particle masses). In this basis the symplectic matrix $\Omega$ is taken to be
\begin{equation}
    \Omega_{IJ}=
\begin{pmatrix}
0 & \mathds{1}_2\\
-\mathds{1}_2 & 0\\
\end{pmatrix}
\end{equation}
and we will raise and lower $R$-symmetry indices with $\Omega$. Then we can take as independent scalars $\Phi_{12},\Phi_{13},\Phi_{23},\Phi_{24},\Phi_{34}$, with $\Phi_{13}=-\Phi_{24}$.
Expanding the mass term for the scalars we find the mass assignment given in \cref{tab:R-sym_SYM}. We note that for the choice $m_1=m=m_2$ one obtains a theory of one massless vector, four massless scalars, four massive fermions of mass $m$ and two massive scalars of mass $2m$. We will refer to this theory as $\mathcal{N}=0^*$ SYM.

\begin{table}[ht]
\centering
\begin{tabular}{*{4}{>{$}c<{$}} }
    \toprule
    \textbf{Mass}  & 0 & |m_1+m_2| & |m_1-m_2|\\
    \midrule
 \text{scalars} & \phi,\, \Phi_{13} & \Phi_{12}, \, \Phi_{34} & \Phi_{14}, \, \Phi_{23}\\
 \text{gluon} & F_{\alpha\beta},\,\tilde{F}_{\dot\alpha\dot\beta} &  &\Bstrut\\[20pt]
    \toprule
 \textbf{Mass}  && |m_1| & |m_2| \\
 \midrule
 \text{Dirac spinors} && \Psi_1+i\Psi_3 & \Psi_2+i\Psi_4
\end{tabular}
\caption{States in the generic R-symmetry reduction of maximal SYM. The scalars $\Phi_{13}$ and $\phi$, coming from the extra component of the five dimensional vector, remain massless.}
\label{tab:R-sym_SYM}
\end{table}

\paragraph{Supersymmetry}
It is clear from these mass assignments that the original $SO(5)$ R-symmetry is broken. Counting on shell degrees of freedom one sees that there can be no residual supersymmetry in this theory if both $m_{1,2}$ are non-vanishing. However, it is possible to preserve $\mathcal{N}=2$ supersymmetry by performing the reduction only along one of the two directions in the Cartan, i.e. taking for example $m_2=0,\,m_1=m$. In this case half of the fermions are massless and we can group the on shell degrees of freedom into one massive ultrashort matter multiplet $\Psi^m$ and one massless vector multiplet $\mathcal{V}^0$, with:
\begin{center}
\begin{tabular}{ l l }
 $\Psi^m$ \quad = & 1 massive Dirac fermion $\Psi_1^B+i\Psi_3^B$,\\
                  & 2 massive complex scalars $\Phi_{12}+i\Phi_{23}$ and $\Phi_{14}+i\Phi_{34}$\\ 
	              & 			\\ 
 $\mathcal{V}^0$\quad = & 1 vector $(F_{\alpha\beta},F_{\dot\alpha\dot\beta})$\\
 						& 2 Weyl fermions $(\Psi_2,\Psi_4)$\\
 						& 1 complex scalar $\phi+i\Phi_{24}$\,,
\end{tabular}
\end{center}
where the 5d gauge field decomposes into its 4d components $F_{\alpha\beta},F_{\dot\alpha\dot\beta}$ and a scalar component $\phi$. Then, in the basis of \eqref{Cartan-R-sym} with $m_2=0$, we find that $\mathcal{N}=2$ supersymmetry is preserved, since for $J=2,3$
\begin{equation}
\begin{aligned}
  \left[Q_{AJ},P^5\right]&\,(\Psi_1^B+i\Psi_3^B,\Phi_{12}+i\Phi_{23},\Phi_{14}+i\Phi_{34})=0\\
  \left[Q_{AJ},P^5\right]&\,(F_{\alpha\beta},F_{\dot\alpha\dot\beta},\Psi_2,\Psi_4,\phi+i\Phi_{24})=0  \,
  .
\end{aligned}
\end{equation}
 Thus the states of the reduced theory sit in $\mathcal{N}=2$ supermultiplets generated in this basis by the action of $Q_{2,3}$, whereas the two remaining supercharges of the higher dimensional theory transform massless to massive states and vice versa.

We recognize the reduced theory as the $\mathcal{N}=2^*$ theory that one obtains by giving a mass to the adjoint $\mathcal{N}=2$ hypermultiplet sitting inside the $\mathcal{N}=4$ massless vector.
More generally, for a reduction with $p$ non-vanishing mass parameters we obtain a theory with $\cN=4-2p$ residual supersymmetries.

\paragraph{Worldsheet model.} The operators:
\begin{equation}
J_{IJ}=\eta_{(I}\cdot\eta_{J)}
\end{equation}
are generators for the $\mathrm{Sp}(4)$ R-symmetry acting on worldsheet operators. We can then take the current:
\begin{equation}
j_H=\eta_{(I}\cdot\eta_{J)} H^{IJ}=m_1\eta_{(1}\cdot\eta_{3)}+m_2\eta_{(2}\cdot\eta_{4)}\,,
\end{equation}
to construct a massive ambitwistor string model for the class of R-symmetry reductions of sYM introduced above. This element spans the Cartan subalgebra for different values of $m_{1,2}$ and does not spoil the closure of the algebra of constraints
\footnote{The only non-trivial OPE is:
\begin{equation}
(\lambda^2-j_H)\circ(\mathcal{Z}^{(a}\cdot\mathcal{W}^{b)})\sim 0 \, .
\end{equation}}.

The mass assignment for various states in the multiplet is then determined by the OPE
\begin{equation}
j_H(\sigma)\circ\Phi_{\kappa,q}(\sigma_i)\sim\frac{1}{\sigma-\sigma_i}\sum_{s=1}^2m_s\big(q_{i;s}\frac{\partial}{\partial q_{i;s}}-q_{i;s+2}\frac{\partial}{\partial q_{i;s+2}}\big)\Phi_{\kappa,q}\, ,
\end{equation}
where $\Phi_{\kappa,q}$ is the supersymmetric plane wave representative \eqref{super-representative} that forms the basic building block of the vertex operator. Using the general OPE $\mathrm{Res} (j_H\circ V)=MV$ of \eqref{group-action-on-vo}, this gauging assigns masses
\begin{equation}
M = m_1\big(q_{1}\frac{\partial}{\partial q_{1}}-q_{3}\frac{\partial}{\partial q_{3}}\big)+m_2\big(q_{2}\frac{\partial}{\partial q_{2}}-q_{4}\frac{\partial}{\partial q_{4}}\big)\,.
\end{equation}
Here, the mass parameter $M$ is understood as an operator containing derivative with respect to the supermomenta. Different states in the original 5d multiplet are thereby assigned different masses; e.g. gluons (at the top and the bottom of the multiplet) remain massless. We can easily convince ourselves that the resulting spectrum indeed agrees with the Lagrangian discussion above.

\paragraph{Formulae.} From these models we obtain amplitude formulae for theories with various amounts of residual supersymmetry, such as the $\mathcal{N}=2^*$ theory discussed above. The peculiarity in these expressions is that the full superamplitude remains expressed as an expansion in the original (now broken) $\mathcal{N}=4$ superspace. One can then extract amplitudes involving the desired massive or massless particles by reading them off from the appropriate coefficients. As we have seen above, the mass operator contains derivatives with respect to specific components of supermomenta, so massive and massless multiplets of the reduced theory can be embedded together in the larger (broken) superspace, at the price of introducing derivatives in the scattering equations.

\subsection{CSS gauged supergravities}
The same kind of symmetry reduction can be carried out on the gravitational model, exploiting the $\mathrm{Sp}(8)$ R-symmetry of five dimensional maximal supergravity. In the supergravity literature this procedure goes by the name of CSS reduction, after Cremmer, Scherk and Schwarz \cite{Scherk:1979zr,Cremmer:1979uq}, see also the companion paper \cite{upcoming:part1}. It has been shown \cite{Andrianopoli_2002} that the result of a CSS reduction of five dimensional supergravity by an element $H$ of the $E_{6(6)}$ 
is a gauged supergravity in four dimensions. We will equivalently refer to such models as CSS reductions or CSS gaugings. When taking the element $H$ in the maximal compact subgroup $\mathrm{USp}(8)$ of $E_{6(6)}$, the gauge group is called `flat' and the theory has Minkowski vacua. It depends on four independent mass parameters, corresponding to the four elements of the Cartan subgroup of $\mathrm{Sp}(8)$. These fix the scale of the spontaneous supersymmetry breaking, which produces a theory with residual $\mathcal{N}=8-2p$ supersymmetry, with $p$ the number of vanishing mass parameters.

Maximal supergravity in five dimensions has a gravity multiplet whose content is summarized in table \ref{table:1}.
\begin{table}[ht]
\centering
\begin{tabular}{c  c  c}
Spin & dof & Sp$(4)_R$	\Bstrut\\
\midrule
$2$ & $5_B$ & \textbf{1} \Tstrut\\
$\frac{3}{2}$ & $8\times4_F$ & \textbf{8}\\
$1$ & $27\times3_B$ & \textbf{27}\\
$\frac{1}{2}$ & $48\times2_F$ & \textbf{48}\\
$0$ & $42\times1_B$ & \textbf{42}
\end{tabular}
\caption{Degrees of freedom and R-symmetry representation for the states in the gravity multiplet in $5$d $N=4$ supergravity.}
\label{table:1}
\end{table}
All particles but the graviton transform non trivially as $\mathrm{Sp}(8)$ antisymmetric traceless\footnote{This is intended as $\Phi_{I_1I_2\dots I_k}\Omega^{I_iI_j}=0$.} tensors. As a consequence, under an R-symmetry reduction of the form
\begin{equation}
\partial_4\Phi_{I_1I_2\dots I_k}=H_{[I_1}^J\Phi_{I_2\dots I_k]J}\,,
\end{equation}
some of the gravitinos, graviphotons, gravi-photinos and scalars acquire a mass. For $\mathrm{Sp}(8)$, choosing an analogous basis as above for $\mathrm{Sp}(4)$, we can write a generic element of the four dimensional Cartan subalgebra as:
\begin{equation}\label{eq:Cartan_masses}
H=\diag(m_1,m_2,m_3,m_4,-m_1,-m_2,-m_3,-m_4)\,.
\end{equation}
Then, for a $k-$index tensor, the mass assignment is:
\begin{equation}
    m=\Big|\sum_{s=1}^{k}m_{I_s}\Big|\,.
\end{equation}
These theories are described by worldsheet models of the form:
\begin{equation}
S=\int_{\Sigma} \mathcal{Z}^a \cdot \bar{\partial}_e \mathcal{Z}_a+A_{a b} \mathcal{Z}^a \cdot \mathcal{Z}^b+a\left(\lambda^2-j^{H_R}\right)+\tilde{a}\left(\tilde{\lambda}^2-j^{H_R}\right)+S_{\rho_1}+S_{\rho_2}\,,
\end{equation}
where the $\mathrm{Sp}(8)$ current is given by:
\begin{equation}
j^{H_R}=\eta_I\cdot\eta_JH^{IJ}\,.
\end{equation}
Similarly to the massive ``CSS'' gauge models of section \ref{SYM-Rsymred}, this current generates masses only for part of the original supermultiplet by acting as a derivative in the supermomenta:
\begin{equation}
j^{H_R}(\sigma)\cdot\Phi_{\kappa,q}(\sigma_i)\sim\frac{1}{\sigma-\sigma_i}\sum_{s=1}^4m_s\left(q_{i;s}\frac{\partial}{\partial q_{i;s}}-q_{i;s+4}\frac{\partial}{\partial q_{i;s+4}}\right)\Phi_{\kappa,q}=:\frac{\mathcal{D}q_i}{\sigma-\sigma_i}\Phi_{\kappa,q}\,,
\end{equation}
As a consequence, the formulae we obtain are superamplitudes containing both massive and massless component amplitudes. It is only once we specify the external states that we can talk about the scattering equations that localise the correlator being massive or massless. This is best seen for the vectorial scattering equations, which take the form
\begin{equation}\label{diff-SE}
\delta\left(\sum_{j=1}^n\frac{k_i\cdot k_j-\mathcal{D}q_i\mathcal{D}q_j}{\sigma_{ij}}\right)\e^{F_\cN}\,.
\end{equation}
As in the gauge theory R-symmetry reduction, models with $p$ non-vanishing mass parameters $m_l$ preserve $\mathcal{N}=8-2p$ supersymmetry. Below, we discuss some cases of special interest.

\paragraph{CSS gauging with $\mathcal{N}=6$.}
We begin by considering the case $m_1=m$, $m_{2,3,4}=0$.
Taking into account the tracelessness conditions, we summarize below the massive and massless spectrum of the reduced theory. We find that the reduction preserves $\mathcal{N}=8-2=6$ supersymmetry and the states make up one massless gravity multiplet and one massive ultrashort gravitino multiplet, c.f. \cref{tab:table2}. This corresponds to a model with $j^H=m\,\eta_{(1}\cdot\eta_{5)}$.

\begin{table}[ht]
\centering
\begin{tabular}{*{3}{>{$}c<{$}} }
    \toprule
\multirow{2.4}{*}{\textbf{Spin}}
    & \multicolumn{2}{c}{\textbf{Mass}}   \\
    \cmidrule{2-3}
    &  0     & m              \\
    \midrule
2 & 2_B & \Tstrut\\
\frac{3}{2} & 6\times2_F & 2\times4_F\\
1 & 16\times2_B & 12\times3_B\\
\frac{1}{2} & 26\times2_F & 28\times2_F\\
0 & 30\times1_B & 28\times1_B\Bstrut\\
    \bottomrule
& \mathcal{H}^0_{\mathcal{N}=6} & 2\mathcal{X}^m_{\mathcal{N}=6}\Tstrut
\end{tabular}
\caption{Spectrum of the reduced theory under the choice $H=\diag(m,0,0,0,-m,0,0,0)$.}
\label{tab:table2}
\end{table}

\paragraph{CSS gauging with $\mathcal{N}=4$.}
By the same procedure we can obtain theories with residual $\mathcal{N}=4$ supersymmetry. Taking $m_{1}=m=-m_2$, $m_{3,4}=0$, we obtain one massless graviton multiplet, four massless vector multiplets, four massive gravitino multiplets with mass $m$ and two massive vector multiplets of mass $2m$, see \cref{tab:table3}. All massive multiplets are ultrashort representations. The corresponding model has $j^H=m\,(\eta_{(1}\cdot\eta_{5)}-\eta_{(2}\cdot\eta_{6)})$.

\begin{table}[ht]
\centering
\begin{tabular}{*{4}{>{$}c<{$}} }
    \toprule
\multirow{2.4}{*}{\textbf{Spin}}
    & \multicolumn{3}{c}{\textbf{Mass}}   \\
    \cmidrule{2-4}
    &  0     & m      & 2m          \\
    \midrule
2 		  & 2_B 		  & 				&				 \Tstrut\\
\frac{3}{2} & 4\times2_F  & 4\times4_F	&				\\
1 		  & 10\times2_B & 16\times3_B	&	2\times3_B	\\
\frac{1}{2} & 20\times2_F & 24\times2_F	&	8\times2_F	\\
0 		  & 26\times1_B & 16\times1_B	&	10\times1_B	\Bstrut	\\
    \bottomrule
    & \mathcal{H}^0_{\mathcal{N}=4}\oplus4\mathcal{V}^0 & 4\mathcal{X}^m_{\mathcal{N}=4} & 2\mathcal{V}^{2m}\Tstrut
\end{tabular}
\caption{Spectrum of the reduced theory under the choice $H=\diag(m,-m,0,0,-m,m,0,0)$.}
\label{tab:table3}
\end{table}

\subsection{Double copy}
The gauged supergravities described above have been the object of recent work by Chiodaroli, G\"unaydin, Johansson and Roiban \cite{Chiodaroli:2015wal,Chiodaroli:2017ngp,Chiodaroli:2018dbu}, who have studied how they can be obtained as double copies of massive gauge theories.
Worldsheet models in the ambitwistor string and the formulae they produce have an explicit double copy structure, whereby one chooses a left and a right systems, which can be combined in any pairing. Having constructed models for gauged supergravities, we observe an  instance of double copy where one supergravity theory can be arise from several different left/right pairs. Here we describe this novel `worldsheet' double copy, we illustrate it with examples and we relate it to the momentum space double copy of \cite{Chiodaroli:2018dbu}.

On the worldsheet, we establish a prescription for double copying gauge theory models. We start with two models that are composed of one set of worldsheet fermions $S_\rho$ and one current algebra $S_j$. We also consider the $\eta$-system to come as part of the matter action\footnote{This is justified by the criticality of the models.} and to incorporate supersymmetry breaking terms such as those in the R-symmetry reductions discussed above. Altogether the models take the form:
\begin{equation}
 S = S_{4d}^0+\left(S^{\eta}_{\mathcal{N}}(H^R)+S_{\rho}\right)+\textcolor{gray}{\left(S_j-\int_\Sigma (aj^H_{CB}+\tilde aj^H_{CB})\right)}\,.
\end{equation}
Here, $S_{4d}^0$ is the trivially reduced bosonic model \eqref{symred-model} with $j^H=0$,\footnote{This 4d massless model will also play an important role for loop amplitudes from the worldsheet, as we will see in the next section.} and we have absorbed the currents $j^H$ into their respective matter systems. As such, $j^H_{CB}$ is associated to an element of the color group for the Coulomb Branch, and we group the fermionic components of the twistors with the $S_\rho$-system;
\begin{equation*}
S^{\eta}_{\mathcal{N}}(H^R)=\int_{\Sigma}\eta_I\cdot\dbar \eta^I+A_{ab}\eta_I^a\eta^{bI}-\big(a+\tilde a)\eta_I\cdot\eta_JH_R^{IJ}\,.
\end{equation*}
Note that this pairing is also suggested by the anomaly cancellations. the resulting model has $\mathcal{N}=4-2p$ residual supersymmetry after spontaneous symmetry breaking, with $p$ the number of non-vanishing mass parameters, and the R-symmetry indices run up to $4$.

From two models of this type we can form a gravitational model by replacing the current algebrae (and associated Coulomb-Branch-like gaugings) with the worldsheet fermion systems to write:
\begin{equation}
S^{\eta}_{\mathcal{N}_1+\mathcal{N}_2}\left(\scalebox{0.8}{$
\left( \begin{array}{cc}
H_1^R & 0 \\
0 & H_2^R 
\end{array} \right)$}
\right)+S_{\rho_1}+S_{\rho_2}\,.
\end{equation}
We identify $\eta^{\cI}=(\eta_1^I,\eta_2^I)$ as well as $\Omega_{\mathcal{IJ}}=\scalebox{0.8}{$\left( \begin{array}{cc}
\Omega_1^R & 0 \\
0 & \Omega_2^R 
\end{array} \right)$}$, so that schematically we can write:
\begin{equation}
\mathrm{SYM}\big(H_1^{\scalebox{0.6}{CB}}\oplus H_1^{R}\big)\,\otimes\, \mathrm{SYM}\big(H_2^{\scalebox{0.6}{CB}}\oplus H_2^{R}\big)\sim \mathrm{sugra}\big(H_1^R\oplus H_2^R\big)
\end{equation}
The charge associated to the symmetry reduction is the sum of the R-symmetry charges of the two gauge theories, through $j^H_{\mathrm{sugra}}=j^{H_1^R}_{SYM}+j^{H_2^R}_{SYM}$. This indicates that on the worldsheet the double copy as prescribed here doesn't need the mass spectra of the \textit{left} and \textit{right} theories to match. As indicated by the notation, there are in fact a lot of different pairings that produce the same double copy. Not only can we have different values of $\mathcal{N}_1$ and $\mathcal{N}_2$ summing to $\mathcal{N}$, but we are also free to add any Coulomb-branch-like reductions to both gauge models, see Table \ref{table:DC}.

The reason for this greater flexibility of the double copy on the worldsheet is ultimately the role the currents $j^H$ and scattering equations play in determining the masses and propagators: $j^H$ both determines the mass assignment, and ensures via the symmetry reduction procedure that all propagators have the correct form. On the other hand, the currents $j^H$ play a key part in the worldsheet double copy prescription, with Coulomb-branch currents $j^H_{CB}$ -- dropped during the double copy --- never contributing to the mass assignment in the gravitational theory. Since the R-symmetry charges are additive, there is no need to match the mass spectra of the gauge theories. On momentum space on the other hand, the propagator structure dictates that a double copy can only be defined for matching mass spectra.

As an example, let us consider the CSS gauged supergravity with residual $\mathcal{N}=6$ supersymmetry. Here we have no choice but to take one of the two models to be $\mathcal{N}=4$ SYM and the other the $\mathcal{N}=2^*$ massive theory of section \ref{SYM-Rsymred}. Both of these models are in principle free to be on the Coulomb branch. The multiplets and respective R-symmetry charges of the left and right theories combine as follows:
\begin{equation}
\mathcal{V}^0\otimes\mathcal{V}^0\rightarrow\mathcal{H}^0\qquad\mathcal{V}^0\otimes\Psi^{\pm m}\rightarrow\mathcal{X}^{\pm m}\,,
\end{equation}
so that overall they double copy to $\mathcal{N}=6^*$ supergravity with one gravity multiplet and two massive gravitino multiplets:
\begin{equation}
\underbrace{\mathcal{V}^0}_{\mathcal{N}=4}\otimes\underbrace{(\mathcal{V}^0\oplus\Psi^{+m}\oplus\Psi^{-m})}_{\mathcal{N}=2^*}\rightarrow\underbrace{(\mathcal{H}^0\oplus\mathcal{X}^{+m}\oplus\mathcal{X}^{-m})}_{\mathcal{N}=6^*}\,.
\end{equation}
We insist on the fact that this is a double copy prescription \textit{on the worldsheet}. In order to understand what this implies for spacetime amplitudes, it is easier to consider a component amplitude instead of the full superamplitude, so that the equations \eqref{diff-SE} become proper massive or massless scattering equations. We pick a component amplitude involving two massive gravitinos and all massless top state gravitons, so that the kinematic variables involved in the scattering equations are massive for $i=1,2$ and massless for the rest. We obtain the desired amplitude on spacetime by evaluating the correlator on the solutions to the scattering equations:
\begin{equation}
\sum_{\{\sigma_i,u_i,v_i\}}U_{12}\frac{(\sigma_{ij}\sigma_{jk}\sigma_{ki})^2}{\det\Phi_{ijk}^{ijk}}\det{}'\HH\det{}'\HH\,,
\end{equation}
where the factor $U_{12}$ comes from the supersymmetry exponential factor.
This expression still presents a double copy structure, where each of the two reduced determinants can be taken as a contribution from a gauge theory model. However, both sub-integrands are evaluated on the solutions to scattering equations that are massive in particles $1$ and $2$, so that the mass spectrum has to match between the two gauge theories, contrary to the worldsheet double copy. In particular, here a spin $3/2$ massive state has to come as a double copy of a massive spin $1/2$ with a \textit{massive} spin $1$. Repeating the same reasoning with the rest of the states, we find that \textit{on spacetime} the amplitude comes from the double copy:
\begin{equation}
\underbrace{(\mathcal{V}^0\oplus\mathcal{W}^m)}_{\mathrm{CB }\,\,\mathcal{N}=4}\otimes\underbrace{(\mathcal{V}^0\oplus\Psi^{+m}\oplus\Psi^{-m})}_{\mathcal{N}=2^*}\rightarrow\underbrace{(\mathcal{H}^0\oplus\mathcal{X}^{+m}\oplus\mathcal{X}^{-m})}_{\mathcal{N}=6^*}\,.
\end{equation}
That is, the momentum space double copy, after evaluating the massive scattering equations, requires the $\mathcal{N}=4$ theory to be on the Coulomb branch. In other words, evaluating the correlator on spacetime selects one pair of theories out of all the candidates for the worldsheet double copy. This result corresponds to what was observed in \cite{Chiodaroli:2018dbu} using BCJ numerators.

This phenomenon is even more explicit in the case of the $\mathcal{N}=4^*$ CSS supergravity. Here the worldsheet double copy allows both:
\begin{equation}
\underbrace{(\mathcal{V}^0\oplus\Psi^{+m}\oplus\Psi^{-m})}_{\mathcal{N}=2^*}\otimes\underbrace{(\mathcal{V}^0\oplus\Psi^{+m}\oplus\Psi^{-m})}_{\mathcal{N}=2^*}\rightarrow\underbrace{(\mathcal{H}^0\oplus4\mathcal{V}^0\oplus4\mathcal{X}^{m}\oplus2\mathcal{V}^{2m})}_{\mathcal{N}=4^*}\,.
\end{equation}
and
\begin{equation}
\underbrace{(\mathcal{V}^0\oplus\mathcal{W}^m\oplus\mathcal{W}^{2m})}_{\mathrm{CB}\,\,\mathcal{N}=4}\otimes\underbrace{(4\psi^m\oplus A^0\oplus2\phi^{2m}\oplus4\phi^0)}_{\mathcal{N}=0^*}\rightarrow\underbrace{(\mathcal{H}^0\oplus4\mathcal{V}^0\oplus4\mathcal{X}^{m}\oplus2\mathcal{V}^{2m})}_{\mathcal{N}=4^*}\,.
\end{equation}
where $\mathcal{N}=0^*$ is an R-symmetry reduction of Super-Yang Mills with no residual supersymmetry as described in section \ref{SYM-Rsymred}.
On spacetime, on the other hand, double copying two $\mathcal{N}=2^*$ we couldn't possibly produce states of mass $2m$ because of the requirement of mass matching. The only way we can obtain the desired spectrum is by double copying the $\mathcal{N}=0^*$ with $\mathcal{N}=4$ on the Coulomb branch with color symmetry breaking pattern $SU(3N)\rightarrow SU(N)\times SU(N)\times SU(N)$. We expect this to hold in the formalism of \cite{Chiodaroli:2018dbu}.

\begin{table}[ht]
\centering
\begin{tabular}{*{5}{c}}\toprule
\multirow{2.4}{*}{\textbf{Left}}
    & \multicolumn{4}{c}{\textbf{Right}}   \\[5pt]
    \cmidrule(lr){2-5}
    &  $\cN=2^*$  &  $\cN=2^*$ on CB & $\cN=4$ & $\cN=4$ on CB \\
    \cmidrule{1-5}
     $\cN=0^*$\phantom{ on CB}&  \tikzmark{top left 1}$\cN=2^*$   & $\cN=2^*$    &   \tikzmark{top left 2}$\cN=4^*$ &   $\pmb{\cN=4^*}$       \\
     $\cN=0^*$ on CB    & $\cN=2^*$ & $\cN=2^*$\tikzmark{bottom right 1}   &   $\cN=4^*$   &   $\cN=4^*$\tikzmark{bottom right 2}     \\[10pt]
    $\cN=2^*$\phantom{ on CB} &   \tikzmark{top left 3}$\cN=4^*$ & $\cN=4^*$ &  \tikzmark{top left 4}$\cN=6\phantom{^*}$ &   $\cN=6\phantom{^*}$     \\
    $\cN=2^*$ on CB  &   $\cN=4^*$ & $\cN=4^*$\tikzmark{bottom right 3} &  $\cN=6\phantom{^*}$ &   $\pmb{\cN=6}\phantom{^*}$\tikzmark{bottom right 4}     \\[5pt]
     \bottomrule
\end{tabular}
\caption{Double copy on the worldsheet. Within each coloured block, all resulting CSS supergravities are the same, so multiple left/right gauge theories double copy to the same supergravity on the world-sheet. The (unique) space-time double copy is highlighted in  bold-face.}
\label{table:DC}
\DrawBox[ultra thick, SteelBlue1 , fill=SteelBlue1 , opacity=0.25]{top left 1}{bottom right 1}
\DrawBox[ultra thick, DodgerBlue2, fill=DodgerBlue2, opacity=0.25]{top left 2}{bottom right 2}
\DrawBox[ultra thick, DodgerBlue2, fill=DodgerBlue2, opacity=0.25]{top left 3}{bottom right 3}
\DrawBox[ultra thick, DodgerBlue3, fill=DodgerBlue3, opacity=0.35]{top left 4}{bottom right 4}
\end{table}

\section{Loops from the gluing operator in four dimensions}\label{chapter:loops}
Whereas loop-level correlators have long been an active field of research in the RNS ambitwistor string \cite{Adamo:2013tsa, Casali:2014hfa,  Geyer:2015bja, Geyer:2015jch, He:2015yua, Feng:2016nrf,Cachazo:2015aol, Baadsgaard:2015twa,  Geyer:2017ela, He:2016mzd,  He:2017spx, Adamo:2015hoa, Geyer:2016wjx, Geyer:2018xwu, Geyer:2019hnn, Edison:2020uzf, Yu:2017bpw, Casali:2020knc, Feng:2019xiq,Cardona:2016bpi,Cardona:2016wcr,Gomez:2016cqb, Gomez:2017lhy, Gomez:2017cpe, Ahmadiniaz:2018nvr, Farrow:2020voh}, progress for the twistorial models has been rather limited  \cite{Dolan:2007vv, Farrow:2017eol, Wen:2020qrj}. The cause for this discrepancy is two-fold; firstly, in the $\mathcal{N}=4$ (ambi-)twistor string, superconformal gravity states propagating in the loop mix with  the $\mathcal{N}=4$ super Yang-Mills modes  when presented on the torus, whereas for $\mathcal{N}=8$ supergravity the absence of a $bc$-system complicates the calculation of torus correlators \cite{Skinner:2013xp}. Furthermore, the progress in the RNS ambitwistor string model relied on going to the \emph{nodal sphere} and then separating out the desired modes to run in the loop at the nodes, but this required relaxing the $P^2=0$ constraint which is automatically solved in the twistorial models.

This relies on a property specific to the ambitwistor string models: the one-loop integrands, in addition to being modular invariant, are fully localized on a loop-level extension of the scattering equations \cite{Adamo:2013tsa}. They can therefore be simplified  by a residue theorem on the moduli space \cite{Geyer:2015bja}, effectively trading  one of the scattering equations for a localization on the non-separating boundary divisor, where the torus degenerates to a nodal sphere. The resulting amplitude formul{\ae} over the nodal sphere are compact, manifestly rational, and can be extended from 10d supergravity to a variety of other theories and dimensions \cite{Geyer:2015jch,He:2015yua, Feng:2016nrf}. Moreover, extensions of this argument remain valid at higher loop order, recasting the loop integrand as a moduli integral over the $g$-nodal sphere \cite{Geyer:2016wjx, Geyer:2018xwu, Geyer:2021oox}.

An extension of this idea has been implemented successfully for twistorial amplitude expressions:  Wen and Zhang \cite{Wen:2020qrj} recently constructed D3-brane and Yang-Mills loop-amplitude formulae from a back-to-back forward limit of 6d tree-level twistorial amplitudes \cite{Heydeman:2017yww, Albonico:2020mge}.This back-to-back forward limit construction has its origin precisely in the nodal sphere description of the loop integrand, but a similar derivation of the formalism from a higher genus description of the twistor string is still lacking for the reasons discussed above.

However, there does exist an alternative derivation of the nodal-sphere loop amplitude expressions from the worldsheet model: At the level of the CFT, the  simple structure of the loop correlators originates from the presence of a so-called \emph{gluing operator} $\Delta$, which encodes the propagator of the target-space field theory \cite{Roehrig:2017gbt}. As an off-shell object, $\Delta$ cannot be a local operator in the CFT,\footnote{to be precise, it cannot be local and BRST-invariant, but we'd like to retain the latter} and it contains (in addition to a genuinely non-local factor) a pair of local operators --- corresponding to the off-shell states of the propagator ---  inserted at two special points $\sigma_+$ and $\sigma_-$. If the two marked points lie on different worldsheets, $\Delta$ functions as a standard tree-level propagator, and can be used to formulate the BCFW recursion at the level of the underlying CFT \cite{Roehrig:2017gbt}. However, if both $\sigma_\pm$ lie on the same sphere\footnote{corresponding to a non-separating boundary divisor of the genus-one moduli space} the correlators reproduce precisely the one-loop integrand formul{\ae} localized on the nodal sphere. In the ambitwistor string, one-loop integrands can thus be recovered from $g=0$ correlators in the presence of a gluing operator.

Here we propose that many of the issues plaguing the twistorial models at loop level can be resolved by following this latter strategy of defining a gluing operator and working directly on the nodal sphere. In an important distinction from the RNS model however, it turns out that the inherently on-shell nature of the 4d twistorial ambitwistor strings hinders our ability to define a gluing operator \cite{Witten:2003nn, Berkovits:2004hg, Skinner:2013xp, Geyer:2014fka}.\footnote{Here, we mean by `on-shell' that the constraint $P^2=0$, which is gauged in the RNS ambitwistor string, is explicitly solved in the twistor models, with $P_{\alpha\dot\alpha}=\lambda_\alpha\tilde\lambda_{\dot\alpha}$.} This can be understood intuitively from the degeneration from genus one to the nodal sphere: as discussed above,  the residue theorem trades one of the scattering equations for the localization on the nodal sphere, and therefore $P^2\neq 0$ on the nodal sphere. At the level of the CFT, this arises from the non-local component of $\Delta$, which modifies the effective gauged current from $P^2$ to $P^2-\ell^2\omega_{+-}^2$. The twistorial models, on the other hand, \emph{solve} the constraint $P^2=0$, and thus cannot account for the deformations away from $P^2=0$ necessary for the definition of the gluing operator.

The models of section \ref{sec:2TT-SymRed} give an alternative massless model in four dimensions that allows for more degrees of freedoms. While all external particles remain on-shell in 4d, we will see that this introduces enough `off-shell' aspects into the model to allow for a gluing operator to be defined in close analogy with \cite{Roehrig:2017gbt}. 
We begin this section by discussing the action for the massless  worldsheet model, its origin from the massive model in \S\ref{sec:4d-model}, and its relation to the familiar 4d ambitwistor string. The extra currents in this model play an important role in the construction of the gluing operator, and we highlight the differences and similarities to $\Delta_{\scalebox{0.7}{RNS}}$ of \cite{Roehrig:2017gbt}. We conclude by calculating $n$-point correlators involving $\Delta$, which give rise to the twistorial one-loop integrand formul{\ae} of \cite{Wen:2020qrj}.

\subsection{The model}
Let us first introduce an ambitwistor worldsheet model that (i) agrees with the familar 4d ambitwistor string models for tree-level correlators involving only vertex operators and (ii) allows for the definition of a gluing operator, i.e. contains gauged currents that allow for deformations with $P_{\scalebox{0.7}{4d}}^2\neq 0$. One way of achieving this in the twistorial models is to reduce a higher-dimensional model -- e.g. the 5d worldsheet model reviewed in \S\ref{subsec:5dmodels} -- to 4d. While in these models $P^2=0$ is still solved exactly, the 4d part $P_{\text{4d}}$ can now satisfy $P_{\text{4d}}^2\neq 0$. 
Equivalently, we may start with the massive model of \S\ref{sec:4d-model}, but restrict to the massless case where  $j^{\scalebox{0.7}{$H$}}=0$. This gives the action:
\begin{equation}\label{symred-model-massless}
S_{\scalebox{0.7}{4d}}^0 =\int_\Sigma \mathcal{Z}^a\cdot\dbar_e\mathcal{Z}_a+A_{ab}\mathcal{Z}^a\cdot\mathcal{Z}^b + a\lambda^2+ \tilde a \tilde{\lambda}^2+S_m\,,
\end{equation}
where $\cZ$ are as before the equivalent of the Dirac supertwistors $\mathcal{Y}=(\lambda_A,\mu^A,\eta^I)$ we considered in the models of \cite{Albonico:2022pmd}.
That this reduces the target space to $\mathbb{A}_4$ is intuitively clear from the mass relations \eqref{LG-kappa}. Alternatively, this can be seen by explicitly integrating out the additional degrees of freedom in the path integral, see \cite{Albonico-thesis}.
We also include the same worldsheet matter  $S_m$ as in the massive case, with
\begin{equation}
 S_m^{\scalebox{0.7}{sYM}}=S_\rho+S_j\,,\qquad\qquad S_m^{\scalebox{0.7}{sugra}}=S_\rho+S_{\tilde \rho}\,.
\end{equation}
where $S_j$ is a current algebra, and $S_\rho$ is the fermionic system that lifts the worldsheet gauge algebra to the super gauge algebra $\mathfrak{sl}_2 \ltimes H(0,2p)$, with $p=1$ for super Yang-Mills and $p=2$ for supergravity respectively.


Since the model $S_{\scalebox{0.7}{4d}}^0$ is a special case of the massive model discussed in the previous sections, it is clear that the BRST gauge-fixing is only modified trivially so that:
\begin{equation}
 Q=Q_{\scalebox{0.7}{m}}\bigg|_{j^{\scalebox{0.5}{$H$}}=\tilde j^{\scalebox{0.5}{$\tilde H$}}=0}
\end{equation}
The anomaly counting is also unaffected by taking  $j^{\scalebox{0.7}{$H$}}=\tilde j^{\scalebox{0.7}{$\tilde H$}}=0$, so the SL$(2)$ gauge anomaly vanishes for maximal supersymmetry, 
\begin{align}
  &\mathfrak{a}_{\scalebox{0.6}{$\mathrm{SL}(2)$}}^{\scalebox{0.6}{sYM}}=\frac{3}{4}\left(4-\mathcal{N}\right)\,,
  &&   \mathfrak{a}_{\scalebox{0.6}{$\mathrm{SL}(2)$}}^{\scalebox{0.6}{sugra}}=\frac{3}{4}\left(8-\mathcal{N}\right)\,.
\end{align}
The models also maintain the same central charges as in the massive case, and are therefore critical if we include the central charge from six compactified dimensions, as well as a current algebra of central charge $\mathfrak{c}_j=16$ for super Yang-Mills.

\paragraph{Vertex operators and tree-level correlators.} As before, the vertex operators take the  form \eqref{super-representative}. Because we take $j^H=0$, the BRST cohomology only contains massless states
\begin{equation}
 Q\circ V(\sigma_i)\sim\big(a \kappa^2+\tilde a\tilde \kappa ^2\big)\, V(\sigma_i)=0\,,
\end{equation}
which is reflected in the delta functions produced by the picture changing operators as in \eqref{vo-matter}.
Tree-level correlators then take a very similar form as in the massive case,
\begin{equation}
\mathcal{A}_{n}=\prod_{i=2}^n\delta(\kappa_i^2)\; \delta(\tilde\kappa_i^2)\int d\mu_{n}^{\mathrm{pol}} \; \mathcal{I}_{n}\, e^{F_\mathcal{N}}\,.
\end{equation}
These are trivial dimensional reductions of the six and five dimensional formulae to four dimensional massless kinematics: in \cite{Albonico:2020mge} they have been shown to agree with the familiar amplitudes in maximal super Yang-Mills and maximal supergravity, as obtained from the twistor or ambitwistor string.

\subsection{Gluing operator}\label{sec:gluing_op}
While the 4d ambitwistor string  and the worldsheet model \eqref{symred-model-massless} are equivalent at tree-level, a gluing operator may readily be defined in the latter, but not the former. The reason for this is that the gluing operator plays the role of a target-space propagator, and is therefore an inherently off-shell object. The constraint $P^2=0$ is solved rather than gauged in the original twistorial models, thus preventing any deformation that leads to $P^2\neq 0$. For $S_{\scalebox{0.7}{4d}}^0$ on the other hand, $P_{\sA\sB}$ is reduced from six to four dimensions via the gauged currents $\lambda^2$ and $\tilde\lambda^2$, so a non-local operator can lead to  $P_{\scalebox{0.7}{$4d$}}^2\neq 0$ by deforming them.

We may also see the need for a non-local operator from a different perspective, as stressed in \cite{Roehrig:2017gbt}: since the propagator is off-shell, the gluing operator cannot both be local and BRST invariant. We will follow the approach of ref.~\cite{Roehrig:2017gbt}, and construct the gluing operator as a non-local, but BRST invariant object. We thus require that the gluing operator $\Delta$:
\begin{enumerate}
 \item[(i)] encodes a target-space propagator, i.e. includes two local operators $\mathcal{O}_\pm$, which are extensions of the vertex operators $V$ to off-shell momentum $\pm \ell$, as well as the appropriate sum over states 
 \item[(ii)] is BRST invariant.
\end{enumerate}
From these requirements, we write the following general form of the gluing operator (c.f. the original construction in \cite{Roehrig:2017gbt}),
\begin{equation}\label{gluing-op-def}
 \Delta (\sigma_+,\sigma_-) = \int \frac{d^{\scalebox{0.6}{$D$}}\ell}{\ell^2}\, W (\sigma_+,\sigma_-)\,\sum_{\mathrm{states}}\mathcal{O}_+(\sigma_+) \mathcal{O}_-(\sigma_-)\,.
\end{equation}
For the operators $\mathcal{O}_\pm$, we will need two off-shell, back-to-back momenta $\pm\ell$, which we parametrize in the massive spinor-helicity formalism as\footnote{To be explicit, the last equation implies that we use the following convenient choice for the relation between the spinors of the back-to-back momenta: $\kappa_{-\sA}^0 = \kappa_{+\sA}^0$ and $\kappa_{-\sA}^1 = -\kappa_{+\sA}^1$.}
\begin{equation}
 \ell_{\alpha\dot\alpha}= (\kappa_{+\alpha}\tilde\kappa_{+\dot\alpha})\,,
 \qquad \qquad 
 -\ell_{\alpha\dot\alpha}= (\kappa_{-\alpha}\tilde\kappa_{-\dot\alpha})\,, 
 \qquad\qquad \kappa_{-\sA}^a = (-1)^a\kappa_{+\sA}^a\,,
\end{equation}
where the mass parameter $M_\ell=:L$ is defined as usual via 
\begin{align}\label{eq:def_loop}
 &\ell^2 = L^2\,,
 &&\det (\kappa_\pm)=\pm L\,.
\end{align}
We may then define the operators $\mathcal{O}_\pm$ as the (trivial) extension of the vertex operators $V$ to an off-shell momentum,
\begin{equation}\label{eq:def_O_vs_V}
 \mathcal{O}_\pm= V\bigg|_{k\rightarrow\pm \ell}\,.
\end{equation}
While this is enough to satisfy condition (i) above, other choices of $\mathcal{O}_\pm$ may in principle be possible that also satisfy $\mathcal{O}_\pm\big|_{\pm\ell\rightarrow k }= V$. We will verify below that for the choice \eqref{eq:def_O_vs_V}, there exists a $W$ such that the gluing operator $\Delta$ is BRST-invariant.
To be explicit, the operators $\mathcal{O}_\pm$ are given by
\begin{align*}
 \mathcal{O}_\pm(\sigma_\pm) &= \int d^2u\,d^2v\;  \bar\delta^4\big((u\lambda_\sA)-(v\,\kappa_{\pm \sA})\big) \,\bar\delta\big((\epsilon_\pm v)-1\big)\,w\, 
 \e^{i u_{a}\left(\mu^{\sA a} \epsilon_{\pm \sA}+q_{\pm\,I} \eta^{I a}\right)-\frac{1}{2}(\xi v) q_\pm^{2}}\, .
\end{align*}
The sum over states depends on the model in question. For both super Yang-Mills and supergravity, it will be convenient to take  $(\epsilon_+\epsilon_-)=1$. For the $S_\rho$ matter system, the sum over states can then conveniently be performed by a fermionic integral over the supermomenta $q_\pm$ of the propagating particle, whereas for the current algebra $S_j$ the colour-flow through the propagator takes the form $\delta_{\mathfrak{ab}}$. For super Yang-Mills, we thus have
\begin{equation}\label{eq:sum_states_loops}
 \sum_{\mathrm{states}}\mathcal{O}_+(\sigma_+) \mathcal{O}_-(\sigma_-) 
 =\int d^{\scalebox{0.7}{$\mathcal{N}$}}q_+\, d^{\scalebox{0.7}{$\mathcal{N}$}}q_-\;\delta_{\mathfrak{ab}}\;
 \mathcal{O}_+^{\mathfrak{a}}(\sigma_+) \mathcal{O}_-^{\mathfrak{b}}(\sigma_-)  \,e^{iq_+\cdot q_-}\,,
\end{equation}
and similarly for supergravity.
At this stage, it is easy to verify explicitly that the operator \eqref{eq:sum_states_loops} is not BRST-closed due to the off-shell momentum $\ell$. The failure to be BRST-closed must be compensated by the operator $W (\sigma_+,\sigma_-)$ in \eqref{gluing-op-def}, which is therefore genuinely non-local. Using the BRST-closure to define $W$, we find
\begin{equation}
 W(\sigma_+,\sigma_-) = \mathrm{exp}\left(\pm\int_\Sigma (L\, a+ L \,\tilde a) \,\omega_{+-}\right)\,,
\end{equation}
where  $\omega_{ij}$ is the differential with simple poles at the marked points, $\omega_{ij}=\frac{\sigma_{ij}\,d\sigma}{(\sigma-\sigma_i)(\sigma-\sigma_j)}$.
Let us see explicitly that this achieves the objective, and that $\Delta$ is now in the BRST cohomology. Since $W$ depends on the gauge fields  $a$ and $\tilde a$, it modifies the BRST operator to an \emph{effective BRST operator} $Q_{\mathrm{eff}}$. After BRST quantization and integrating out the gauge fields in the presence of the gluing operator, this effective BRST operator takes the form\footnote{The calculation here mirrors \cite{Roehrig:2017gbt} closely, and many additional details can be found there.}
\begin{equation}
 Q_{\mathrm{eff}}\supset\oint cT + t \left(\lambda^2 -L\,\omega_{+-}\right)+ \tilde t \left(\tilde\lambda^2 -\tilde L\,\omega_{+-}\right)\,,
\end{equation}
where we have only given the currents affected by the presence of $W$. We see that the effective BRST operator contains precisely the correct terms to render  the gluing operator BRST-closed,
\begin{equation}
  Q_{\mathrm{eff}}\circ \Delta=0\,.
\end{equation}

\subsection{One-loop amplitudes}
Having constructed the gluing operator $\Delta$ as the BRST-closed operator encoding the propagator, we can now calculate loop amplitudes (here for super Yang-Mills) as correlators including $\Delta$ on a single Riemann sphere:
\begin{equation}\label{one-loop-correlator}
    \int_{\mathfrak{M}_{1, n}}\left\la\mathcal{V}_1\left(\sigma_1\right) \cdots \mathcal{V}_n\left(\sigma_n\right)\right\ra_{\Sigma}=\int_{\mathfrak{M}_{0, n+2}}\left\la\Delta\left(\sigma_{+}, \sigma_{-}\right) \mathcal{V}_1\left(\sigma_1\right) \cdots \mathcal{V}_n\left(\sigma_n\right)\right\ra_{\Sigma}\,.
\end{equation}
As proposed in \cite{Roehrig:2017gbt} for the RNS ambitwistor string, this will calculate one-loop amplitudes. We will see that the expressions precisely match the one-loop amplitudes  obtained in \cite{Wen:2020qrj} from a back-to-back forward limit of the 6d spinorial amplitude formul{\ae}.

From the form of the gluing operator in the previous section we can write the amplitude \eqref{one-loop-correlator} as:
\begin{equation}\label{n+2point}
    \int\frac{\rd^4\ell}{\ell^2}\rd^{\mathcal{N}}q_+\rd^{\mathcal{N}}q_-\e^{iq_+\cdot q_-}
    \int_{\mathfrak{M}_{0, n+2}}
    W(\sigma_+,\sigma_-)\delta_{\mathfrak{a}\mathfrak{b}}
    \left\la\mathcal{V}_1\left(\sigma_1\right) \cdots \mathcal{V}_n\left(\sigma_n\right)\mathcal{O}^{\mathfrak{a}}_+\left(\sigma_+\right)\mathcal{O}^{\mathfrak{b}}_-\left(\sigma_-\right)\right\ra_{\Sigma}\,.
\end{equation}
Here the factor $W(\sigma_+,\sigma_-)$ acts as we described in the previous section to provide an effective `mass term' for punctures $\sigma_\pm$ associated with the on-shell momentum. Then the correlator is computed as an $(n+2)-$point correlator with two off-shell particles with back-to-back momenta. This formula is an analogue of the ones we derived for tree-level scattering in the \cref{sec:2TT-SymRed}, with adjacent particles $\pm$ in the color-ordering because of the sum over states dictated by the gluing operator. Because of the special kinematic configuration involved, the scattering equation and the spin $1$ integrand can be simplified as follows.

\paragraph{Polarized scattering equations.} We embed the spinors $\kappa_\pm$ and $\tilde\kappa_\pm$ into 6d kinematics $ \kappa_{\pm A}^a$ as in \S\ref{spinor-helicity-54d}, 
so that the $4d$ part of the loop momentum $\ell$ is now off-shell, c.f. \eqref{eq:def_loop}. Similarly, we embed the external momenta (massless in $4d$) via  
\begin{equation}
    \kappa_\alpha^a = \left(0,-\kappa_\alpha\right)\,,\qquad  
   \tilde \kappa_{\dot \alpha}^a = \left(\tilde\kappa_{\dot \alpha},0\right)\,,
\end{equation}
and $4d$ polarization data can be incorporated naturally via $\epsilon_{ia} = (0,-1)$ for negative helicity, and $\epsilon_{pa} = (1,0)$ for positive helicity eigenstates.
In particular, this implies that
\begin{subequations}\label{eq:epsilon_4d}
\begin{align}
 &\epsilon_i^\alpha:=(\epsilon_i\, \kappa_i^\alpha ) = \kappa_i^\alpha
 &&\tilde\epsilon_i^{\dot \alpha}:=(\epsilon_i\, \tilde\kappa_i^{\dot\alpha} ) =0
 &&  i\in -\,, \\
 &  \epsilon_p^\alpha:=(\epsilon_p \kappa_p^\alpha ) = 0
 &&\tilde\epsilon_p^{\dot \alpha}:= (\epsilon_p \tilde\kappa_p^{\dot\alpha} ) =\tilde\kappa_p^{\dot\alpha}
 && p\in +\,.
\end{align}
\end{subequations}
The polarized scattering equations  for any particle $i\in\{1,2,\dots,n,+,-\}$ are then given by
\begin{equation}\label{loop-6dSE+}
    \mathcal{E}_{i\alpha}:=\left(u_{i}\lambda_\alpha(\sigma_i)\right) -\left(v_{i}\kappa_{i\alpha}\right)=0\,, \qquad\qquad\tilde{\mathcal{E}}_{i \dot \alpha}:=(u_{i}\tilde\lambda_{\dot \alpha}(\sigma_i) )-\left(v_{i}\tilde\kappa_{i \dot \alpha}\right)=0\, .
\end{equation}
where, as before, $\lambda$ and $\tilde\lambda$ are defined by
\begin{subequations}
\begin{align}
\lambda_{\alpha}^a(\sigma)&=\sum_{i\in h_-}\frac{u_{i}^a\epsilon_{i\alpha}}{\sigma-\sigma_i}+\frac{u_+^a \epsilon_{+\alpha}}{\sigma-\sigma_+}+\frac{u_-^a \epsilon_{-\alpha}}{\sigma-\sigma_-}\,,
\\
\tilde \lambda_{\dot\alpha}^a (\sigma)&=\sum_{p\in h_+}^n\frac{u_{p}^a\tilde\epsilon_{p\dot\alpha}}{\sigma-\sigma_p}+\frac{u_+^a \tilde\epsilon_{+\dot\alpha}}{\sigma-\sigma_+}+\frac{u_-^a \tilde\epsilon_{-\dot\alpha}}{\sigma-\sigma_-}\, .\label{loop-lambda-def}
\end{align}
\end{subequations}
Here, we have used that half the $\epsilon$'s vanish, \eqref{eq:epsilon_4d}, 
and that without loss of generality, we can choose the polarization of the loop momentum to be
\begin{equation}
 \epsilon_{+a} = (1,0)\,,\qquad \epsilon_{-a} = (0,1)\,,
\end{equation}
i.e. in the conventions of \cite{Wen:2020qrj}:
\begin{subequations}
 \begin{align}
  &  \epsilon_{+}^{\alpha}:=(\epsilon_+\, \kappa_{+}^{\alpha} ) = \kappa_{+}^{\alpha 0}
 &&\tilde\epsilon_{+}^{\dot \alpha}:= (\epsilon_+ \,\tilde\kappa_+^{\dot\alpha} )=\tilde\kappa_+^{\dot\alpha 0}
 && \mathrm{for}\;\;+\ell\,, \\
 &\epsilon_{-}^{\alpha}:=(\epsilon_-\, \kappa_-^\alpha ) = \kappa_-^{\alpha 1} = -\kappa_+^{\alpha 1} 
 &&\tilde\epsilon_-^{\dot \alpha}:=(\epsilon_-\, \tilde\kappa_-^{\dot\alpha} ) =\tilde\kappa_-^{\dot\alpha 1} = -\tilde\kappa_+^{\dot\alpha 1}
 &&  \mathrm{for}\;\;-\ell\,. 
 \end{align}
\end{subequations}
This implies in particular that we can express the loop momentum $\ell$ as
\begin{equation}
 \ell^{\alpha\dot\alpha}= (\kappa_{+}^{\alpha}\tilde\kappa_{+}^{\dot\alpha}) = \epsilon_-^\alpha \tilde\epsilon_+^{\dot\alpha}-\epsilon_+^\alpha \tilde\epsilon_-^{\dot\alpha}\,.
\end{equation}

\paragraph{Integrands.} The only non-trivial part comes from $\det{}'H$, which we can simplify in this one-loop set-up. With data as above, $H$ is given by
\begin{subequations}
\begin{align}
 H_{ij}=&H_{ij}^- =\frac{\la \epsilon_i\epsilon_j\ra}{\sigma_{ij}} & 
  H_{pq}=&H_{pq}^+ =\frac{[ \tilde\epsilon_p\tilde\epsilon_q ]}{\sigma_{pq}} &
  &H_{ip} = 0 \\
  &  H_{i\pm}
  =\frac{\la \epsilon_i\epsilon_\pm\ra}{\sigma_{i\pm}} &
  &H_{p\pm}
  =\frac{[ \tilde\epsilon_p\tilde\epsilon_\pm]}{\sigma_{p\pm}}\,, &&
\end{align}
\end{subequations}
Here, we take  $i,j\in h_-$ and $p,q\in h_+$. 
We have the freedom to define the reduced determinant by removing the $\pm$ rows and columns from $H$, corresponding to choosing the operators $\mathcal{O}_\pm$ in the $(-1,-1)$ picture. With this choice, the resulting determinant is block-diagonal, and the result has the appealing form
\begin{equation}
 \det{}'H = \frac{1}{(u_+u_-)^2}\det{}H^{\scalebox{0.6}{$[+-]$}}_{\scalebox{0.6}{$\,[+-]$}} = \frac{1}{(u_+u_-)^2}\det H^+\,\det H^-\,,
\end{equation}
reminiscent of tree-level. Indeed, this form makes it obvious that the integrand behaves correctly on a single cut. 

Putting everything together, we can write the loop amplitude \eqref{n+2point} as:
\begin{equation*}
    \mathcal{A}_n^{\mathrm{1-loop}}=\int\frac{\rd^4\ell}{\ell^2}\rd^{\mathcal{N}}q_+\rd^{\mathcal{N}}q_-\e^{iq_+\cdot q_-}
    \int\rd\mu^{\mathrm{pol}}_{n+2}
    \frac{1}{(u_+u_-)^2} \det H^+ \det H^-
    \mathrm{PT}(\alpha,\sigma_+,\sigma_-)\,
    \e^{F_\mathcal{N}}
    \,,
\end{equation*}
where the polarisation measure is familiar from the six-dimensional tree level formulae, with the polarised scattering equations as described above. A similar expression can be found for supergravity and both agree with the formulae presented in \cite{Wen:2020qrj}.

\subsection{Comparison to the gluing operator in the RNS ambitwistor string}
In \cref{sec:gluing_op}, we constructed the gluing operator $\Delta_{4d}$ following the same guiding principles used for  $\Delta_{\scalebox{0.7}{RNS}}$ in  ref.~\cite{Roehrig:2017gbt}; both are built from two local operators that trivially extend the vertex operators off-shell, and are BRST-closed. In this section, we compare the two gluing operators, and discuss similarities and differences. 
As we will see below in more detail, $\Delta_{\scalebox{0.7}{RNS}}$ can be constructed directly in the 10d model, but requires the constraint $P^2$ to be gauged rather than solved, and thus does not exist in spinorial models. 
For clarity, we will compare the two  gluing operators in the RNS model reduced to $d<10$, where both constructions are well-defined and lead to equivalent gluing operators.

$\pmb{\Delta_{\scalebox{0.7}{RNS}}}:$ Let us start by reviewing briefly the  gluing operator $\Delta_{\scalebox{0.7}{RNS}}$ as constructed in \cite{Roehrig:2017gbt}. 
Following the same motivation as given above, the gluing operator takes the form 
\begin{equation}\label{eq:Delta_RNS}
 \Delta_{\scalebox{0.7}{RNS}} (\sigma_+,\sigma_-) = \int \frac{d^{\scalebox{0.6}{$D$}}\ell}{\ell^2}\, W (\sigma_+,\sigma_-)\,\sum_{\mathrm{states}}\mathcal{O}_+(\sigma_+) \mathcal{O}_-(\sigma_-)\,,
\end{equation}
where $\mathcal{O}_\pm$ are again off-shell extensions of the vertex operator, obtained  by replacing the on-shell momentum $k$ by the off-shell $\pm\ell$ respectively as in \eqref{eq:def_O_vs_V}. \footnote{For the bi-adjoint scalar $\mathcal{O}_+^{a\dot a} = c\tilde{c}\,j^a\tilde{j}^{\dot a}\,e^{i\ell\cdot X}$, with the sum over states implemented via $\Delta_{ab\dot a \dot b} = \delta_{ab}\delta_{\dot a\dot b}$.}
In the RNS ambitwistor string, BRST invariance requires $W$ to be the following Wilson-line-like operator, 
\begin{equation}\label{eq:W_RNS}
 W_{\scalebox{0.7}{RNS}} (\sigma_+,\sigma_-) = \mathrm{exp}\left(\frac{\ell^2}{2}\int_\Sigma\tilde{e}\,\omega_{+-}^2\right)\,.
\end{equation}
After BRST quantization, this leads to an effective BRST operator of the form
\begin{equation}\label{Q_eff_RS}
 Q_{\mathrm{eff}} = \oint c\,T+\frac{\tilde{c}}{2}\left(P^2-\ell^2\omega_{+-}^2\right)\,.
\end{equation}
Note that this operator is well-defined in $D=10$ dimensions, and no dimensional reduction has been necessary in its derivation. As discussed in the beginning of the section, this reflects that in the RNS ambitwistor string, $P^2=0$ is a \emph{gauged} constraint, which can be deformed by the Wilson-line-like operator $W$.

In order to compare $\Delta_{\scalebox{0.7}{RNS}} $ to the gluing operator in the twistorial model, we reduce it to 4d. Due to the absence of non-trivial winding modes, the toroidal compactification is trivial in the RNS ambitwistor string \cite{Geyer:2015jch}, and the formula \eqref{eq:Delta_RNS} remains valid, but with the loop momentum $\ell_{\scalebox{0.7}{$(4d)$}}$ reduced to 
4d. This extends straightforwardly to the BRST operator:
\begin{equation}\label{Q_eff_RS_4d}
 Q_{\mathrm{eff}}^{\scalebox{0.7}{$(4d)$}} = \oint c\,T+\frac{\tilde{c}}{2}\left(P_{\scalebox{0.7}{$(4d)$}}^2-\ell_{\scalebox{0.7}{$(4d)$}}^2\omega_{+-}^2\right)\,.
\end{equation}

$\pmb{\Delta_{\scalebox{0.7}{4d}}}:$ It is helpful to transpose the construction of the last section from the twistorial to the RNS model. In analogy with \eqref{symred-model-massless}, we toroidally compactify five dimensions, and \emph{gauge} the reduction from 5d to 4d by including the following term in the action, 
\begin{equation}
 S\supset \int_\Sigma a\, P\cdot \Omega\,. 
\end{equation}
Here $\Omega_1$ is the vector pointing in the `fifth' dimension, and the constraint both restricts tangent vectors to 4d and identifies different parallel 4-planes as explained in \S\ref{subsec:5dmodels}. While we may still define $W$ as in \eqref{eq:W_RNS}, we can now alternatively achieve BRST invariance of the gluing operator by taking 
\begin{equation}\label{W_4d_RNS}
 W_{\scalebox{0.7}{RNS}} ^{{\scalebox{0.7}{4d}} } (\sigma_+,\sigma_-) = \mathrm{exp}\left(|\ell|\int_\Sigma a\,\omega_{+-}\right)\,.
\end{equation}
This is the RNS equivalent of $W_{\scalebox{0.7}{4d}}$ in the twistorial model. Note that in contrast to $W_{\scalebox{0.7}{RNS}}$,
this construction is only possible when dimensionally reducing to  $D<10$. On the other hand, it has the advantage of being applicable in models where the $P^2=0$ constraint is solved  rather than gauged, as we have seen explicitly in the preceding section.

Despite the slightly differing constructions, both gluing operators give the same effective BRST operator after quantization;
\begin{equation}\label{Q_eff_4d}
 Q_{\mathrm{eff}}^{\scalebox{0.7}{$(4d)$}} = \oint c\,T+\frac{\tilde{c}}{2} P_{\scalebox{0.7}{$(5d)$}}^2
  = \oint c\,T+\frac{\tilde{c}}{2}\left(P_{\scalebox{0.7}{$(4d)$}}^2-\ell_{\scalebox{0.7}{$(4d)$}}^2\omega_{+-}^2\right)\,.
\end{equation}
In the second equality, we have integrated out the gauge field $a$ to find $P\cdot\Omega = |\ell_{\scalebox{0.7}{$(4d)$}}|\, \omega_{+-}$, as dictated by the inclusion of the effective term in \eqref{W_4d_RNS}.

\section{Summary and discussion}\label{sec:disc-chap5}
We have shown how masses can be implemented in ambitwistor strings via a symmetry reduction that gauges the corresponding generators on the worldsheet, supplemented by gauge or $R$-symmetry generators.  This article focuses on the twistorial models and gives a systematic consistent framework to obtain models and amplitude formulae for theories in 4d involving massive particles that include fermions and supersymmetry. This gives an alternative derivation of  the two-twistor string of \cite{Geyer:2022cey}, derived there as a chiral string whose target is the  twistorial representation of the phase space of massive particles (complexified), but here derived  as a symmetry reduction from twistorial ambitwistor models for maximally supersymmetric theories in five dimensions. The theories obtained via these  reductions include both Coulomb-branch-type theories and CSS supergravities. 
 This provides an important extension to a framework that seemed until recently to be intrinsically massless.  In  the accompanying paper \cite{upcoming:part1} we follow the same strategy for the RNS ambitwistor string which is more awkward for the description of fermions and supersymmetry, but which gives a geometrically somewhat simpler description and is to a certain extent dimension agnostic.

From the path integral of these models, we have arrived at the  compact formulae \eqref{final-formula}, supplemented by \eqref{measure} and \eqref{integrands}, supported on a massive version \eqref{massive-PSE} of the polarised scattering equations and with manifest supersymmetry for appropriate gauge and gravity  theories including massive particles. Like all twistor-string, CHY and ambitwistor-string amplitude formulae, all the integrations are saturated against delta functions so that these are really residue formulae summing contributions from the $(n-3)!$ solutions to a massive extension of the scattering equations \eqref{massive-SE-vectorial} that we also discuss further below. The extra data in the polarised extension is uniquely obtained by linear equations on the support of these scattering equations and the amplitude formulae are linear in the polarization data as  shown in \cite{Albonico:2020mge}.  

Contrary to the massless four-dimensional formulae of \cite{Roiban:2004ka, Geyer:2014fka}, in which the double copy properties are hidden in the measure, the expressions derived here present the standard structure with two half integrands that can be combined to form amplitudes for scalars, spin-one and spin-two particles as in the CHY formulae and corresponding RNS models of \cite{Mason:2013sva, Casali:2015vta}.  In the context of R-symmetry reductions we presented a novel instance of worldsheet double copy between gauge theories with massive matter and various degrees of supersymmetry and gauged supergravities.

To conclude, we give some open research directions.

\paragraph{Spectrum of the original 4d twistor and ambitwistor string.}
This two-twistor string is most naturally regarded as an extension  of the massless 4d ambitwistor string of \cite{Geyer:2014fka} described in \S\ref{4d-models-intro}, but in a framework in which a massless field can be \emph{deformed} to go off-shell. Despite the elegance of the amplitude formulae to which they give rise,  the original 4d models of \S\ref{4d-models-intro} have a number of conceptual problems that are resolved by this extension.  For example, the ambitwistor models of \S\ref{4d-models-intro} have gauge anomalies if not maximally supersymmetric, but if they are maximally supersymmetric, we find a doubling of the spectrum if the $(Z,\tilde Z)$ remain as spinors on the worldsheet.  In the twistor string on the other hand, the same redundancy appears due to the sum over line bundles of $(Z,\tilde Z)$; the same particle can be encoded either in the original vertex operator, or the `dual' vertex operator with a different degree line bundle.
However, when the models of \S\ref{4d-models-intro} are extended to the two-twistor string, different degrees are no longer possible as the gauge group becomes $\SL(2,\C)$ which has no winding number and hence no degree, and there is no longer any doubling of spectrum.   Thus these models seem altogether healthier, even if more complicated.  
Unlike the 4d massless models, they also have manifest worldsheet double copy.  It is shown in \cite{Albonico:2020mge,Albonico-thesis} how the polarized scattering equations reduce to those in 4d so that the two-twistor formulae reduce to those of \cite{Geyer:2014fka} when massless 4d vertex operators are substituted into the various formulae that we use here. In light of the discussion above, we expect that a more careful treatment of this reduction \emph{at the level of the models} would be capable of resolving the doubling of the spectrum in the 4d models.

\paragraph{Reductions along several dimensions and `spectrum' of accessible theories.} 
While here we have only considered reductions from $5$ to $4$ dimensions, one might hope to extend the formalism  to more complicated reductions from higher dimensions. This would allow for the study of more generic massive theories in 4d, as well as massive theories in 5d along the lines of \cite{Chiodaroli:2022ssi}. In the twistorial models of \S\ref{sec:2TT-SymRed} we are limited in this regard. While in principle we could perform reductions along two extra dimensions (i.e. coming down from six), as we have mentioned,  the six dimensional models of \cite{Geyer:2020iwz} do not contain ordinary gravity or gauge theory---they only contain higher derivative versions of gauge and gravity theories.  This issue arises because in 6d  the matter system \eqref{action-WS-fermions} needs both chiralities  $\lambda_A^a$ and $\lambda^A_{\dot a}$ and these are not both present in the models we have been considering. On reduction to 5d, these two chiralities are identified for massless particles so the problem evaporates, as exploited in the 5d models used here.  However, when a massive reduction from 6d is considered; the massive little group in 5d is again $\SL(2,\C)\times \SL(2,\C)$ and the chirality issue is likely to remain. One way of circumventing this issue would be to start with the larger  ten or eleven dimensional frameworks described in \cite{Geyer:2019ayz}, but these are much more cumbersome and have not been completed to fully consistent models either.  Another avenue might be to use the pure spinor models of \cite{Berkovits:2013xba,Gomez:2013wza}.

\paragraph{R-symmetry and double copy.} 
In section \ref{sectionRsymred} we have given symmetry reductions along the $R-$symmetry generators in maximally supersymmetric Yang-Mills and gravity. For gauged supergravities obtained as CSS symmetry reductions, we have given a novel form of massive double copy on the worldsheet. 
In  the table \ref{table:DC}, we have seen that there is a  family of left and right pairs for one given supergravity theory and that the amplitude formula automatically selects one of these pairs on space-time via the scattering equations. However,  for all the other possible pairs we don't have a way of deriving the double copy constructions \textit{on momentum space}. It seems obvious that for the pairs that don't have matching mass spectra there should not be a valid momentum space double copy but it would be interesting to see if this extends to all the pairs of gauge theories. 

Naively, this  suggests that the space of theories reachable by the momentum space double copy is a strict subset of the corresponding space for the worldsheet. However, this only holds for theories with known worldsheet models and this excludes most  known theories. A more detailed comparison of the webs of double copy, both on momentum space and the worldsheet, would thus be interesting.

Finally, there is a well established formalism for the construction of gauged supergravities, see e.g.\ \cite{Samtleben:2008pe} for a review. For further investigations of these theories in the context of the ambitwistor string, it would be good to make contact with the relevant objects in that description such as the embedding tensor and the symplectic frame.

\paragraph{Solutions to the polarised scattering equations and Ward identities.} 
The massless polarised scattering equation in four dimensions exhibit a special structure (see section \ref{4d-models-intro}) whereby solutions are split into MHV sectors, so that for instance at $n-$point and MHV degree only one solution for $\{\sigma_i\}$ contributes to the MHV amplitude and this can be evaluated explicitly \cite{Du:2016blz}. For massive particles, even in the case of two massive particles and $n-2$ massless gluons of the same helicity, we checked numerically that the amplitude has support on all the $(n-3)!$ solutions for $\{\sigma_i\}$. 

In recent years several $n$-point formulae have been derived by BCFW recursion \cite{Ochirov_2018,Ochirov:2022nqz}.
We have seen in \S\ref{sec:ward-id} that the supersymmetry of the full superamplitude can be exploited to obtain Ward identities relating the different component amplitudes. This can be done for any superamplitude formula and one obtains relations where the coefficients are the polynomials in the moduli $U_{ij}$. What is interesting is that we have found a case where we could easily solve for one particular coefficient $U_{ij}$ without the need to solve the full system, thus relating different $n-$point amplitudes that had previously appeared in the literature \cite{Forde:2005ue,Ochirov_2018}. We hope that further investigation of special configurations could lead to more cases where we can solve for a subset of coefficients $U_{ij}$ to find more Ward identities.

\paragraph{Loops.}
The fact that, as  described above, the two-twistor string is  naturally regarded as an extension  of the massless 4d ambitwistor string \cite{Geyer:2014fka}, but 
with better anomaly properties suggests that loops might be approachable by extending the formulae to higher genus curves as in conventional string theory; such a strategy was implemented for the RNS ambitwistor-string at one loop in \cite{Adamo:2013tsa} leading to a modular invariant one-loop integrand formulae on the torus. If the conformal anomaly is cancelled by appropriate additional worldsheet matter, we would expect our models here also to allow such a formulation at one loop, but there are further technicalities for the twistorial model.  In particular, the fact that all the main fields are spinors on the worldsheet means that there would need to be a sum over spin structures for all the ingredients in the model affecting also the polarized scattering equations. In practice one would expect that the additional  dependence on the spin structure would only affect the linear equations satisfied by the $(u_a,v_a)$  variables, so that the formulae would still be underpinned by the bosonic scattering equations on the torus proposed in \cite{Adamo:2013tsa}.

In a framework in which a massless field can be \emph{deformed} to go off-shell the techniques of \cite{Geyer:2015bja,Geyer:2015jch} can be considered to transform such higher genus formulae to potentially give formulae on nodal Riemann spheres at all loop orders. At one loop this can be conveniently expressed in terms of a  gluing operator analogous to that of  \cite{Roehrig:2017gbt}; we have seen that this can be formulated in our context,   allowing us to obtain loop formulae for theories with fermions and supersymmetry such as maximal super Yang-Mills theory. 
However, extending this construction to higher loop order is challenging for similar reasons as in the RNS models; important new features appear for $g\geq 2$ loops \cite{Geyer:2016wjx, Geyer:2018xwu}. A more promising direction would be to construct a gluing operator for super Yang-Mills theory on the Coulomb branch; however, this requires a twistorial model in 6d as a starting point; one extra dimension being needed for masses and loop order each.

\section*{Acknowledgements}
We would like to thank Alex Ochirov, Congkao Wen and Fernando Alday for discussions.  GA was supported by the EPSRC under grant EP/R513295/1 and by the Mathematical Institute. YG's work is supported by Thailand NSRF via PMU-B [grant number B05F650021 and B01F650006].  LJM would also like to thank the IHES and ENS in Paris  for hospitality while this was being written up and the STFC for financial support from  grant number ST/T000864/1.

\appendix

\section{The Coulomb Branch as a symmetry reduction from the Lagrangian}\label{CB-SR-appendix}
In this picture, mass terms for the Coulomb branch arise in the symmetry reduction from the kinetic terms in the action of $N=2$ SYM:
\begin{equation}\label{action5dsym}
\begin{split}
    S=\int d^5x\Tr[&-\frac{1}{4}F_{mn}F^{mn}-\frac{1}{2}D_m\Phi^j D^m\Phi^j
    \\&+\frac{1}{4}[\Phi^j,\Phi^k]^2 + \frac{i}{2}\bar{\Psi}^{AI}(\gamma^m)_{A}^B D_m\Psi_{BI} -\frac{1}{2}\bar{\Psi}^{AI} (\Gamma^{j})_{I}^{K}[\Phi_{j},\Psi_{AK}]]\,,
\end{split}
\end{equation}
where $m$ runs from $0$ to $4$ and $i$ runs from $1$ to $5$. $\gamma^m$ and $\Gamma^i$ are respectively Lorentzian and Euclidean gamma matrices in five dimensions. The spinors result from the reduction of the $16_+$ representation of $SO(1,9)$ which decomposes under the subalgebrae $SO(1,4)\times SO(5)$ as
\begin{equation}
    16_+\rightarrow(4,4)\quad:\quad\psi_A^I\,,
\end{equation}
where the $4$ is the fundamental of $Sp(4)\simeq SO(5)$. They are subject to a Majorana condition.\\
One can easily check that the condition:
\begin{equation}
\partial_4(A_\mu,\Phi^{IJ},\Psi_A^I)=[H,(A_\mu,\Phi^{IJ},\Psi_A^I)]\,,
\end{equation}
produces equivalent masses to the ones generated by vev'd scalars in $N=4$ SYM. 

\subsection*{Scalars in maximal SYM in five and four dimensions}\label{scalar-reps}
In this appendix we describe the representations of scalars in maximal super Yang Mills upon reduction from $10$ dimensions. When reducing this theory from $10$ to $d$ dimensions, the $l=10-d$ extra components of the connection are reduced to $l$ real scalars fields $\{\Phi_a\}_{a=1}^l$. The theory is invariant under rotations of the scalars, i.e. these transform in the fundamental representation of $SO(l)$ R-symmetry transformations.
We can alternatively write these scalars in terms of representations of the spin covering group $\mathrm{Spin}(l)$.\\
\\
In four dimensions we have $\mathrm{Spin}(6)\simeq SU(4)$ and the $6$ scalars transform in the antisymmetric tensor $\Phi_{IJ}=-\Phi_{JI}$ ($I,J=1,\dots4$), satisfying the self-duality condition
$\star\Phi=\Phi^\dagger$, with:
\begin{equation}
(\star\Phi)^{IJ}=\frac{1}{2}\epsilon^{IJKL}\Phi_{KL}\,.
\end{equation}
We can construct such a representation from $\{\Phi^a\}_{a=1}^6$ as follows. We label the six components $\Phi^a=(\phi^1,\phi^2,\phi^3,\tilde{\phi}^1,\tilde{\phi}^2,\tilde{\phi}^3)$ and construct $\Phi_{IJ}$ via:
\begin{equation}
\Phi_{mn}=\epsilon_{mnp}(\tilde\phi^p-i\phi^p)\quad\quad\phi_{m4}=\tilde\phi_m+i\phi_m\quad\quad m,n=1,2,3\,.
\end{equation}
\\
In five dimensions $\mathrm{Spin}(5)\simeq Sp(4)$ and we can use Euclidean gamma matrices to relate the fundamental of $SO(5)$ to the antisymmetric tensor of $Sp(4)$:
\begin{equation}
\Phi_I^J=(\Gamma^a)_I^J\Phi_a\,.
\end{equation}
$Sp(4)$ indices are raised and lowered via the matrix $\Omega_{IJ}$.

\section{R-symmetry reduction}\label{R-sym-red-appendix}
In this appendix we describe the R-symmetry reduction of maximal super Yang-Mills from five to four dimensions at the level of the lagrangian.
We begin expanding the kinetic terms under \eqref{Rsymred}. Starting with the field strength:
\begin{equation}
    F_{mn}F^{mn}=F_{\mu\nu}F^{\mu\nu}+2D_\mu\phi D^\mu\phi\,,
\end{equation}
with $\phi=A_4$. This is the kinetic term for a four dimensional vector plus the kinetic term for an extra scalar. 
The kinetic term for the scalars $\Phi_{IJ}$ is: 
\begin{equation}
\begin{split}
    \frac{1}{\alpha}D_m\Phi^iD^m\Phi^i= D_m\Phi_{IJ}D^m\Phi^{IJ}=&D_\mu\Phi_{IJ}D^\mu\Phi^{IJ}
    +H_{[I}^M\Phi_{J]M}H^{[I}_P\Phi^{J]P}
    -iH_{[I}^M\Phi_{J]M}[\phi,\Phi^{IJ}]\\
    &-i[\phi,\Phi_{IJ}]H^{[I}_M\Phi^{J]M}
    -[\phi,\Phi_{IJ}][\phi,\Phi^{IJ}]\,.
\end{split}
\end{equation}
Here we have the kinetic term for the $5$ scalars in four dimensions, a mass term for the $5$d scalars, a cubic and a quartic interaction term between the $5$d scalars and the extra scalar $\phi$.

Now the kinetic term for the fermions:
\begin{equation}
    \bar{\Psi}^{AI}(\gamma^m)_{A}^B D_m\Psi_{BI} = \bar{\Psi}^{AI}(\gamma^\mu)_{A}^B D_\mu\Psi_{BI}
    +H_{I}^{J}\bar{\Psi}^{AI}(\gamma^4)_{A}^{B} \Psi_{BJ} 
    -i\bar{\Psi}^{AI}(\gamma^4)_{A}^{B}[\phi,\Psi_{BI}]\,,
\end{equation}
corresponding to the four dimensional kinetic term, mass terms for the spinors and a Yukawa interaction term between the spinors and the extra scalar coming out of the vector.

However, the fermions here are still in the spinor representation of $SO(1,4)$, whereas we'd like to write these as spinors in four dimensions.
First, let's remind ourselves that the original $10-$dimensional Weyl spinors obey a Majorana condition:
\begin{equation}
    \Psi_+^\mathrm{T}C=\bar\Psi_+\,.
\end{equation}
This condition reads
\begin{equation}\label{4dMajorana}
    \Psi_{AI}C^{AB}\Omega^{IJ}=\Psi^{AI}(\gamma_0)_A^B\delta_I^J=\bar\Psi^{BJ}\,,
\end{equation}
for spinors in $5$ spacetime dimensions.
Now we want to further bring this down to four dimensions. We need make a choice for the $4$d gamma matrices in terms of the five dimensional ones, and in particular we will do that so that the charge conjugation matrix is the same as the five dimensional one. This is possible because both in $4$ and $5$ dimensions the $C$-matrix is antisymmetric. However in five dimensions the $C$-matrix is $C_-$, i.e. the five dimensional gamma matrices have the following symmetry:
\begin{equation}
    C\gamma_\alpha C^{-1}=\gamma_\alpha^\mathrm{T}\qquad\alpha=0,\dots4\,.
\end{equation}
In four dimensions there are in principle two choices for the $C$-matrix but only $C_+$ is compatible with the Majorana condition, so that the symmetry property of gamma matrices in this basis differs in four and five dimensions:
\begin{equation}\label{4d-symmetry-gammas}
    C G_\mu C^{-1}=-G_\mu^\mathrm{T}\qquad\mu=0,\dots3\,.
\end{equation}
On the other hand the chiral matrix $G_5=iG_0\cdots G_3$ in four dimensions has the same symmetry properties as the five dimensional gamma matrices so we can take $G_5=\gamma_4$.
Then for the other four dimensional gamma matrices it's easy to verify that $G_\mu=-i\gamma_\mu\gamma_4$ satisfy \eqref{4d-symmetry-gammas}.

We can further choose to write the four dimensional gamma matrices in the Weyl basis:
\begin{equation}
G_\mu=\begin{pmatrix}
0 & \sigma_\mu\\
\bar{\sigma}_\mu &0
\end{pmatrix}
\qquad\qquad
G_5=\begin{pmatrix}
\mathds{1} & 0\\
0 & -\mathds{1}
\end{pmatrix}\,.
\end{equation}
In this basis the charge conjugation matrix can be written
\begin{equation}
    C=iG_0G_2=-i\begin{pmatrix}
\sigma_2 & 0\\
0 & -\sigma_2
\end{pmatrix}\,.
\end{equation}
The Majorana condition \eqref{4dMajorana} can then be written:
\begin{equation}
    \Psi_{AI}C^{AB}\Omega^{IJ}=i\bar\Psi^{CJ}(G_5)_C^B\,,
\end{equation}
or equivalently
\begin{equation}
    \Psi_{AI}=iC^{-\mathrm{T}}\cdot G_5^{-\mathrm{T}}\cdot G_0^{-\mathrm{T}}\Psi^{*}=-(G_0\cdot G_2\cdot G_5\cdot G_0)_{AC}\cdot\Psi_{CJ}^{*}\Omega_{JI}
    =\begin{pmatrix}
0 &\sigma_2 \\
 \sigma_2 & 0
\end{pmatrix}_{AC}\Psi_{CJ}^{*}\Omega_{JI}\,,
\end{equation}
where we kept the notation $A,...$ for the four dimensional Dirac spinor indices. We can further look at the condition on the projected left and right components of the fermion:
\begin{equation}
\begin{split}
\begin{pmatrix}
0 &\sigma_2 \\
 \sigma_2 & 0
\end{pmatrix}_{AC}((\frac{\mathds{1}\pm G_5}{2})^D_{C}\Psi_{DJ})^{*}\Omega_{JI}
&=(\frac{\mathds{1}\mp G_5}{2})^B_{A}    \begin{pmatrix}
0 &\sigma_2 \\
 \sigma_2 & 0
\end{pmatrix}_{BD}(\Psi_{DJ})^{*}\Omega_{JI}\\
    &=(\frac{\mathds{1}\mp G_5}{2})^D_{A}\Psi_{DI}
    \,,
\end{split}
\end{equation}
So we have:
\begin{equation}
    \Psi_{R/L;I}=\sigma_2\Psi_{L/R;J}^*\Omega_{JI}\,.
\end{equation}
Now the mass term for the fermions reads:
\begin{equation}
\begin{split}
    H_{I}^{J}\bar{\Psi}^{AI}(G_5)_{A}^{B} \Psi_{BJ}&=
    -iH_{I}^{J}\Omega^{KI}\Psi_{LK}\cdot\sigma_2\cdot \Psi_{LJ}-iH_{I}^{J}\Omega^{KI}\Psi_{RK}\cdot\sigma_2\cdot \Psi_{RJ}\\
    &=-iH_{I}^{J}\Omega^{KI}(\Psi_{LK}\cdot\sigma_2\cdot \Psi_{LJ}-\Psi_{LM}^*\cdot\sigma_2\cdot \Psi_{LN}^*\Omega_{MK}\Omega_{NJ})\,,
\end{split}
\end{equation}
in terms of Majorana spinors in four dimensions $\Psi_I=(\chi_I\,\,\,\sigma_2\chi_J^*\Omega_{JI})^T$.

If we write out the scalar mass terms explicitly we find:
\begin{equation}
   \frac{1}{2} H_{[I}^M\Phi_{J]M}H^{[I}_P\Phi^{J]P}=\Phi_{12}\Phi^{12}(m_1+m_2)^2+\Phi_{13}\Phi^{13}(m_1-m_2)^2+\Phi_{24}\Phi^{24}(m_1-m_2)^2+\Phi_{34}\Phi^{34}(m_1+m_2)^2\,,
\end{equation}
whereas for the spinors we simply have:
\begin{equation}
\begin{split}
    H_{I}^{J}\Omega^{KI}\Psi_{LK}\cdot\sigma_2\cdot \Psi_{LJ} =&
    - m_1(\Psi_{L1}\cdot\sigma_2\cdot \Psi_{L4}+\Psi_{L4}\cdot\sigma_2\cdot \Psi_{L1})\\
    &+m_2(\Psi_{L2}\cdot\sigma_2\cdot \Psi_{L3}+\Psi_{L3}\cdot\sigma_2\cdot \Psi_{L2})\,,
\end{split}
\end{equation}
and similarly for the conjugate term.

Overall the lagrangian describes one massless vector $A_\mu$, two massless scalars $\phi,\Phi_{14}$, four massive scalars and four massive Majorana fermions:
\begin{equation}
\begin{split}
    S=\int d^4x\,
    \Tr[&-\frac{1}{4}F_{\mu\nu}F^{\mu\nu}
    -\frac{1}{2}D_\mu\phi D^\mu\phi
    -\frac{\alpha}{2}D_\mu\Phi_{IJ}D^\mu\Phi^{IJ}
    +\frac{i}{2}\bar{\Psi}^{AI}(\gamma^m)_{A}^B D_m\Psi_{BI}\\
    &
    +\frac{1}{2}H_{I}^{J}\Omega^{KI}(\Psi_{LK}\cdot\sigma_2\cdot \Psi_{LJ}-\Psi_{LM}^*\cdot\sigma_2\cdot \Psi_{LN}^*\Omega_{MK}\Omega_{NJ})\\
    &-\frac{\alpha}{2}H_{[I}^M\Phi_{J]M}H^{[I}_P\Phi^{J]P}\\
    &-\frac{1}{2}\bar{\Psi}^{AI} (\Gamma^{j})_{I}^{K}[\Phi_{j},\Psi_{AK}]
    -\frac{1}{2}\bar{\Psi}^{AI}(G_5)_{A}^{B}[\phi,\Psi_{BI}]\\
    &+\frac{\alpha}{2}iH_{[I}^M\Phi_{J]M}[\phi,\Phi^{IJ}]
    +\frac{\alpha}{2}i[\phi,\Phi_{IJ}]H^{[I}_M\Phi^{J]M}\\
    &+\frac{1}{4}[\Phi^j,\Phi^k]^2 
    +\frac{\alpha}{2}[\phi,\Phi_{IJ}][\phi,\Phi^{IJ}])
    ]\,.
\end{split}
\end{equation}

\bibliography{twistor-bib}        
\bibliographystyle{JHEP}  

\end{document}